\documentclass[prd,twocolumn,superscriptaddress,floatfix,nopacs,preprintnumbers,nofootinbib]{revtex4-2}
\usepackage[utf8x]{inputenc}
\usepackage{mathrsfs}
\usepackage{amssymb,bm}
\usepackage{amsmath}
\usepackage{mathtools}
\usepackage{slashed}
\usepackage{graphics}
\usepackage{graphicx}
\usepackage{dsfont}
\usepackage{longtable}
\usepackage{float}

\usepackage{tcolorbox}
\usepackage{lipsum}

\newcommand{\vect}[1]{\boldsymbol{#1}_{\perp}}

\usepackage[caption = false]{subfig}
\usepackage[normalem]{ulem}
\usepackage{xcolor}

\definecolor{lcolor}{rgb}{0.5,0,0}
\definecolor{citcolor}{rgb}{0,0.3,0.0}

\newlength{\mycol}
\usepackage[breaklinks,colorlinks,citecolor=citcolor,urlcolor=blue,linkcolor=lcolor]{hyperref}

\newcommand{\xt}{{\vect{x}}}
\newcommand{\bt}{\vect{b}}
\newcommand{\bti}{\vect{b}^{i}}
\newcommand{\yt}{{\vect{y}}}
\newcommand{\rt}{\vect{r}}
\newcommand{\zt}{{\vect{z}}}

\newcommand{\xpom}{{x_\mathbb{P}}}

\newcommand{\gev}{\ \textrm{GeV}}
\newcommand{\fm}{\ \textrm{fm}}

\newcommand{\M}[1]{\widetilde{\mathcal{M}}_{#1}^{\gamma^* A,V}}

\newcommand{\Mgamma}[1]{\widetilde{\mathcal{M}}_{#1}^{\gamma^* A,\gamma}}

\newcommand{\Mgammastar}[1]{\widetilde{\mathcal{M}}_{#1}^{\gamma^* A,\gamma^*}}

\newcommand{\Lep}[1]{\widetilde{L}_{#1}}

\newcommand{\xbj}{{x_\text{Bj}}}

\newcommand{\as}{\alpha_\mathrm{s}}
\newcommand{\der}{\mathrm{d}}

\newcommand{\qso}{Q_\mathrm{s0}}
\newcommand{\jpsi}{\mathrm{J}/\psi}
\newcommand{\jpsim}{\mathrm{J}/\psi}

\newcommand{\phik}{\phi_{k\Delta}}
\newcommand{\phirb}{\phi_{r b}}

\newcommand{\Tr}{\mathrm{Tr}}

\begin{document}

\author{Heikki M\"{a}ntysaari}
\email{heikki.mantysaari@jyu.fi}
\affiliation{
Department of Physics, University of Jyväskylä,  P.O. Box 35, 40014 University of Jyväskylä, Finland
}
\affiliation{
Helsinki Institute of Physics, P.O. Box 64, 00014 University of Helsinki, Finland
}
\author{Kaushik Roy}
\email{kauroy@mpp.mpg.de}
\affiliation{Max-Planck-Institut f\"{u}r Physik, F\"{o}hringer Ring 6, 80805 M\"{u}nchen, Germany}
\affiliation{Department of Physics and Astronomy, Stony Brook University, Stony Brook, NY 11794, USA}
\affiliation{Physics Department, Brookhaven National Laboratory, Bldg. 510A, Upton, NY 11973, USA}
\author{Farid Salazar}
\email{farid.salazarwong@stonybrook.edu}
\affiliation{Department of Physics and Astronomy, Stony Brook University, Stony Brook, NY 11794, USA}
\affiliation{Physics Department, Brookhaven National Laboratory, Bldg. 510A, Upton, NY 11973, USA}
\affiliation{Center for Frontiers in Nuclear Science (CFNS), Stony Brook University,
Stony Brook, NY 11794, USA}
\author{Bj\"{o}rn Schenke}
\email{bschenke@bnl.gov}
\affiliation{Physics Department, Brookhaven National Laboratory, Bldg. 510A, Upton, NY 11973, USA}

\title{Gluon imaging using azimuthal correlations in diffractive scattering\\ at the Electron-Ion Collider}


\preprint{}

\begin{abstract}
We study coherent diffractive photon and vector meson production in electron-proton and electron-nucleus collisions within the Color Glass Condensate effective field theory. 
We show that electron-photon and electron-vector meson azimuthal angle correlations are sensitive to non-trivial spatial correlations in the gluon distribution of the target, and perform explicit calculations using spatially dependent McLerran-Venugopalan initial color charge configurations coupled to the numerical solution of small $x$ JIMWLK evolution equations.
We compute the cross-section differentially in $Q^2$ and $|t|$ and find sizeable anisotropies in the electron-photon and electron-$\jpsi$ azimuthal correlations ($v_{1,2} \approx 2 - 10 \%$) in electron-proton collisions for the kinematics of the future Electron-Ion Collider. In electron-gold collisions these modulations are found to be significantly smaller ($v_{1,2} <0.1 \%$). We also compute incoherent diffractive production where we find that the azimuthal correlations are sensitive to fluctuations of the gluon distribution in the target.

\end{abstract}

\maketitle

\section{Introduction}

Revealing the internal structure of protons and nuclei is one of the central motivations behind the future Electron-Ion Collider (EIC) in the US~\cite{Accardi:2012qut,Aschenauer:2017jsk,Aidala:2020mzt} and similar projects proposed at CERN~\cite{Agostini:2020fmq} and in China~\cite{Chen:2018wyz}. These high energy deeply inelastic scattering experiments provide access to hadron and nuclear structure at small Bjorken-$x$, where non-linear effects~\cite{Gribov:1984tu,Mueller:1985wy} tame the growth of gluon densities and generate an emergent semi-hard scale known as the saturation momentum $Q_{\mathrm{s}}$. The existence of this scale allows for a weak coupling description of the gluon dynamics in this high density regime of Quantum Chromodynamics (QCD) in a semi-classical effective field theory (EFT) framework known as the Color Glass Condensate (CGC)~\cite{McLerran:1993ni,McLerran:1993ka,McLerran:1994vd,Iancu:2000hn,Iancu:2001ad,Ferreiro:2001qy,Iancu:2003xm,Gelis:2010nm,Kovchegov:2012mbw,Blaizot:2016qgz}.

Proton and nuclear structure in terms of fundamental quark and gluon constituents can be described in terms of parton distribution functions (PDFs). For the proton these are well known from precise structure function measurements at the HERA electron-proton collider~\cite{Abramowicz:2015mha}. In case of nuclei, the partonic content is not as precisely known due to the lack of nuclear DIS experiments in collider kinematics~\cite{Eskola:2016oht}.

To move beyond one dimensional PDFs that describe the parton density as a function of longitudinal momentum fraction, one can define, for instance, the transverse momentum dependent (TMD) parton distributions~\cite{Collins:1981uw,Mulders:2000sh,Meissner:2007rx,Petreska:2018cbf} that also include the dependence on the intrinsic transverse momentum carried by the partons, and describe their orbital angular momentum within the hadron (see e.g.~\cite{Aschenauer:2015ata}). Similarly, Generalized Parton Distributions (GPDs)~\cite{Ji:1996ek,Radyushkin:1997ki,Mueller:1998fv} describe the spatial distribution of partons inside the hadron or nucleus~\cite{Burkardt:2000za, Ralston:2001xs, Diehl:2002he} (see Refs.~\cite{Goeke:2001tz,Diehl:2003ny, Belitsky:2005qn} for reviews on the subject). An ultimate goal in mapping the proton and nuclear structure is to determine the Wigner distribution~\cite{Ji:2003ak,Belitsky:2003nz,Lorce:2011kd} that encodes all partonic quantum information. To access these distributions, differential measurements such as two-particle or two-jet correlations are needed. These measurements have been shown to allow access to the elliptic part of the gluon Wigner distribution at small $x$~\cite{Hatta:2016dxp,Hagiwara:2017fye,Mantysaari:2019csc}, the Weizsäcker-Williams unintegrated gluon distribution~\cite{Dumitru:2018kuw}, multi-gluon correlations~\cite{Mantysaari:2019hkq} and gluon saturation~\cite{Lappi:2012nh,Albacete:2018ruq,Salazar:2019ncp,Kolbe:2020tlq}.

Exclusive processes are powerful probes of partonic structure, since at lowest order in the QCD coupling an exchange of two gluons is necessary, rendering the cross-section approximately proportional to the squared parton (or, at small $x$, gluon) distributions\,\cite{Ryskin:1992ui}. Additionally, as the total momentum transfer can be measured, these processes provide access to the spatial structure of the target, because the impact parameter is the Fourier conjugate to the momentum transfer. Measurements of exclusive vector meson production in (virtual) photon-hadron scattering at HERA~\cite{Aktas:2005xu,Chekanov:2002xi,Chekanov:2002rm,Aktas:2003zi,Chekanov:2009ab,Breitweg:1999jy,Alexa:2013xxa} and in ultra peripheral collisions~\cite{Bertulani:2005ru,Klein:2019qfb} at the LHC~\cite{Aaij:2013jxj,TheALICE:2014dwa,Khachatryan:2016qhq,Acharya:2018jua} have been used to probe non-linear QCD dynamics in nuclei~\cite{Lappi:2013am,Guzey:2013qza} and to
determine the event-by-event fluctuating shape of protons and nuclei, see Refs.~\cite{Mantysaari:2019jhh,Mantysaari:2018zdd,Mantysaari:2017dwh,Mantysaari:2016jaz,Mantysaari:2016ykx,Cepila:2016uku,Cepila:2017nef,Cepila:2018zky,Mitchell:2016jio} and Ref.~\cite{Mantysaari:2020axf} for a review. Similarly, exclusive photon production, or Deeply Virtual Compton Scattering (DVCS) processes, have been measured at HERA~\cite{Adloff:2001cn,Aktas:2005ty,Chekanov:2008vy,Aaron:2007ab} and were shown to be particularly sensitive to the orbital angular momentum of quarks and gluons \cite{Ji:1996ek,Mueller:1998fv,Ji:1996nm,Radyushkin:1997ki}.\\

The study of azimuthal correlations between the produced photon and the scattered electron in DVCS is sensitive to the gluon structure of the target. The first theoretical studies within the dipole framework in the small-x regime have been considered in Ref.~\cite{Hatta:2017cte} featuring $\cos(\phi)$ and $\cos(2\phi)$ modulations; the latter being generated by the elliptic gluon Wigner distribution (see also Ref.~\cite{Zhou:2016rnt} for azimuthal correlations in electron quasi-elastic scattering  on  a  large  nucleus). The authors in Ref.~\cite{Hatta:2017cte} also showed the consistency of the dipole framework with the standard collinear factorization approach. In the present work, we evaluate the azimuthal correlations in DVCS and $J/\psi$ production in the CGC EFT. In addition, we study incoherent diffractive production and demonstrate that at moderate values of $|t|$ the azimuthal correlations are sensitive to event-by-event fluctuations of the spatial gluon distribution in the target.

This paper is organized as follows. First, in Section~\ref{sec:electroproduction}, we review the calculation of exclusive cross-sections, and express the lepton-vector particle azimuthal correlations in terms of the hadronic scattering amplitudes for the production of a vector particle with polarization $\lambda'$ in the scattering of an unpolarized nucleus with a virtual photon with polarization $\lambda$.

In Section~\ref{sec:DVCS-JPsi-amplitude}, we compute these amplitudes for DVCS and exclusive $\jpsi$ production following the CGC EFT momentum space Feynman rules. We begin by considering the off-forward $\gamma^*A \rightarrow \gamma^*A$ amplitude, we work in the dipole picture, in which the virtual photon fluctuates into a quark-antiquark pair, scatters eikonally with the dense gluon field of the nucleus, and then recombines into a virtual photon. We express our results in terms of convolutions of light-cone wave-functions with the impact parameter dependent dipole-target scattering amplitude.  
We then present the explicit expressions for DVCS in Section\,\ref{sec:DVCS_amplitudes} and for exclusive vector meson production in Section\,\ref{sec:vmproduction}. The modifications needed to compute incoherent diffraction are discussed in Section\,\ref{sec:incoherent_xs}.

Our numerical setup is presented in detail in  Section~\ref{sec:setup}. We introduce both a simple impact parameter dependent Golec-Biernat-W\"usthoff (GBW) model~\cite{GolecBiernat:1998js} with angular correlations, and a more realistic CGC dipole computed from the spatially dependent McLerran-Venugopalan (MV)~\cite{McLerran:1993ni,McLerran:1993ka,McLerran:1994vd} initial color charge configurations coupled to small $x$ JIMWLK\footnote{JIMWLK is an acronym for  Jalilian-Marian, Iancu, McLerran, Weigert, Leonidov, Kovner.} evolution~\cite{JalilianMarian:1997jx,JalilianMarian:1997gr,JalilianMarian:1997dw,Iancu:2001ad,Iancu:2000hn,Ferreiro:2001qy,Weigert:2000gi}. Our numerical results with predictions for the azimuthal angle correlations at the EIC are shown in Section~\ref{sec:results}. As a baseline study, we present the numerical results using the GBW model in Section\,\ref{sec:gbw_numerics}. Our numerical predictions for electron-proton and electron-gold collisions from the impact parameter dependent CGC dipole amplitude are shown in Sections\,\ref{sec:CGC_numerics} and \ref{sec:large_nucleus} for different values of virtuality $Q^2$ and momentum transfer $|t|$. 

We also present predictions for the nuclear suppression factor for DVCS and exclusive $\jpsi$ production at $t=0$. In Section\,\ref{sec:incoh_numerics}, we study incoherent diffraction with and without substructure fluctuations. We summarize our main results in Section\,\ref{sec:conclusions} and include multiple appendices (\ref{appendix:polflip}-\ref{appendix:mode_decomposition}) that include supplemental details. 

\begin{figure}[tb] 
    \centering
    \includegraphics[scale=0.85]{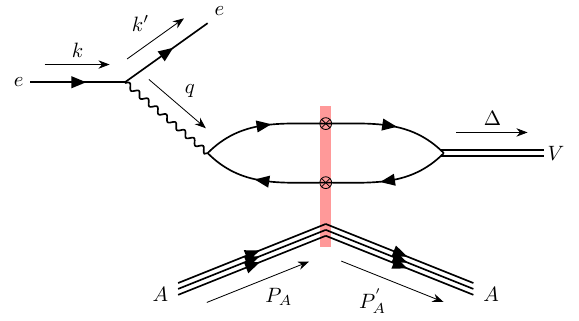}
    \caption{Exclusive electroproduction of a vector particle $V$ at small $x$. The produced particle can be a photon or a vector meson.}
    \label{fig:DVCS_cartoon}
\end{figure}

\section{Exclusive vector particle electroproduction}
\label{sec:electroproduction}
In this section we review general aspects of the computation of the vector particle (V) electroproduction cross-section in the exclusive process
\begin{equation}
e(k)+A(P_A) \rightarrow e(k')+V(\Delta)+A(P'_A) \, , 
\end{equation}
shown in Fig.~\ref{fig:DVCS_cartoon}, where $A$ can be a hadron or nucleus target with initial and final state momenta $P_A$ and $P'_A$, respectively. We begin by specifying the reference frame and defining the kinematic variables of interest (see Table\, \ref{tab:varDef} for a complete list of kinematic quantities). We will express the amplitude squared in the usual way, as a product of lepton and hadron tensors, and then decompose these in a basis spanned by the longitudinal and transverse circular polarization vectors of the virtual photon exchanged in the scattering process. 
The advantage of using this basis is that (besides the lepton  and  hadron  tensors having only 6 independent components, instead of 10) there is a direct connection between the polarization changing and helicity flip amplitudes and the modulations in the azimuthal angle between the electron and vector particle.

We next evaluate the components of the lepton tensor in this basis, and use symmetry arguments to deduce the generic structure of the hadron tensor and the associated amplitude for the hadronic subprocess. Using these elements, we derive a general expression for the differential cross-section of coherent diffractive vector particle electroproduction at small $x$, with matrix elements to be evaluated within the CGC EFT. In particular, the cross-section is differential in the azimuthal angle between the electron and the produced vector particle. The study of modulations of the cross section in this angle is the main goal of this work.

\subsection{Reference frame and kinematics}

We work in a frame in which the virtual photon and nucleon have large $-$ and $+$ light-cone\footnote{We define light-cone coordinates as: $x^{\pm}=\frac{1}{\sqrt{2}}( x^{0} \pm x^{3} )$. Four vectors are expressed as $v=(v^{+},v^{-},\bm{v}_{\perp})$ where $\bm{v}_{\perp}$ is the two dimensional transverse vector with components $v^{1}$ and $v^{2}$. The magnitude of the 2D vector $\bm{v}_{\perp}$ is denoted by $v_{\perp}$. The metric contains the following non-zero entries $g^{+-}=g^{-+}=1$ and $g^{ij}=-\delta^{ij}$ for $i,j=1,2$, so that the scalar product reads $a\cdot b=a^{+}b^{-}+a^{-}b^{+}-\bm{a}_{\perp} \cdot \bm{b}_{\perp}.$}
momentum components, respectively, and have zero transverse momenta:
\begin{align}
    P^\mu = \left(P^+, \frac{m^2_N}{2P^+}, \vect{0} \right) , \label{eq:Pvector} \\
    q^\mu = \left( -\frac{Q^2}{2 q^-}, q^-, \vect{0} \right)\,, \label{eq:qvector}
\end{align}
where $m_N$ is the nucleon mass.
The momentum of the produced vector particle $V$ is:
\begin{align}
    \Delta^\mu = \left( \frac{M^2_V + \vect{\Delta}^2}{2 \Delta^-}, \Delta^-, \vect{\Delta} \right) .
\end{align}
In this frame, the incoming and outgoing electron carry transverse momentum $\vect{k}$ that satisfies the relation
\begin{align}
    \vect{k}^2=\frac{(1-y)}{y^2}Q^2  ,
\end{align}
where $y$ and $Q^2$ are the inelasticity of the collision and virtuality of the exchanged photon, respectively.

Note that requiring the form of Eqs.\,\eqref{eq:Pvector} and \eqref{eq:qvector} does not uniquely define the reference frame, due to the residual azimuthal symmetry. To completely fix the frame, we specify the azimuthal angle for the incoming electron:
\begin{align}
    \vect{k} = (k_\perp \cos \phi_k, k_\perp \sin \phi_k) .
    \label{eq:electron-transverse-momentum}
\end{align}
To characterize the collision, we introduce two invariant quantities $\xpom$ and $t$ (see Table~\ref{tab:varDef}). At high energies, the momentum transfer squared $t$, from the target to the projectile is mostly transverse:
\begin{align}
    t= -\frac{\vect{\Delta}^2+\xpom^2 m^2_N}{1-\xpom} \approx-\vect{\Delta}^2.
\end{align}

Here $\xpom$ describes the longitudinal momentum fraction of the nucleon carried by the effective color neutral \emph{pomeron} exchange in the infinite momentum frame (IMF):
\begin{align}
\label{eq:xpom}
    \xpom = \frac{Q^2 +M_V^2 -t}{W^2 + Q^2 - m^2_N}.
\end{align}

\begin{longtable*}[H]{ll}
\caption{ Kinematic variables \label{tab:varDef} 
} \\ \hline \hline
$P_A \ (P'_A)$ & incoming (outgoing) nucleus four-momentum \\
$P \ (P')$ & incoming (outgoing) nucleon four-momentum \\
$k \ (k')$ & incoming (outgoing) electron four-momentum \\ 
$q=k-k'$ & virtual photon momentum \\
$\Delta$ & outgoing four-momentum of produced vector particle $V$ \\ 
$s=(P+k)^2$ & nucleon-electron system center of momentum squared  \\
$W^2=(P+q)^2$ & nucleon-virtual photon system center of momentum squared squared \\
$m^2_N = P^2$ & nucleon invariant mass squared\\
$M^2_V = \Delta^2$ & invariant mass squared of produced particle $V$\\
$t=(P-P')^2$ & momentum transfer squared\\
$Q^2=-q^2$ & virtuality of exchanged photon \\
$y = \frac{P.q}{P.k}$ & inelasticity: in $\vec{P}=0$ frame the fraction of the electron momentum carried by the virtual photon \\ 
$\xpom = \frac{(P-P').q}{ P.q}$ & Pomeron-$x$: in IMF 
fraction of the nucleon longitudinal momentum carried by the exchanged ``pomeron''   \\
\hline \hline
\end{longtable*}

\subsection{Lepton and hadron tensor decomposition: virtual photon polarization basis}

The amplitude for the production of a vector particle $V$ in electron-nucleus scattering is given by
\begin{align}
    \left \langle P'_A \big| M^{eA,V}_{\lambda'} \big| P_A \right \rangle = & \left[ i e \bar{u}(k') \gamma^\mu u(k) \right] \nonumber \\ &\times \frac{i\Pi_{\mu \alpha}(q)}{Q^2} 
    \left \langle P'_A \big| M^{V,\alpha}_{\lambda'} \big|P_A \right \rangle , \label{eq:AmplitudeVEP}
\end{align}
where $\lambda'$ labels the polarization of the produced vector particle. The nuclear matrix element
$
\left \langle P'_A \big| M^{V,\alpha}_{\lambda'} \big|P_A \right \rangle
$
represents the amplitude for the hadronic subprocess producing a vector particle.

In light-cone gauge\footnote{This is the light-cone condition with respect to the left moving projectile \cite{Roy:2018jxq}. The conventional choice of light-cone gauge $A^{+}=0$,  defined for the right moving target allows for the number density interpretation of PDFs.} $A^- = 0$, the tensor structure of the photon propagator has the form:
\begin{align}
    \Pi^{\mu \alpha} = - g^{\mu \alpha} + \frac{q^\mu n^\alpha + n^\mu q^\alpha}{q^-}, \quad n^\mu = \delta^{\mu +} .
    \label{eq:photon-propagator-num}
\end{align}
By squaring the amplitude, averaging over the spin of the incoming electron, and summing over the spin of the outgoing electron and the polarizations of the particle $V$, we obtain the well-known decomposition for the unpolarized amplitude squared\footnote{We have used here the Dirac equation and the Ward identity $q_{\alpha}X^{\alpha \beta}=q_{\beta} X^{\alpha \beta}=0$.}:
\begin{align}
    \frac{1}{2}  \sum_{\substack{ \rm spins \\ \rm pol \ \lambda'}}  \left| \left \langle P'_A \big| M^{eA,V}_{\lambda'} \big| P_A \right \rangle \right|^2 &= \frac{1}{Q^4} L^{\mu \nu} X_{\mu\nu} \ , \label{eq:HadronLepton1}
\end{align}
where the lepton and hadron tensors are defined by
\begin{align}
    L^{\mu \nu} &= 2 e^2 \left(k'^\mu k^\nu + k^\mu k'^\nu - g^{\mu\nu} \frac{Q^2}{2} \right) \label{eq:LeptonTensor1} , \\
    X^{\mu \nu} &= \sum_{\lambda'}   \left \langle P_A \big| M^{V,\mu\dagger}_{\lambda'} \big|P'_A \right \rangle \left \langle P'_A \big| M^{V,\nu}_{\lambda'} \big|P_A \right \rangle \label{eq:HadronTensor1} .
\end{align}

\subsubsection{Polarization basis}
It is convenient to express the decomposition in Eq.\,\eqref{eq:HadronLepton1} in the basis of the polarization vectors of the virtual photon. In $A^{-}=0$ gauge (and for $\bm{q}_{\perp}=0$) they read
\begin{align}
    \epsilon^\mu(\lambda=0,q) &= \left( \frac{Q}{q^-},0,\vect{0} \right), \nonumber \\
    \epsilon^\mu(\lambda=\pm 1,q) &= \left(0,0,\vect{\epsilon}^\lambda \right). \label{eq:PolVecInPhoton}
\end{align}
with two-dimensional transverse vectors $\vect{\epsilon}^{\pm 1} = \frac{1}{\sqrt{2}} \left(1,\pm i \right)$. Here $\lambda =0 $ refers to the longitudinal polarization, and $\lambda = \pm 1$ denote the two transverse circular polarization states of the virtual photon.
The polarization vectors satisfy the following completeness relation:
\begin{align}
    \Pi_{\mu \alpha}(q) = \sum_{\lambda} (-1)^{\lambda +1} \epsilon^*_\mu(\lambda,q) \epsilon_\alpha(\lambda,q)
    \label{eq:Completeness_Relation} \ ,
\end{align}
where $\Pi_{\mu \alpha}$ was defined for $A^{-}=0$ gauge in Eq.\,\eqref{eq:photon-propagator-num}. We can now insert\footnote{One can think of this as a resolution of the identity in terms of a sum over a complete set of states.}  Eq.\,\eqref{eq:Completeness_Relation} on the r.h.s of   Eq.\,\eqref{eq:HadronLepton1} and apply the Ward identity ($q_{\alpha}X^{\alpha \beta}=q_{\beta} X^{\alpha \beta}=0$) to obtain an alternative expression for the unpolarized amplitude squared as
\begin{align}
   \frac{1}{2}  \sum_{\substack{ \rm spins \\ \rm pol \ \lambda'}} \left| \left \langle P'_A \big| M^{eA,V}_{\lambda'} \big| P_A \right \rangle \right|^2 = \frac{1}{Q^4} \sum_{\lambda \bar{\lambda}} (-1)^{\lambda+\bar{\lambda}} L_{\lambda \bar{\lambda}} X_{\lambda \bar{\lambda}} \ , \label{eq:HadronLepton2}
\end{align}
where we define the lepton and hadron tensor in the polarization basis as \footnote{From now on, unless otherwise specified, we will only work with the lepton and hadron tensor in the polarization basis.}
\begin{align}
    L_{\lambda \bar{\lambda}} = L^{\mu \nu} \epsilon_\mu(\lambda,q) \epsilon^*_\nu(\bar{\lambda},q) \label{eq:LeptonTensor2} \ , \\
    X_{\lambda \bar{\lambda}} = X^{\alpha \beta} \epsilon^*_\alpha(\lambda,q) \epsilon_\beta(\bar{\lambda},q) \ . \label{eq:HadronTensor2}
\end{align}

An immediate advantage of this relation is that it reduces the number of independent components of the lepton and hadron tensors from 10 to 6, which follows from the application of the Ward identities in deriving Eq.\,\eqref{eq:HadronLepton2}. In the remainder of the section we will see that the lepton-hadron tensor decomposition in this basis has a clear connection to the azimuthal angle correlations between the electron and the vector particle $V$.

\subsubsection{Lepton tensor in the polarization basis} \label{sec:lepton-tensor-pol-basis}

The lepton tensor in the polarization basis is obtained from Eqs.\,\eqref{eq:LeptonTensor1} and \eqref{eq:LeptonTensor2}:  
\begin{align}
    L_{\lambda \bar{\lambda}} = 2 e^2 \epsilon_\mu(\lambda,q) \epsilon^*_\nu(\bar{\lambda},q) \left(k'^\mu k^\nu + k^\mu k'^\nu - g^{\mu\nu} \frac{Q^2}{2} \right) \,.
\end{align}
In our choice of reference frame and because of the choice of the circularly polarized basis, the lepton tensor acquires a phase dependence on the azimuthal angle of the electron, defined in Eq.\,\eqref{eq:electron-transverse-momentum}:
\begin{align}
    L_{\lambda \bar{\lambda}} &= e^2\  \widetilde{L}_{\lambda \bar{\lambda}} \ e^{i(\lambda-\bar{\lambda}) \phi_k} \ ,
\label{eq:LeptonTensor3}
\end{align}
where the elements  $\widetilde{L}_{\lambda \bar{\lambda}}$ are given by
\begin{align}
    \Lep{00} &= \frac{4 Q^2 (1-y)}{y^2}, \ \ \ \ 
    \Lep{\pm 1 \pm 1} = \frac{Q^2 \left[1 + (1-y)^2 \right]}{y^2} \, , \nonumber  \\
    \Lep{0 \pm 1} &= \Lep{\pm 1 0} = -\frac{\sqrt{2} Q^2 (2-y)\sqrt{1-y}}{y^2} \, ,  \nonumber \\
    \Lep{\pm 1 \mp 1} &= \frac{2Q^2(1-y)}{y^2} \, .
\end{align}
Note that these phase structures depend on the difference between the polarization states of the virtual photon in the amplitude and the complex conjugate amplitude.

\subsubsection{Hadron tensor in the polarization basis}
\label{sec:Hadron_tensor}

The hadron tensor in the polarization basis follows from Eqs.\,\eqref{eq:HadronTensor1} and \eqref{eq:HadronTensor2}:
\begin{align}
    X_{\lambda \bar{\lambda}} = \sum_{\lambda'} \left \langle P_A \big| M_{\lambda, \lambda'}^{\gamma^* A, V \dagger} \big| P'_A \right \rangle \left \langle P'_A \big| M_{\bar{\lambda}, \lambda'}^{\gamma^* A, V} \big| P_A \right \rangle  , \label{eq:HadronTensor3} 
\end{align}
where we define
\begin{align}
    \left \langle P_A \big| M_{\lambda, \lambda'}^{\gamma^* A,V} \big| P'_A \right \rangle  = \epsilon_{\alpha}(\lambda,q)   \left \langle P_A \big| M^{V,\alpha}_{\lambda'} \big| P'_A \right \rangle   . \label{eq:Amplitude_Vproduction}
\end{align}
The expression in Eq.\,\eqref{eq:Amplitude_Vproduction} is the amplitude for the production of a vector particle $V$ with polarization $\lambda'$ in the collision of a nucleus with mass number $A$ and a virtual photon with polarization $\lambda$.

In order to obtain an explicit expression for the hadron tensor, we must compute the 9 amplitudes corresponding to the different polarization states $(\lambda,\lambda')$ of the virtual photon and the produced vector particle. This will be the subject of the next section.

However, it is possible to deduce, without explicit computation, the phase structure of the hadron tensor $X_{\lambda \bar{\lambda}}$. Requiring the azimuthal rotational invariance of the decomposition in Eq.\,\eqref{eq:HadronLepton2} and noticing the phase structure of the lepton tensor in  Eq.\,\eqref{eq:LeptonTensor3}, we conclude that the hadron tensor in the polarization basis decomposition must have the form:
\begin{align}
    X_{\lambda \bar{\lambda}} &= (q^-)^2 e^{-i(\lambda - \bar{\lambda)} \phi_{\Delta}} \widetilde{X}_{\lambda \bar{\lambda}} \ , \label{eq:HadronTensor4}
\end{align}
where $\phi_{\Delta}$ is the azimuthal angle of the produced vector particle.
For convenience we have also factored $(q^-)^2$. 

Furthermore, by requiring the phase structure of Eq.\,\eqref{eq:HadronTensor4}, one can deduce from Eq.\,\eqref{eq:HadronTensor3} that the amplitude must have the form
\begin{align}
    \left \langle P'_A \big| M_{\lambda, \lambda'}^{\gamma^* A,V} \big| P_A \right \rangle &= q^- e^{i(\lambda - \lambda') \phi_{\Delta}} \left \langle P'_A \big| \widetilde{M}_{\lambda, \lambda'}^{\gamma^* A,V} \big| P_A \right \rangle \ . 
\label{eq:HadronAmplitude1}
\end{align}
such that
\begin{align}
    \tilde{X}_{\lambda \bar{\lambda}} = \sum_{\lambda'} \left \langle P_A \big| \widetilde{M}_{\lambda, \lambda'}^{\gamma^* A,V\dagger} \big| P'_A \right \rangle \left \langle P'_A \big| \widetilde{M}_{\bar{\lambda}, \lambda'}^{\gamma^* A,V} \big| P_A \right \rangle
\end{align}
The matrix elements on the right-hand side of Eq.\,\eqref{eq:HadronAmplitude1} are independent of $\phi_\Delta$. 

\subsection{Cross-section, CGC average and azimuthal correlations}

\label{sec:Coherent}

We will now use the results derived in the previous subsections to obtain a general expression for the unpolarized cross-section for the electroproduction of a vector particle $V$, and in particular the azimuthal angle correlations between the electron and the vector particle.

Our discussion thus far has been very general in terms of the nuclear matrix elements. In the coming section, we will compute these matrix elements in the CGC EFT. In the CGC one describes the target state $\vert P_{A} \rangle $ in terms of large $x$ stochastic classical color sources, characterized by a charge density $\rho_{A}$, which generate the small $x$ dynamical color fields. For a given operator $O$, one calculates the observable quantity in the CGC EFT via a double averaging procedure:
\begin{enumerate}
    \item Compute the quantum expectation value of the operator $O$ for a given configuration of color sources $\rho_A$: 
    \begin{equation}
        \mathcal{O}[\rho_A] = \left \langle O \right \rangle_{\rho_A} .
    \end{equation} 
    We will see in the next section that this amounts to using modified momentum space Feynman rules that embody the effects of strongly correlated gluons inside the target at small $x$.
    
    \item Perform a classical statistical average over the different color source configurations using a gauge invariant weight functional $W_Y[\rho_A]$, at a certain rapidity scale $Y$. The resulting quantity is the CGC average of the observable, formally defined as~\cite{Iancu:2000hn,Ferreiro:2001qy}
    \begin{equation}
    \big \langle \mathcal{O} \big \rangle_{Y}= \int [\mathcal{D}\rho_{A}] \, W_{Y}[\rho_{A}]  \mathcal{O} [\rho_{A}] . 
    \label{eq:CGC-averaged-amplitude}
\end{equation}
The rapidity scale $Y=\ln (1/x)$ controls the separation of the small $x$ and large $x$ degrees of freedom, where $x$ is typically chosen to be the  longitudinal momentum fraction probed by the process (see also the discussion in Ref.~\cite{Ducloue:2019ezk}); we choose $x=\xpom$.
\end{enumerate}

Using the following normalization for the initial and final states of the target
\begin{multline}
\label{eq:normalization}
    \big \langle P'_A | P_A \big \rangle = (2 P_A^+) (2 \pi)^3 \\
    \times \delta(P'^{+}_A-P^{+}_A) \delta^{(2)}(\vect{P'_{A}}-\vect{P_{A}}) \, ,
\end{multline}
it is easy to identify the correspondence between the nuclear matrix elements entering  the amplitude squared in Eq.\,\eqref{eq:HadronLepton2} (see also Eq.\,\eqref{eq:HadronAmplitude1}) and the CGC averaged amplitude\footnote{Note the use of $\mathcal{M}$ instead of $M$ to denote the counterparts of the general amplitude definitions in the CGC EFT.}:
\begin{align}
    \frac{\left \langle P'_A \big| M^{\gamma^* A,V}_{\lambda,\lambda'} \big| P_A \right \rangle }{\big \langle P'_A \big| P_A \big \rangle} \leftrightarrow \left \langle \mathcal{M}^{\gamma^* A,V}_{\lambda, \lambda'} \right \rangle_Y \, .
    \label{eq:CGC-average-hadron-amplitude}
\end{align}
Note that this prescription~\cite{Buchmuller:1997eb,Hebecker:1997gp,Buchmuller:1998jv,Kovchegov:1999kx} of performing the CGC averaging at the level of the amplitude is specific to the case when the target remains intact after the scattering.

With these definitions in mind, it can be shown
that the differential cross-section is given by 
\begin{align}
    \mathrm{d} \sigma^{e A \rightarrow e A V} \! &= 
   \frac{1}{2} \!\! \sum_{\substack{ \rm spins \\ \rm pol \ \lambda'}} \left|\Big \langle   \mathcal{M}_{\lambda'}^{e A,V} \Big \rangle_{Y} \right|^2 \! 2\pi \delta(q^- - \Delta^-) \frac{\widetilde{\mathrm{d} k'} \widetilde{\mathrm{d} \Delta} }{F} \, , 
\end{align}
where $\Big \langle   \mathcal{M}_{\lambda'}^{e A,V} \Big \rangle_{Y}$ follows from Eqs.\,\eqref{eq:AmplitudeVEP},\eqref{eq:Completeness_Relation},\eqref{eq:Amplitude_Vproduction},\eqref{eq:HadronAmplitude1} and \eqref{eq:CGC-average-hadron-amplitude} as
\begin{align}
  \Big \langle   \mathcal{M}_{\lambda'}^{e \, A,V} \Big \rangle_{Y}&= -\frac{e \overline{u}(k') \gamma^{\mu} u(k)}{Q^2} \,\sum_{\lambda} (-1)^{\lambda+1} \epsilon^{*}_{\mu}(\lambda,q) \nonumber \\
  & \times q^{-}e^{i(\lambda-\lambda')\phi_{\Delta}} \Big \langle   \widetilde{\mathcal{M}}_{\lambda,\lambda'}^{\gamma^{*} \, A,V} \Big \rangle_{Y} \, .
\end{align}
Above $F=2k^- $ is the electron flux factor, $\widetilde{\mathrm{d} k'}$ and $\widetilde{\mathrm{d} \Delta}$ are the phase space measures of the scattered electron, and produced vector particle, respectively, defined as
\begin{align}
    \widetilde{\der k'} &= \frac{\der k'^-  \der ^2 \vect{k}'}{(2\pi)^3 2 k'^-} , \nonumber \\ \widetilde{\der \Delta} &= \frac{\der \Delta^-  \der ^2 \vect{\Delta}}{(2\pi)^3 2 \Delta^-} \, .
\end{align}
Recall that in our choice of reference frame $\vect{k}=\vect{k}'$.

It is useful to express the phase space measure of the electron in terms of the DIS invariants:
\begin{align}
    \widetilde{\der k'} = \frac{y}{\xpom} \frac{\der \xpom \der Q^2 \der \phi_k}{4 (2 \pi)^3} \, \label{eq:electron-ps} . 
\end{align}
Typically, this expression is given in terms of $\xbj$. Here we used $d \ln \xpom = d \ln \xbj$.

The delta function $(2\pi) \delta(q^- - \Delta^-)$ is a manifestation of the eikonal approximation used to describe the high-energy scattering process.

Integrating the cross-section over the overall azimuthal angle and $\Delta^-$ using the delta function, we obtain
\begin{align}
    \frac{\mathrm{d} \sigma^{e A \rightarrow e A V}}{\mathrm{d} |t| \mathrm{d} \phik \mathrm{d} \xpom \mathrm{d} Q^2 } =&  \frac{y^2}{ \xpom} \frac{1}{(4 \pi)^4 (q^-)^2}  \frac{1}{2} \!\! \sum_{\substack{ \rm spins \\ \rm pol \ \lambda'}} \left|\Big \langle   \mathcal{M}_{\lambda'}^{e A,V} \Big \rangle_{Y} \right|^2,
    \label{eq:diff-CS-VCS-1}
\end{align}
where $\phik = \phi_k - \phi_\Delta$ is the azimuthal angle of the produced particle $V$ with respect to the scattered electron. Using the results for the lepton tensor in the polarization basis derived in Section~\ref{sec:lepton-tensor-pol-basis}, we finally obtain the differential cross-section for exclusive vector particle production correlated with the scattered electron (or electron plane):
\begin{widetext}
\begin{align}
   & \frac{\mathrm{d} \sigma^{e A \rightarrow e A V}}{\der \xpom \der Q^2 \der |t| \der \phi_{k\Delta}} = \frac{\alpha_{em}}{32 \pi^3 Q^2 \xpom} \sum_{\lambda'=0,\pm 1} \left \{ (1-y)  \left|\left\langle \widetilde{\mathcal{M}}^{\gamma^*A,V}_{0,\lambda'}\right \rangle_Y \right|^2 + \frac{1}{4} \left[ 1 + (1-y)^2 \right] \sum_{\lambda=\pm 1} \left| \left \langle \widetilde{\mathcal{M}}^{\gamma^*A,V}_{\lambda,\lambda'} \right \rangle_Y \right|^2 \right. \nonumber \\
    & \left. +  \frac{(2-y)\sqrt{2-2y}}{2}  \!\! \sum_{\lambda=\pm 1} \mathrm{Re}  \left(\left\langle\widetilde{\mathcal{M}}^{\gamma^*A,V}_{0,\lambda'} \right \rangle_Y \! \left \langle \widetilde{\mathcal{M}}^{\gamma^*A,V}_{\lambda,\lambda'} \right \rangle_Y^*  \right) \cos( \phi_{k\Delta}) \! + \!  (1-y) \mathrm{Re} \left( \left \langle \widetilde{\mathcal{M}}^{\gamma^*A,V}_{+1,\lambda'} \right \rangle_Y  \! \left \langle \widetilde{\mathcal{M}}^{\gamma^*A,V}_{-1,\lambda'} \right \rangle_Y^* \right) \cos(2 \phi_{k\Delta})  \! \right \}
    \label{eq:xs}
\end{align}
\end{widetext}
Here
$\alpha_{\rm em}$ is the QED fine structure constant and `$\rm{Re}$' stands for the real part of the products of the CGC averaged amplitudes in the second and third lines. The explicit expressions for $\langle \M{\lambda, \lambda'} \rangle_Y$ will be obtained using the CGC EFT in Section~\ref{sec:DVCS-JPsi-amplitude}.

The expression for the differential cross-section above has been previously obtained at small $x$ for DVCS in \cite{Hatta:2016dxp}, and a similar expression for vector meson production in terms of its spin density matrix and helicity amplitudes has been known \cite{Schilling:1973ag}.

\section{DVCS and $\jpsi$ production at small $x$ in the CGC}
\label{sec:DVCS-JPsi-amplitude}
In this section we compute the amplitudes $\M{\lambda,\lambda'}$ for DVCS and for $\jpsi$ production in virtual photon-nucleus collisions at small $x$ within the CGC EFT.

\begin{figure}[!htbp]
    \centering
    \includegraphics{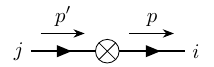}
    \caption{Dressed quark propagator in the CGC EFT. The effective vertex represented by a crossed dot embodies all possible scatterings with the dense target including the case of ``no scattering". Color indices $i$ and $j$ are in the fundamental representation of SU($N_{\mathrm{c}}$). \label{fig:dressed-quark-propagator}}
    \label{fig:my_label}
\end{figure}

It is convenient to visualize the electroproduction of a particle $V$ at small $x$ in the dipole picture, in which the virtual photon $\gamma^*$ fluctuates into a color neutral quark$-$antiquark dipole $q\bar{q}$, which subsequently scatters eikonally with the dense gluon field of the nucleus, and finally recombines into the observed final state particle $V=\gamma, \jpsim$ (see Fig.\,\ref{fig:DVCS_cartoon}). In the CGC EFT, the eikonal interactions (parametrically of $\mathcal{O}$(1) in the QCD coupling) with the strong background classical color field are resummed into a ``dressed" quark propagator, which is diagrammatically represented in Fig.~\ref{fig:dressed-quark-propagator}, and can be written in momentum space as
\begin{equation}
    S_{ij}(p,p')=S^{0}_{ik}(p) \mathcal{T}_{kl}(p,p') S^{0}_{lj}(p') \, ,
    \label{eq:dressed-quark-prop}
\end{equation}
where the free Feynman propagator for a quark with flavor $f$ and momentum $p$ is given by
\begin{align}
    S^0_{ij}(p) = \frac{i (\slashed{p}+m_f)}{p^2-m_f^2 + i \epsilon} \delta_{ij} \ .
\end{align}
Here $m_f$ is the mass of the quark, and $i,j$ denote the color indices in the fundamental representation of SU$(N_{\mathrm{c}})$.
The effective vertex for this interaction has the following expression in momentum space \cite{McLerran:1998nk,Balitsky:2001mr}:
\begin{align}
    \mathcal{T}_{ij}(p,p') &=  (2\pi) \delta(p^--p'^-) \gamma^- \text{sign}(p^-) \nonumber \\
    &\times \int \der^2 \vect{z} e^{-i (\vect{p}-\vect{p'})\cdot \vect{z}} V_{ij}^{\text{sign}(p^-)}(\vect{z}) \ ,
    \label{eq:effective-vertex}
\end{align}
with light-like Wilson lines defined as 
\begin{align}
\label{eq:wline}
    V_{ij} (\vect{z}) = \mathrm{P}_{-} \Big\{ \exp \left( -\ ig \!\! \int_{-\infty}^\infty  \!\! \der z^- \ \frac{\rho^a_A (z^-,\vect{z})}{\nabla^2_\perp  }  \ t_{ij}^a \right) \Big\} ,
\end{align}
where $\rho_{A}(z^{-},\bm{z}_{\perp})$ is the charge density of the large $x$ color sources, $t^a$'s represent generators of SU$(N_{\mathrm{c}})$ in the fundamental representation and $\mathrm{P}_{-}$ denotes the path ordering of the exponential along the $-$ light-cone direction. The superscript $\mathrm{sign}(p^-)$ in Eq.\,\eqref{eq:effective-vertex} denotes matrix exponentiation: $V^{+1}(\vect{z}) = V(\vect{z})$ and $V^{-1}(\vect{z}) = V^\dagger(\vect{z})$, where the latter follows from the unitarity of $V(\vect{z})$.
The eikonal nature of the interaction is manifest in the conservation of the longitudinal momentum $p^-$ and the diagonal nature of the effective vertex in transverse coordinate space.\\

\begin{figure}[htbp]
    \centering
    \includegraphics[scale=1.2]{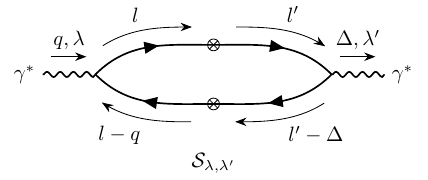}
    \caption{Leading order (in the QCD coupling $\alpha_{\mathrm{s}}$) Feynman diagram for the off-forward scattering of a virtual photon with a nucleus. The crossed circles indicate the dressed quark and antiquark propagators as defined in Eq.\,\eqref{eq:dressed-quark-prop} with the effective vertex given by Eq.\,\eqref{eq:effective-vertex}. \label{VCS_diagram}} 
\end{figure}

We start by computing the off-forward virtual photon-nucleus amplitude  $\mathcal{M}^{\gamma^*A,\gamma^*}_{\lambda ,\lambda'}$ (see Fig.~\ref{VCS_diagram}) and note that:
\begin{enumerate}
\item We recover the DVCS process by taking the virtuality of the final state photon equal to zero.
\item We obtain the results for $\jpsi$ production by substituting the light-cone wave-function of the final state photon by that of the $\jpsi$ (following the procedure in \cite{Kowalski:2006hc}).
\end{enumerate}

\subsection{The off-forward $\gamma^* A \rightarrow \gamma^* A$ Amplitude}
\label{sec:off_forward_amp}
Let the $4$-momentum vector of the outgoing virtual photon be
\begin{align}
    \Delta^\mu = \left( \frac{-Q'^2 + \vect{\Delta}^2}{2 \Delta^-}, \Delta^-, \vect{\Delta} \right) ,
\end{align}
where we define the  $Q'^2 = - \Delta^2$. Recall that in our choice of frame, the virtual photon has zero transverse momentum $\vect{q}=0$, while the outgoing virtual photon has non-zero transverse momentum $\vect{\Delta}$. 

The polarization vectors for the outgoing virtual photon are then:
\begin{align}
    \epsilon^\mu(\lambda=0,\Delta) &= \left( \frac{Q'}{\Delta^-},0,\vect{0} \right), \nonumber \\
    \epsilon^\mu(\lambda=\pm 1,\Delta) &= \left(\frac{\vect{\epsilon}^\lambda \cdot \vect{\Delta}}{\Delta^-},0,\vect{\epsilon}^\lambda \right). \label{eq:PolVecOutPhoton}
\end{align}

The momentum space expression for the virtual photon-nucleus scattering matrix at small $x$ can be written as
\begin{multline}
    \mathcal{S}_{\lambda ,\lambda'}[\rho_A] =(e q_f)^2  \int_{l,l'} \Tr[S^0(l) \slashed{\epsilon}(\lambda,q) S^0(l-q)  \\
    \times \mathcal{T}(l-q,l'-\Delta) S^0(l'-\Delta)  \slashed{\epsilon}^*(\lambda',\Delta) S^0(l') \mathcal{T}(l',l)  ] , \label{eq:Scattering_amplitude}
\end{multline}
where $\lambda$ and $\lambda'$ are the polarizations of the initial and final state photon. The trace $\Tr$ is performed both over spinor and color indices, and $q_f$ denotes the fractional charge of the quark in the loop.

In writing Eq.\,\eqref{eq:Scattering_amplitude} we introduced the short-hands:
\begin{align}
    \int_{l} = \int_{\vect{l}} \int_{l^+} \int_{l^-}, \quad \int_{\vect{l}} = \int \frac{\der^2 \vect{l}}{(2\pi)^2}, \quad \int_{l^\pm} = \int \frac{\der l^{\pm}}{2\pi} .
\end{align}

The amplitude $\mathcal{M}^{\gamma^*A,\gamma^*}_{\lambda ,\lambda'}$ is obtained by subtracting the non-scattering contribution $(\rho^a_A=0)$ and factoring out an overall longitudinal momentum conserving delta function: \begin{align}
(2 \pi) \delta(q^- - \Delta^-) \mathcal{M}^{\gamma^*A,\gamma^*}_{\lambda, \lambda'} = \mathcal{S}_{\lambda, \lambda'}[0] - \mathcal{S}_{\lambda, \lambda'}[\rho^a_A] \ .   
\end{align}\\\\\\
The amplitude can be written as follows: 
\begin{align}
    \mathcal{M}^{\gamma^*A,\gamma^*}_{\lambda ,\lambda'} =& (2q^-) N_c \int_{\vect{b}} \!\!\!\! e^{-i \vect{\Delta} \cdot \vect{b}}  \!\! \int_{\vect{r}} \!\!\!\! \mathcal{D}(\vect{r},\vect{b}) \   \mathcal{A}_{\lambda ,\lambda'}(\vect{r}) , \label{eq:AVCS}
\end{align}
where we introduce the dipole and impact parameter vectors:
\begin{align}
    \vect{r} = \vect{x}-\vect{y}, \quad \vect{b} = \frac{1}{2} (\vect{x}+\vect{y}) .
\end{align}\\
In writing Eq.\,\eqref{eq:AVCS} we used the short-hands:
\begin{align}
    \int_{\vect{b}} = \int \der^2 \vect{b} , \quad \int_{\vect{r}} = \int \der^2 \vect{r} .
\end{align}
The impact parameter dependent Wilson line operator in Eq.\,\eqref{eq:AVCS} is given by
\begin{align}
\label{eq:dipole_tr}
    \mathcal{D}(\vect{r},\vect{b}) &= 1- \frac{1}{N_c} \Tr \left[ V\left(\vect{b} + \frac{\vect{r}}{2} \right) V^\dagger \left( \vect{b} - \frac{\vect{r}}{2} \right)  \right],
\end{align}
and the color-independent sub-amplitude reads
\begin{widetext}
\begin{align}
    \mathcal{A}_{\lambda ,\lambda'}(\vect{r}) =- 2\pi \left(e q_f \right)^2 \int_{l,l'} \frac{(2q^-) \delta(l^- - l'^-)  e^{i (\vect{l}-\vect{l}' + \frac{1}{2}\vect{\Delta}) \cdot \vect{r}} A_{\lambda, \lambda'}(l,l')}{\left[ l^2 -m_f^2 + i \epsilon \right]\left[ (l-q)^2 -m_f^2+ i \epsilon \right] \left[ l'^2  -m_f^2 + i \epsilon \right] \left[ (l'-\Delta)^2 -m_f^2 + i \epsilon \right]} \ , \label{NfullVCS}
\end{align}
\end{widetext}
with Dirac structure
\begin{multline}
    A_{\lambda, \lambda'}(l,l') = \frac{1}{(2q^-)^2} \Tr[ (\slashed{l}+m_f)\slashed{\epsilon}(\lambda,q)(\slashed{l}-\slashed{q} +m_f)\\ \\
    \times \gamma^- (\slashed{l}' -\slashed{\Delta} +m_f) \slashed{\epsilon}^*(\lambda',\Delta)(\slashed{l}' +m_f)\gamma^- ] \ . \label{NVCS}
\end{multline}

Performing the $l^-$ integration with the help of the delta function, and the $l^+$ and $l'^+$ via contour integration (see Appendix\,\ref{sec:UsefulIntegrals}), we obtain the simpler expression:
\begin{align}
    \mathcal{A}_{\lambda, \lambda'}(\vect{r})\! =\! (e q_f)^{2} \!\! \int_{z} \! \int_{\vect{l},\vect{l}'}   \!\!\!\!\!\!\!\!\!\!\! \frac{e^{i (\vect{l} -\vect{l}' +\frac{1}{2} \vect{\Delta}) \cdot \vect{r} }  \! A_{\lambda, \lambda'}(\vect{l},\vect{l}',z)}{N_1(\vect{l}) N_2(\vect{l}')} , \label{eq:Alambdalamda'}
\end{align}
where we introduce the light-cone longitudinal momentum fraction $z=l'^-/q^-$, and the denominators:
\begin{align}
    N_1(\vect{l}) &=  z\bar{z} Q^2 + m^2_f + \vect{l}^2 , \nonumber \\
    N_2(\vect{l}') &= z\bar{z} Q'^2 + m^2_f+ \left(\vect{l}' - z \vect{\Delta} \right)^2\,,
\end{align}
where we define $\bar{z}=1-z$.

Furthermore, following~\cite{Kowalski:2006hc} we adopted the normalization \footnote{Note that the integration over the light-cone fraction $z$ is restricted to values between 0 and 1; otherwise, the result of the $l^+$ and $l'^+$ is identically zero due to the location of the poles (See Appendix\,\ref{sec:UsefulIntegrals}).}:
\begin{align}
    \int_{z} = \int_0^1 \frac{\der z}{4 \pi} \, .
\end{align}
The result of the transverse integration over $\vect{l}$ and $\vect{l}'$ will depend on the Dirac structure $A_{\lambda,\lambda'}$.

Let us consider the case $\lambda=+1$, $\lambda'=-1$, where the evaluation of the Dirac structure results in (see Appendix\,\ref{sec:DiracStructure}):
\begin{align}
    A_{+1,-1}(\vect{l},\vect{l}',z) = -8 z\bar{z} \vect{l}^i (\vect{l}'^m - z \vect{\Delta}^m) \vect{\epsilon}^{+1,i} \vect{\epsilon}^{-1,m*} \ . \label{eq:A-Dirac-p-m}
\end{align}
Inserting Eq.\,\eqref{eq:A-Dirac-p-m} into Eq.\,\eqref{eq:Alambdalamda'}, and evaluating the corresponding transverse integrals (See Appendix~\ref{sec:UsefulIntegrals}) we find the sub-amplitude
\begin{multline}
    \mathcal{A}_{+1,-1}(\vect{r}) = -8 \left(\frac{ e q_f}{2 \pi}\right)^2  \left[ \int_{z}  e^{-i  \left(\frac{z-\bar{z}}{2}\right) \vect{\Delta}  \cdot \vect{r}}  z \bar{z} \frac{\varepsilon_f \vect{r}^i}{r_\perp} \right. \\
    \times \left. K_1(\varepsilon_f  r_\perp) \frac{\varepsilon'_f \vect{r}^m}{r_\perp} K_1(\varepsilon'_f r_\perp) \right] \vect{\epsilon}^{+1,i} \vect{\epsilon}^{-1*,m} , \label{eq:eq:A-sub-p-m_0}
\end{multline}
where $\varepsilon^{2}_f = z \bar{z}\ Q^{2} + m_f^2$, $\varepsilon'^{2}_f = z \bar{z}\ Q'^{2} + m_f^2$ and $K_{1}$ is a modified Bessel function of the second kind. Using the contraction $\frac{\vect{r}^i \vect{\epsilon}^{\lambda,i}}{r_\perp} = \frac{1}{\sqrt{2}} e^{i \lambda \phi_r}$, we obtain
\begin{multline}
    \mathcal{A}_{+1,-1}(\vect{r}) = -4 \left(\frac{ e q_f}{2 \pi}\right)^2  \int_{z}  e^{-i  \left(\frac{z-\bar{z}}{2}\right) \vect{\Delta}  \cdot \vect{r}} e^{2 i  \phi_{r}} z \bar{z} \\
    \times \varepsilon_f  K_1(\varepsilon_f  r_\perp) \varepsilon'_f K_1(\varepsilon'_f r_\perp) \ .
    \label{eq:A-sub-p-m}
\end{multline}
Finally inserting Eq.\,\eqref{eq:A-sub-p-m} into Eq.\,\eqref{eq:AVCS} and factoring out the corresponding phase in $\phi_\Delta$ we obtain the amplitude
\begin{multline}
    \mathcal{M}^{\gamma^*A,\gamma^*}_{+1,-1} = -8 (q^-) N_c \left( \frac{e q_f}{2\pi} \right)^2 e^{2 i \phi_{\Delta}} \Bigg[ \int_{\vect{b}} \!\!\! e^{-i \vect{\Delta} \cdot \vect{b}}  \\ 
    \times \left.  \int_{\vect{r}}  \!\!\! \mathcal{D}(\vect{r},\vect{b})  \int_{z} e^{-i  \left(\frac{z-\bar{z}}{2}\right) \vect{\Delta}  \cdot \vect{r}}  e^{2 i  \phi_{r \Delta} }  z \bar{z} \right. \\
    \times  \varepsilon_f  K_1(\varepsilon_f  r_\perp) \varepsilon'_f K_1(\varepsilon'_f r_\perp) \Bigg] \,, \label{eq:M-p-m}
\end{multline}
where $\phi_{r \Delta} = \phi_r - \phi_{\Delta}$.

Note that this amplitude has the phase factor structure deduced in Section\,\ref{sec:Hadron_tensor}.

The calculations for the remaining amplitudes corresponding to the other polarizations states follow a similar pattern. Factoring out the corresponding phases in $\phi_\Delta$ and a factor of $q^-$, we find explicit expressions for  $ \Mgammastar{\lambda,\lambda'} $ , introduced in Eq.\,\eqref{eq:HadronAmplitude1}:
\begin{widetext}
Helicity preserving amplitudes\footnote{Note that in the forward limit $\vect{\Delta} = 0$, these amplitudes coincide with inclusive DIS cross-section as expected from the optical theorem.}:
\begin{align}
    \Mgammastar{0,0} &= \frac{16 N_c e^2 q_f^2}{(2\pi)^2} \int_{\vect{b}} \!\!\! e^{-i \vect{\Delta} \cdot \vect{b}} \int_{\vect{r}}  \!\!\! \mathcal{D}(\vect{r},\vect{b}) \int_{z} e^{-i  \vect{\delta} \cdot \vect{r}} z^2 \bar{z}^2 \ Q  K_0(\varepsilon_f  r_\perp) Q' K_0(\varepsilon_f' r_\perp) \, , \label{eq:M00} \\
    \Mgammastar{\pm 1,\pm 1} &= \frac{4 N_c e^2 q_f^2}{(2\pi)^2}  \int_{\vect{b}} \!\!\! e^{-i \vect{\Delta} \cdot \vect{b}} \int_{\vect{r}}  \!\!\! \mathcal{D}(\vect{r},\vect{b}) \int_{z} e^{-i  \vect{\delta} \cdot \vect{r}}\left[ \zeta   \ \varepsilon_f   K_1(\varepsilon_f  r_\perp) \varepsilon_f' K_1(\varepsilon_f' r_\perp) + m_f K_0(\varepsilon_f  r_\perp) m_f K_0(\varepsilon_f' r_\perp) \right] \, . \label{eq:M++}
\end{align}
Polarization changing amplitudes ($\rm{T/L \rightarrow L/T}$):
\begin{align}
    \Mgammastar{\pm 1,0} &= \frac{4 \sqrt{2} i N_c e^2 q_f^2}{(2\pi)^2}  \int_{\vect{b}} \!\!\! e^{-i \vect{\Delta} \cdot \vect{b}}  \int_{\vect{r}} \!\!\! e^{\pm i  \phi_{r \Delta}}  \mathcal{D}(\vect{r},\vect{b}) \int_{z} e^{-i  \vect{\delta} \cdot \vect{r}} z\bar{z} \xi  \ \varepsilon_f  K_1(\varepsilon_f  r_\perp)
    Q' K_0(\varepsilon_f' r_\perp) \, , \label{eq:M+0}\\
    \Mgammastar{0,\pm 1} &= -\frac{4 \sqrt{2} i N_c e^2 q_f^2}{(2\pi)^2}  \int_{\vect{b}} \!\!\! e^{-i \vect{\Delta} \cdot \vect{b}}  \int_{\vect{r}}  \!\!\! e^{\mp i  \phi_{r \Delta}}  \mathcal{D}(\vect{r},\vect{b}) \int_{z} e^{-i  \vect{\delta} \cdot \vect{r}} z\bar{z} \xi \ Q K_0(\varepsilon_f  r_\perp)
    \varepsilon_f'  K_1(\varepsilon_f'  r_\perp) \, . \label{eq:M0+}
\end{align}
Helicity flip amplitude ($\rm{T\pm \rightarrow T\mp}$):
\begin{align}
    \Mgammastar{\pm 1, \mp 1} &= -\frac{8 N_c e^2 q_f^2}{(2\pi)^2}  \int_{\vect{b}} \!\!\! e^{-i \vect{\Delta} \cdot \vect{b}} \int_{\vect{r}}  \!\!\! e^{\pm 2 i  \phi_{r \Delta}} \mathcal{D}(\vect{r},\vect{b}) \int_{z} e^{-i  \vect{\delta} \cdot \vect{r}} z \bar{z} \ \varepsilon_f  K_1(\varepsilon_f  r_\perp) \varepsilon_f'  K_1(\varepsilon_f'  r_\perp) \, , \label{eq:M+-}
\end{align}
where we define the short-hands $\zeta = z^2 +\bar{z}^2$, $\xi = z-\bar{z}$, and introduce the off-forward transverse vector $
   \vect{\delta} = \left(\frac{z-\bar{z}}{2}\right) \vect{\Delta} $.
\end{widetext}
To obtain the matrix elements $ \langle \Mgammastar{\lambda,\lambda'} \rangle_Y$ needed to compute the differential cross-section in Eq.\,\eqref{eq:xs}, we need to apply the CGG average (introduced in Eq.\,\eqref{eq:CGC-averaged-amplitude}) to the expressions above. This amounts to replacing the dipole operator $\mathcal{D}_Y(\vect{r},\vect{b})$ by the dipole correlator:
\begin{align}
   &  D_Y(\vect{r},\vect{b}) = \left\langle \mathcal{D}(\vect{r},\vect{b}) \right \rangle_Y \, , \nonumber \\
    &= 1- \frac{1}{N_c} \left \langle \Tr \left[ V\left(\vect{b} + \frac{\vect{r}}{2} \right) V^\dagger \left( \vect{b} - \frac{\vect{r}}{2} \right)  \right] \right \rangle_{Y} 
    \, , 
    \label{eq:dipole_correlator}
\end{align}
with the operation $\langle \ldots \rangle_{Y}$ defined in Eq.\,\eqref{eq:CGC-averaged-amplitude}.

\subsection*{Accessing correlations: kinematic vs intrinsic}
Eqs.\,\eqref{eq:M00}-\eqref{eq:M+-} contain phases $e^{i(\lambda - \lambda')\phi_{r\Delta}}$ that arise from the mismatch in the polarization $\lambda$ of the incoming and the polarization $\lambda'$ of the outgoing photon states. These phases pick up different modes in the angular correlations between the dipole vector $\vect{r}$ and the transverse momentum $\vect{\Delta}$. There are two sources for these correlations:

\begin{enumerate}
    \item the \emph{intrinsic} correlations between the dipole vector $\vect{r}$ and the impact parameter vector $\vect{b}$
    in the correlator $D_Y(\vect{r},\vect{b})$, combined with the Fourier phase $e^{-i \vect{\Delta} \cdot \vect{b}}$,
    \item the off-forward phase $e^{-i \vect{\delta}  \cdot \vect{r}}$.
\end{enumerate}

We refer to the contribution originating from the angular correlations in $D_Y(\vect{r},\vect{b})$ as \emph{intrinsic}, because they are generated as a result of the target's small $x$ structure.

Due to the even symmetry of $D_Y(\vect{r},\vect{b})$ under the exchange $\rt \to -\rt$, the dipole has only even harmonics\footnote{In principle, one could also have a odd harmonics. However, this would require the operator $D_Y(\vect{r},\vect{b})$ to have an odd component in the exchange $\rt \to -\rt$. This contribution corresponds to the exchange of a parity-odd   ``odderon", which would be important in exclusive production of pseudo scalar mesons such as $\eta_c$~\cite{Dumitru:2019qec} and $\pi$~\cite{Boussarie:2019vmk} (see however Ref.~\cite{Lappi:2016gqe}, where it is demonstrated that the high energy evolution suppresses the odderon).
}:
\begin{align}
    D_Y\!\left(\vect{r},\vect{b}\right) \!&=\! D_{Y,0}(r_\perp,b_\perp)\! +\! 2 D_{Y,2}(r_\perp,b_\perp) \cos (2 \phi_{r b})\! +\! ...
\end{align}
The intrinsic angular correlations start only at the second angular mode, and their contribution is dominant for the helicity flip term in Eq.~\eqref{eq:M+-}.

In the absence of intrinsic $\vect{r},\vect{b}$ correlations, it is possible to have non-zero polarization changing and helicity flip amplitudes due to the presence of the off-forward phase. In this case, it is possible to show that in the dilute limit the polarization changing amplitudes are suppressed by $\left( \Delta_\perp /Q \right)$, and the helicity flip amplitude by $\left(\Delta_\perp/Q\right)^2$ relative to the helicity preserving amplitudes. These scalings can also be obtained by considering the overlap between the photon and produced vector particle polarization vectors, noticing that the propagation axis for the produced particle is different from the direction of the incoming photon; therefore, we call these correlations \emph{kinematical}.

We note that in the absence of the off forward transverse vector $\vect{\delta}$, the polarization changing amplitudes vanish due to the anti-symmetry of the integrand around $z=0.5$, unlike the helicity preserving and helicity flip amplitudes which are symmetric around $z=$0.5.

\subsection*{Connection to gluon GPDs}
In Ref.~\cite{Hatta:2017cte} it was shown that at small $x$, the
isotropic $F^0_Y$ and elliptic $F^\epsilon_Y$ modes of the dipole correlator in momentum space
\begin{align}
    F_Y(\vect{q},\vect{\Delta}) &= \!\! \int_{\vect{r}} \!\! \int_{\vect{b}} \!\!\! \frac{e^{i \vect{q}\cdot \vect{r}}}{(2\pi)^2} \frac{e^{i \vect{\Delta}\cdot \vect{b} }}{(2\pi)^2} \left(1 - D_Y(\vect{r},\vect{b}) \right) \nonumber  \\
    &= F^{0}_Y(q_\perp,\Delta_\perp) + 2 F^{\epsilon}_Y(q_\perp,\Delta_\perp) \cos(2\phi_{q\Delta}) \, ,
\end{align}
are related to the unpolarized gluon GPD $xH_g$, and the gluon transversity GPD $ x E_{Tg}$ through
\begin{align}
    x H_g(x,\vect{\Delta}) &= \frac{2 N_c}{\alpha_s} \int d^2 \vect{q} q^2_\perp  F^{0}_Y(q_\perp,\Delta_\perp) \, , \label{eq:unpolGPD}\\
    x E_{Tg}(x,\vect{\Delta})  &= \frac{4 N_c M^2}{\alpha_s \Delta^2_\perp} \int d^2 \vect{q} q^2_\perp  F^{\epsilon}_Y(q_\perp,\Delta_\perp) \,, \label{eq:transversityGPD}
\end{align}
where $Y=\ln(1/x)$, and $M$ is a normalization mass scale. 
More generally, the Fourier transform of the dipole correlator $F_Y$ can also be related to the gluon Wigner distribution at small $x$~\cite{Hatta:2016dxp}.

In the collinear limit the helicity preserving and the helicity flip amplitudes of DVCS are proportional to $x H_g$ and $x E_{Tg}$, respectively, as shown in Ref.~\cite{Hatta:2017cte}; thus, offering the possibility to probe gluon GPDs in exclusive particle production at small $x$. We emphasize that the intrinsic contribution discussed above results in a non-zero elliptic mode $F_Y^\epsilon$, and consequently a non-zero transversity $xE_{Tg}$.

In this work we do not take the collinear limit, and instead calculate the cross-section in general (small $x$) kinematics. Consequently, more general structures than the two GPDs mentioned above enter in the scattering amplitudes.

\subsection{Deeply Virtual Compton Scattering Amplitude}
\label{sec:DVCS_amplitudes}
The amplitudes for DVCS are easily obtained by setting the virtuality of the final state photon $Q'^2 = 0$ in Eqs.~\eqref{eq:M00} through \eqref{eq:M+-}.  

Note that this immediately sets the amplitudes
$\Mgamma{0,0} = \Mgamma{\pm 1, 0} = 0$, since the real final state photon can not be longitudinally polarized. The non-trivial amplitudes are given below.

\begin{widetext}
Helicity preserving amplitudes:
\begin{align}
    \Mgamma{\pm 1, \pm 1} &= \frac{4 N_c e^2 q_f^2}{(2\pi)^2} \int_{\vect{b}} \!\!\! e^{-i \vect{\Delta} \cdot \vect{b}} \int_{\vect{r}} \!\!\! \mathcal{D}(\vect{r},\vect{b}) \int_{z} e^{-i  \vect{\delta} \cdot \vect{r}}\left[ \zeta   \ \varepsilon_f K_1(\varepsilon_f  r_\perp) m_f K_1(m_f r_\perp) + m_f K_0(\varepsilon_f  r_\perp)  m_f K_0(m_f r_\perp) \right] . \label{eq:Mphoton++}
\end{align}
Polarization changing amplitudes ($\rm{T/L \rightarrow L/T}$):
\begin{align}
    \Mgamma{0,\pm 1} &= -\frac{4 \sqrt{2} i N_c e^2 q_f^2}{(2\pi)^2} \int_{\vect{b}} \!\!\! e^{-i \vect{\Delta} \cdot \vect{b}} \int_{\vect{r}}  \!\!\! e^{\mp i  \phi_{r\Delta}}  \mathcal{D}(\vect{r},\vect{b}) \int_{z} e^{-i  \vect{\delta} \cdot \vect{r}} z\bar{z} \xi \ Q K_0(\varepsilon_f  r_\perp)
    m_f K_1(m_f r_\perp) . \label{eq:Mphoton0+}
\end{align}
Helicity-flip amplitudes ($\rm{T\pm \rightarrow T\mp}$):
\begin{align}
    \Mgamma{\pm 1, \mp 1} &= -\frac{8 N_c e^2 q_f^2}{(2\pi)^2} \int_{\vect{b}} \!\!\! e^{-i \vect{\Delta} \cdot \vect{b}} \int_{\vect{r}}  \!\!\! e^{\pm 2 i  \phi_{r \Delta}} \mathcal{D}(\vect{r},\vect{b}) \int_{z} e^{-i  \vect{\delta} \cdot \vect{r}} z \bar{z} \ \varepsilon_f K_1(\varepsilon_f  r_\perp) m_f K_1(m_f r_\perp) . \label{eq:Mphoton+-}
\end{align}
\end{widetext}
The results above are consistent with those obtained in Ref.~\cite{Hatta:2017cte} in the massless quark limit $(m_f \rightarrow 0)$, and when the dipole contains only isotropic and elliptic modulations.

We will include contributions from the light quarks $u,d,s$ and use $m_f=0.14\gev$ when calculating the DVCS cross-section.

The exclusive photon production cross-section consists of the DVCS process computed above, the process of elastic electron-nucleus scattering followed by photon emission off the electron (Bethe-Heitler (BH) process), and the interference terms. In this work we only consider the DVCS process. Experimentally  it is possible to at least partially distinguish the different contributions. For instance, the interference term is charge odd and can be suppressed by performing the same measurement with both electron and positron beams if sufficient precision is achieved in the experiment. The DVCS process dominates at small $\xpom$, but at the EIC energy range the BH contribution may still be non-negligible. At much smaller  $\xpom$ value accessible at the LHeC/FCC-he~\cite{Agostini:2020fmq}, the DVCS contribution should be more dominant.
We also note that the two processes have a different dependence on the inelasticity $y$, which can further be used to enhance  the DVCS signal. For a more detailed discussion related to separation of DVCS and BH contributions at the EIC, the reader is referred to Ref.~\cite{Aschenauer:2013hhw}.

\subsection{Exclusive Vector Meson production}
\label{sec:vmproduction}
Following the discussion\footnote{Note that there was a mistake in the off-forward phase in Ref.~\cite{Kowalski:2006hc}, which was identified and corrected by the authors in Ref.~\cite{Hatta:2017cte}.} in Ref.~\cite{Kowalski:2006hc}, the amplitudes for vector meson production can be obtained by
the following replacements in Eqs.\,\eqref{eq:M00}-\eqref{eq:M+0}:
\begin{align}
    \left(\frac{e q_f}{2\pi}\right) z\bar{z} K_0(\varepsilon_f' r_\perp) &\rightarrow \phi_{L}(r_\perp,z) , \\
    2Q' &\rightarrow M_V + \delta \frac{m^2_f - \nabla^2_\perp}{z\bar{z} M_V} , \label{eq:QprimRepl}
\end{align}
for the outgoing longitudinal polarization, where $M_V$ is the mass of the vector meson, and $\delta$ is a dimensionless constant (not to be confused with the off-forward vector $\bm{\delta}_{\perp}$).

We also need to perform the following replacements in Eqs.\,\eqref{eq:M++}, \eqref{eq:M0+} and \eqref{eq:M+-}
\begin{align}
    \left(\frac{e q_f}{2\pi} \right) z \bar{z} \varepsilon_f' K_1 (\varepsilon_f' r_\perp) &\rightarrow -\partial_\perp \phi_T(r_\perp,z) , \\
    \left(\frac{e q_f}{2\pi} \right) z \bar{z} K_0(\varepsilon_f' r_\perp) &\rightarrow \phi_{T}(r_\perp,z) ,
\end{align}
for outgoing transverse polarization. Here, we have assumed that the vector meson has the same helicity structure as the virtual photon, and introduced scalar functions $\phi_{T,L}(r_\perp,z)$ that describe the vector meson structure.

With these changes we obtain the following amplitudes:
\begin{widetext}
Helicity preserving amplitudes:
\begin{align}
    \M{0,0} &= \frac{8 N_c e q_f}{(2 \pi)}  \int_{\vect{b}} \!\!\! e^{-i \vect{\Delta} \cdot \vect{b}}  \int_{\vect{r}} \!\!\! \mathcal{D}\left(\vect{r},\vect{b} \right) \int_{z} e^{-i \vect{\delta} \cdot \vect{r}} z\bar{z} \ Q K_0(\varepsilon_f r_\perp) \ \left[ M_V + \delta \frac{m^2_f - \nabla^2_\perp}{z\bar{z} M_V}\right]  \phi_L(r_\perp,z) \, , \label{eq:Mjpsi00} \\
    \M{\pm 1,\pm 1} &= \frac{4 N_c e q_f}{(2 \pi)} \int_{\vect{b}} \!\!\! e^{-i \vect{\Delta} \cdot \vect{b}} \int_{\vect{r}} \!\!\! \mathcal{D}\left(\vect{r},\vect{b} \right)
    \int_{z} e^{-i \vect{\delta} \cdot \vect{r}} \frac{1}{z\bar{z}} \ \Big[\!\! -\zeta \varepsilon_f K_1(\varepsilon_f r_\perp) \ \partial_\perp \phi_T(r_\perp,z)  + m^2_f K_0(\varepsilon_f r_\perp) \ \phi_T(r_\perp,z) \Big] \, . \label{eq:Mjpsi++}
\end{align}
Polarization changing amplitudes ($\rm{T/L \rightarrow L/T}$):
\begin{align}
    \M{\pm 1, 0} &=  \frac{i 2\sqrt{2} N_c e q_f}{(2 \pi)} \int_{\vect{b}} \!\!\! e^{-i \vect{\Delta} \cdot \vect{b}}  \int_{\vect{r}}  \!\!\! e^{\pm i  \phi_{r \Delta}} \mathcal{D}\left(\vect{r},\vect{b} \right) \int_{z}  e^{-i \vect{\delta} \cdot \vect{r}}  \xi \ \varepsilon_f K_1(\varepsilon_f r_\perp) \left[ M_V + \delta \frac{m^2_f - \nabla^2_\perp}{z \bar{z} M_V}\right]  \phi_L(r_\perp,z) \, , \label{eq:Mjpsi+0}\\
    \M{0,\pm 1} &=  \frac{i 4\sqrt{2} N_c e q_f}{(2 \pi)} \int_{\vect{b}} \!\!\! e^{-i \vect{\Delta} \cdot \vect{b}} \int_{\vect{r}} \!\!\! e^{\mp i  \phi_{r\Delta}} \mathcal{D}\left(\vect{r},\vect{b} \right) \int_{z}   e^{-i \vect{\delta} \cdot \vect{r}}  \xi \ Q K_0(\varepsilon_f r_\perp) \partial_\perp \phi_T(r_\perp,z)  \, . \label{eq:Mjpsi0+}
\end{align}
Helicity-flip amplitudes ($\rm{T\pm \rightarrow T\mp}$):
\begin{align}
    \M{\pm 1, \mp 1} &= \frac{8 N_c  e q_f}{(2 \pi)} \int_{\vect{b}} \!\!\! e^{-i \vect{\Delta} \cdot \vect{b}} \int_{\vect{r}} \!\!\! e^{\pm 2i  \phi_{r\Delta}} \mathcal{D}\left(\vect{r},\vect{b} \right) \int_{z}  e^{-i \vect{\delta} \cdot \vect{r}} \ \varepsilon_f K_1(\varepsilon_f r_\perp) \partial_\perp \phi_T(r_\perp,z) \, . \label{eq:Mjpsi+-}
\end{align}
\end{widetext}

The precise form of $\phi_{L,T}$ depend on the vector meson in the final state and on the model used for the vector meson structure. In this work, we follow Ref.~\cite{Kowalski:2006hc} and use the Boosted Gaussian ansatz
\begin{align}
    \phi_{L,T} &= \mathcal{N}_{L,T} z \bar{z} \exp\left[- \frac{m^2_f \mathcal{R}^2}{8z \bar{z}} - \frac{2z \bar{z} r^2}{\mathcal{R}^2} + \frac{m^2_f \mathcal{R}^2}{2}\right].
\end{align}
The parameters $\mathcal{R}$ and $\mathcal{N}_{L,T}$ in the above equation and  $\delta$ in Eq.\,\eqref{eq:QprimRepl}, in case of $\jpsi$ production are determined in Ref.~\cite{Kowalski:2006hc}, and we use $m_{f}=m_{c}=1.4\gev$.

We note that the use of phenomenological vector meson wave functions introduces some model uncertainty in the calculation.
In addition to the phenomenological parametrizations such as the Boosted Gaussian model used in this work, there are also other approaches to describe the vector meson wave function. For example, in Ref.~\cite{Lappi:2020ufv} the heavy meson wave function is constructed as an expansion in quark velocity based on Non Relativistic QCD (NRQCD) matrix elements without the need to assume an identical helicity structure with the photon. Another recent approach is to solve the meson structure by first constructing a light front Hamiltonian consisting of a one gluon exchange interaction and an effective confining potential~\cite{Li:2017mlw,Chen:2018vdw}. 
As shown in Ref.~\cite{Lappi:2020ufv}, 
results for the $\jpsi$ production calculated using the phenomenological Boosted Gaussian parametrization are close to those obtained using these more systematic approaches. We note that this parametrization includes both S and D wave modes in the rest frame wave function; however, the D wave contribution results in a negligible contribution.

Exclusively produced $\jpsi$  is experimentally measured through the dilepton channel. 
Similarly to the Bethe-Heitler contribution discussed in case of DVCS above, there are also QED processes with exactly the same final state (for example, elastic $ep$ scattering, and a subsequent emission of virtual photon decaying into two leptons). We do not include this contribution in this work, but it may affect a precise extraction of the azimuthal modulations in exclusive $\jpsi$ production. 

\subsection{Incoherent diffraction}
\label{sec:incoherent_xs}

The cross-section obtained in Eq.~\eqref{eq:xs} corresponds to the processes where the target proton or nucleus remains in the same quantum state, and the cross-section is sensitive to the amplitudes $\left \langle \M{\lambda,\lambda'} \right \rangle_Y$, averaged over the target configurations. The second class of diffractive events, where the target nucleus goes out in a different quantum state $A^*$ than the initial state, is referred to as incoherent diffraction. The cross-section for $eA \to eA
^* V$ scattering can be obtained by first calculating the total diffractive cross-section, i.e., averaging over the target configurations at the cross-section level, and then subtracting the coherent contribution~\cite{Miettinen:1978jb,Frankfurt:1993qi,Frankfurt:2008vi,Caldwell:2009ke}. 
Consequently, the cross-section will depend on the covariances of amplitudes 
\begin{align}
    \left \langle \M{\lambda,\lambda'} \left(\M{\bar{\lambda},\lambda'} \right)^* \right \rangle_Y  -  \left\langle \M{\lambda,\lambda'} \right \rangle_Y \left\langle \M{\bar{\lambda},\lambda'} \right \rangle_Y^* \,,
\end{align}
and thus it is proportional to
\begin{align}
    D_Y^{(2,2)}(\vect{r},\vect{b};\vect{r}',\vect{b}') - D_Y(\vect{r},\vect{b})D_Y(\vect{r}',\vect{b}') \, , 
\end{align}
where the double dipole correlator is defined as
\begin{align}
\label{eq:double_dipole}
    D_Y^{(2,2)}(\vect{r},\vect{b};\vect{r}',\vect{b}') = \left\langle \mathcal{D}(\vect{r},\vect{b}) \mathcal{D}(\vect{r}',\vect{b}') \right \rangle_Y\, .
\end{align}
Unlike the dipole correlator in Eq.\,\eqref{eq:dipole_correlator}, there is no straightforward connection of the double dipole to well-known objects in the collinear framework of QCD. Yet, it provides interesting information, as it is sensitive to the small $x$ gluon fluctuations and event-by-event fluctuations in the color density geometry inside the nuclear target \cite{Caldwell:2009ke,Lappi:2010dd,Mantysaari:2016ykx,Mantysaari:2016jaz,Mantysaari:2019hkq}. See also Ref.~\cite{Mantysaari:2020axf} for a review. 


\section{Numerical setup}
\label{sec:setup}
The necessary ingredient needed to evaluate the DVCS and exclusive $\jpsi$ production cross-sections is the dipole scattering amplitude $D_Y(\rt,\bt)$ introduced in Eq.\,\eqref{eq:dipole_correlator}. 
We perform two separate calculations to evaluate the dipole amplitude, the first using a simple GBW parametrization, the second a more realistic CGC setup. The CGC will also allow us to compute the double dipole operator $D^{(2,2)}_Y(\vect{r},\vect{b};\vect{r}',\vect{b}')$ that appears in the incoherent cross-sections.

\subsection{GBW}
\label{sec:gbw}
To study how the angular modulations in the dipole amplitude affect the observable angular correlations, we use a GBW model~\cite{GolecBiernat:1998js} inspired parametrization as in Refs.~\cite{Altinoluk:2015dpi,Mantysaari:2019csc}, where angular modulation can be controlled by hand:
\begin{equation}
\label{eq:modulatedgbw}
    D_Y(\rt,\bt) = 1-\exp\left[ -\frac{\rt^2 \qso^2}{4} e^{-\bt^2/(2B)} C_\phi(\rt,\bt)\right] \, ,
\end{equation}
with 
\begin{equation}
\label{eq:gbw_c}
    C_\phi(\rt,\bt) = 1 + \frac{\tilde c}{2} \cos (2\phirb) \,.
\end{equation}
Here $\phirb$ is the angle between the impact parameter vector $\bt$ and the dipole vector $\rt$.
We set $\tilde c = 0.5$ to include a certain amount of angular dependence, or $\tilde c=0$ when we compare with the standard GBW dipole parametrization with no dependence on the dipole orientation. The saturation scale $Q^2_s(\vect{b})$, defined as $D_Y(\vect{r}^2=2/Q_s^2(\vect{b}),\vect{b})=1-e^{-1/2}$, and transverse size of the target proton are controlled by (implicitly $Y$ dependent) $\qso^2$ and $B$, respectively. 
As we use the GBW parametrization only to study the effect of azimuthal modulations, we do not attempt to fit the parameters precisely to HERA data, but use $\qso^2=Q_s^2(\vect{b}=0)=1.0 \gev^2$ and $B=4.0 \gev^{-2}$.

The $\cos(2\phirb)$ modulation in \eqref{eq:modulatedgbw} (in case of $\tilde c \neq 0$) results in a non-zero gluon transversity GPD, Eq.\eqref{eq:transversityGPD}, and consequently also an intrinsic contribution to the helicity flip amplitude $\M{\pm 1,\mp1}$ as discussed in Section~\ref{sec:off_forward_amp}.
A particular disadvantage of this \emph{ad hoc} parametrization is that very large dipoles (compared to the target size) can have unrealistic modulations e.g. when both of the quarks miss the target. Additionally, the modulation does not vanish in the homogeneous regime of small $r_\perp$ or $b_\perp$. For a more sophisticated analytical treatment of the angular dependence in the dipole scattering amplitude, see Refs.~\cite{Iancu:2017fzn,Salazar:2019ncp}.

\subsection{CGC}
\label{sec:cgc_setup}

Our main results are obtained using the CGC EFT framework that describes QCD at high energies. 
A particular advantage of the CGC EFT is that it is possible to calculate how the dipole scattering amplitude depends on the dipole orientation (angle $\phirb$), and as such the effect of the elliptic gluon distribution is a prediction. 
For a calculation of the elliptic gluon Wigner distribution from the CGC framework, see Ref.~\cite{Mantysaari:2019csc}. Additionally, it is possible to calculate the energy ($\xpom$) dependence perturbatively.

The CGC framework used in this work is similar to the framework used in Ref.~\cite{Mantysaari:2018zdd} (see also Ref.~\cite{Schenke:2012wb}), and is summarized here. The target structure is described in terms of fundamental Wilson lines $V(\bm{x}_{\perp})$ (see Eq.\,\eqref{eq:wline}), that describe the eikonal  propagation of a quark in the strong color fields of the target. As discussed in Section~\ref{sec:DVCS-JPsi-amplitude}, the dipole amplitude $D_Y$ is obtained as an expectation value of the correlator of Wilson lines, see Eqs.~\eqref{eq:dipole_tr} and~\eqref{eq:dipole_correlator}.
Note that when evaluating the incoherent cross-sections, expectation values for the double dipole operator $D_Y^{(2,2)}$ from Eq.~\eqref{eq:double_dipole} are also needed as discussed in Section~\ref{sec:incoherent_xs} (see also the discussion in Ref.~\cite{Mantysaari:2019hkq}).

The Wilson lines at the initial $\xpom=0.01$ are obtained from the MV  model~\cite{McLerran:1993ni}, in which one assumes that the color charge density of large $x$ sources $\rho_{A}$, is a local random Gaussian variable with expectation value zero and has the following expression for the 2-point correlator (see also Ref.~\cite{Dumitru:2020gla} for recent developments)
\begin{multline}
    g^2 \left \langle \rho^a_{A}(x^-, \xt) \rho^b_{A}(y^-,\yt) \right \rangle_{Y} = g^4 \lambda_A(x^-) \delta^{ab} \\
    \times  \delta^{(2)}(\xt-\yt) \delta(x^- - y^-) \, .   
\end{multline}
Here $a,b$ are color indices in the adjoint representation.
Following Ref.~\cite{Schenke:2012wb}, the local color charge density $\mu^2 = \int \der x^- \lambda_A(x^-)$ is assumed to be proportional to the local saturation scale $Q_s$, extracted from the IPsat parametrization~\cite{Kowalski:2003hm} fitted to HERA structure function data~\cite{Rezaeian:2012ji,Mantysaari:2018nng}. We note that the IPSat model, and with that our initial condition at $x\approx 0.01$, includes leading order DGLAP evolution for gluons.

We use $Q_s=0.8g^2\mu$, where the coefficient is matched in order to reproduce the normalization of the HERA $\jpsi$ production data at $\xpom \approx 10^{-3}$. (see also \cite{Lappi:2007ku} for a more detailed discussion relating the color charge density and the saturation scale)

Given the color charge density $\rho^a_{A}$, one obtains the Wilson lines $V(\xt)$ as a solution to the Yang-Mills equations as shown in Eq.~\eqref{eq:wline}. In the numerical implementation we include an infrared regulator $\tilde m$ to screen gluons at large distance, and write the Wilson lines as
\begin{equation}
  V(\xt) = \mathrm{P}_{-}\left\{ \exp\left({-ig\int_{-\infty}^\infty \der x^{-} \frac{\rho^a(x^-,\xt) t^a}{\boldsymbol{\nabla}^2-\tilde m^2} }\right) \right\}.
  \label{eq:wline_regulated}
\end{equation}
We use $\tilde m=0.4\gev$ for the infrared regulator as determined in Ref.~\cite{Mantysaari:2016jaz}.

In the IPsat parametrization, the local saturation scale $\langle Q_s^2 \rangle$ is proportional to the nucleon transverse density profile $T_p$. When nucleon shape fluctuations are not included, a Gaussian profile is used
\begin{equation}
    T(\bt) = \frac{1}{2\pi B_p} e^{-\bt^2/(2B_p)} \, ,
\end{equation}
where $B_p$ parametrizes the proton size. When shape fluctuations are included following Refs.~\cite{Mantysaari:2016ykx,Mantysaari:2016jaz}, the density is written as
\begin{equation}
\label{eq:Tpfluct}
    T_p(\bt) = \frac{1}{2\pi B_q N_q} \sum_{i=1 }^{N_q} e^{-(\bt-\bti)^2/(2B_q)},
\end{equation}
where the proton density profile is a sum of $N_q=3$ hot spots whose size is controlled by $B_q$. The hot spot positions $\bti$ are sampled from a Gaussian distribution with width $B_{qc}$. After sampling, the hot spot positions are shifted such that the center-of-mass is at the origin. Additionally, the density of each hot spot is allowed to fluctuate following the prescription presented in Ref.~\cite{Mantysaari:2016jaz} (based on Ref.~\cite{McLerran:2015qxa}) as follows. The local saturation scale of each hot spot $Q_s^2$ is obtained by scaling $\langle Q_s^2\rangle$ for each hot spot independently by a factor sampled from the log-normal distribution
\begin{equation}
\label{eq:qsfluct}
P\left( \ln \frac{Q_s^2}{\langle Q_s^2 \rangle}\right) = \frac{1}{\sqrt{2\pi}\sigma} \exp \left[- \frac{\ln^2 \frac{Q_s^2}{\langle Q_s^2\rangle}}{2\sigma^2}\right].
\end{equation}
The sampled values are scaled in order to keep the average saturation scale intact (the log-normal distribution has an expectation value $e^{\sigma^2/2}$). 

Heavy nuclei are generated by first sampling the nucleon positions from the Woods-Saxon distribution. Then, the color charge density in the nucleus is obtained by adding the color charge densities of individual nucleons at every transverse coordinate.   

\begin{figure}[tb]
\centering
\includegraphics[width=0.5\textwidth]{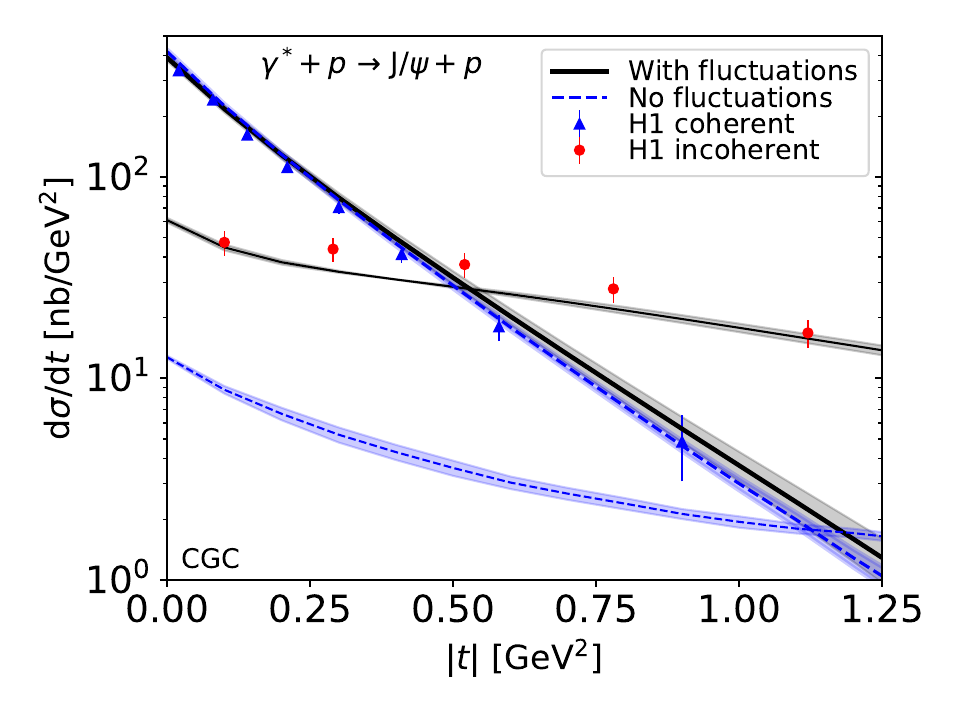}
\caption{Coherent (thick) and incoherent (thin lines) $\jpsi$ photoproduction cross-sections computed from the CGC setup with and without proton shape fluctuations, compared with H1 data~\cite{Alexa:2013xxa} at $W=75\gev$. }
\label{fig:h1_jpsi}
\end{figure}

As originally presented in Refs.~\cite{Mantysaari:2016ykx,Mantysaari:2016jaz}, the parameters controlling the proton and hot spot sizes can be determined by fitting the coherent and incoherent diffractive $\jpsi$ production data from HERA (see also Ref.~\cite{Mantysaari:2017cni} for the description of collective dynamics observed in proton-lead collisions at the LHC using the constrained fluctuating proton shape).  In this work, we parametrize the proton shape at the initial $\xpom=0.01$, requiring that after the JIMWLK evolution discussed below to smaller $\xpom \approx 10^{-3}$, a good description of the HERA data is obtained.\footnote{Note that here, unlike in Ref.~\cite{Mantysaari:2018zdd}, we have evolution over approximately $2.3$ units of rapidity from the initial condition before we compare with the HERA $\jpsi$ data.} When no proton shape fluctuations are included, we use $B_p=3\gev^{-2}$, and the fluctuating proton shape is described by $B_q=0.3\gev^{-2}, B_{qc}=3.3\gev^{-2}$ and $\sigma=0.6$. Note that shifting the center-of-mass to the origin effectively shrinks the proton, and as such both parametrizations result, on average, approximately in protons with the same size. The resulting coherent $\jpsi$ production spectra shown in Fig.~\ref{fig:h1_jpsi} that are sensitive to the average interaction with the target are almost identical.  On the other hand, the incoherent cross-section from the calculation where no substructure fluctuations are included clearly underestimates the HERA data, which suggests that significant shape fluctuations are required to produce large enough fluctuations in the scattering amplitude. 

The energy evolution of the Wilson lines (and consequently that of the dipole amplitude) is obtained by solving the JIMWLK evolution equation separately for each of the sampled color charge configuration. For numerical solutions the JIMWLK equation is written in the Langevin form following Ref.~\cite{Blaizot:2002np}
\begin{equation}\label{eq:langevin1}
\frac{\der V(\xt)}{\der Y} \! = \! V(\xt) (i t^a) \left[
\int \der^2 \zt\,
\varepsilon _{\xt,\zt}^{ab,i} \; \xi_\zt(Y)^b_i  + \sigma_\xt^a 
\right].
\end{equation}
The random noise $\xi$ is Gaussian and local in spatial coordinates, color, and rapidity with expectation value zero and covariance 
\begin{equation}
\label{eq:noice}
\langle \xi_{\xt,i}^a(Y) \xi_{\yt,j}^b(Y')\rangle = \delta^{ab} \delta_{ij} \delta^{(2)} (\xt-\yt) \delta(Y-Y').
\end{equation}
The coefficient of the noise in the stochastic term is
\begin{equation}
\label{eq:noicecoef}
 \varepsilon _{\xt,\zt}^{ab,i} = \left(\frac{\as}{\pi}\right)^{1/2} K_{\xt-\zt}^i
\left[1-U^\dag(\bm{x}_{\perp})  U(\bm{z}_{\perp}) \right]^{ab},
\end{equation}
where
\begin{equation}
\label{eq:jimwlk_k}
K_\xt^i = m x_\perp K_1(m x_\perp) \frac{x^i}{\bm{x}_{\perp}^{2}}.
\end{equation}
The deterministic drift term reads
\begin{equation}
    \sigma_\xt^a = -i\frac{\as}{2\pi^2} \int \der^2 \zt \frac{1}{(\xt-\zt)^2} \mathrm{Tr} \left[ T^a U^\dagger(\xt) U(\zt) \right].
\end{equation}
Here $U(\xt)$ is a Wilson line in the adjoint representation of SU($N_{\mathrm{c}}$) and can be obtained from Eqs.\,\eqref{eq:wline} and \eqref{eq:wline_regulated} by replacing $t^{a}$ with the adjoint generators $T^{a}$. For the details of the numerical procedure, in particular,  related to the elimination of the deterministic drift term and expression of the Langevin step in terms of fundamental Wilson lines only, see the discussion in Refs.~\cite{Lappi:2012vw,Mantysaari:2018zdd}. The JIMWLK kernel in Eq.\,\eqref{eq:jimwlk_k} includes an infrared regulator $m$, first suggested in Ref.\,\cite{Schlichting:2014ipa}, which exponentially suppresses gluon emission at long distances $\gtrsim 1/m$. In this work we use $m=0.2\gev$, along with a fixed strong coupling constant $\as=0.21$, as constrained in Ref.~\cite{Mantysaari:2018zdd}.

As discussed above, the free parameters of the framework are constrained by the HERA $\jpsi$ production data at $W=75\gev$~\cite{Alexa:2013xxa}, which corresponds to $\xpom\approx 10^{-3}$  (or approximately $Y=2.3$ units of JIMWLK evolution from our initial condition; the $|t|$ dependence of $\xpom$ is neglected). 
The agreement of the calculated $\jpsi$ photoproduction ($Q^2=0$) cross-sections with HERA data is demonstrated in Fig.~\ref{fig:h1_jpsi}, where results with and without proton shape fluctuations are shown. Note that here we consider $\jpsi$ production in $\gamma^* + p$ scattering. Later we will study $e+p$ scattering where azimuthal correlations with respect to the outgoing lepton can be accessed.

\begin{figure}[tb]
\includegraphics[width=0.5\textwidth]{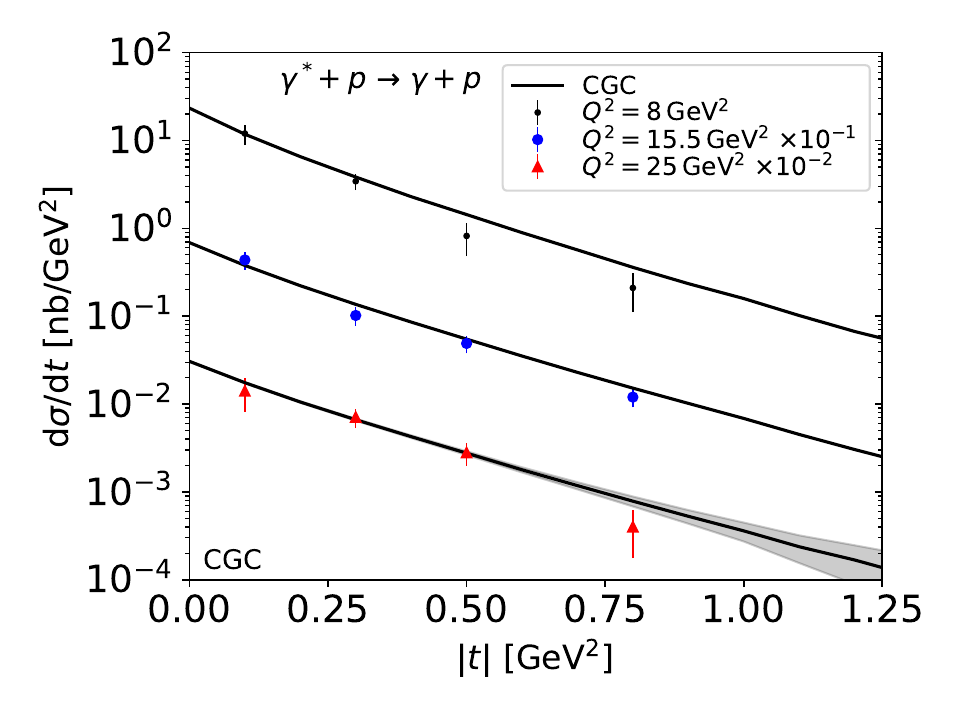}
\caption{Coherent DVCS cross-sections in different $Q^2$ bins at $W=82\gev$ computed from the CGC setup (without proton shape fluctuations) compared to H1 data~\cite{Aaron:2007ab}. }
\label{fig:h1_dvcs}
\end{figure}

In addition to vector meson production, HERA has also measured the (coherent) DVCS  cross-section~\cite{Adloff:2001cn,Aktas:2005ty,Chekanov:2008vy,Aaron:2007ab}. 
The CGC framework also provides a good description of the HERA DVCS data as demonstrated in  Fig.~\ref{fig:h1_dvcs}. When comparing $\jpsi$ and $\gamma$ production cross-sections to the HERA data, we have included an approximately $45\%$ skewness correction~\cite{Shuvaev:1999ce} from Ref.~\cite{Mantysaari:2017dwh}, which takes into account the fact that in the two-gluon exchange limit, the exchanged gluons carry very different fractions of the target longitudinal momentum, and the cross-section can be related to collinearly factorized parton distribution functions. The same constant factor is included in all cross-sections shown in this work. The energy dependence of the total cross-section and $t$ slope has been studied in~\cite{Mantysaari:2018zdd}.


\section{Numerical results}
\label{sec:results}
In this section we present numerical results for azimuthal modulations in DVCS and $\jpsi$ production in electron-proton and electron-nucleus scattering, calculated in the energy range accessible at the future high-energy EIC. 
In case of electron-proton collisions, our results are shown at $\sqrt{s}=140\gev$, and in case of electron-gold scattering we use $\sqrt{s}=90\gev$. We present results at fixed $\xpom$, which results in the inelasticity $y$ to depend on $\xpom$, $Q^2$, and $|t|$. We note that the azimuthal modulations and the angle independent cross-section have a different dependence on $y$, as apparent from Eq.~\eqref{eq:xs}. This $y$-dependence alone results in a suppression of azimuthal modulations with decreasing $\xpom$, especially in the case of $\jpsi$ production. We discuss the separation of this effect from genuine JIMWLK evolution effects in Appendix~\ref{appendix:y}.

\subsection{Effect of angular modulations in the GBW dipole amplitude}
\label{sec:gbw_numerics}

As discussed in Section~\ref{sec:off_forward_amp}, there are two contributions to the polarization changing and helicity flip amplitudes: a kinematical contribution due to the off-forward phase $e^{-i \vect{\delta} \cdot \vect{r}}$, and intrinsic contribution arising from the correlations between $\vect{r}$ and $\vect{b}$ in the amplitude $D_Y(\vect{r},\vect{b})$. 

To isolate the intrinsic contribution and demonstrate the effect of the non-zero gluon transversity GPD, we first study DVCS using the GBW dipole amplitude as presented in Section~\ref{sec:gbw} (note that as we calculate the cross-section in general kinematics, more general objects than the gluon GPDs introduced in Section~\ref{sec:off_forward_amp} are probed). This parametrization allows us to easily include the intrinsic contribution by introducing the dependence on the dipole orientation relative to the impact parameter using the parameter $\tilde c$ introduced in Eq.~\eqref{eq:gbw_c}.

\begin{figure*}
\subfloat[DVCS, no angular dependence]{%
\includegraphics[width=0.48\textwidth]{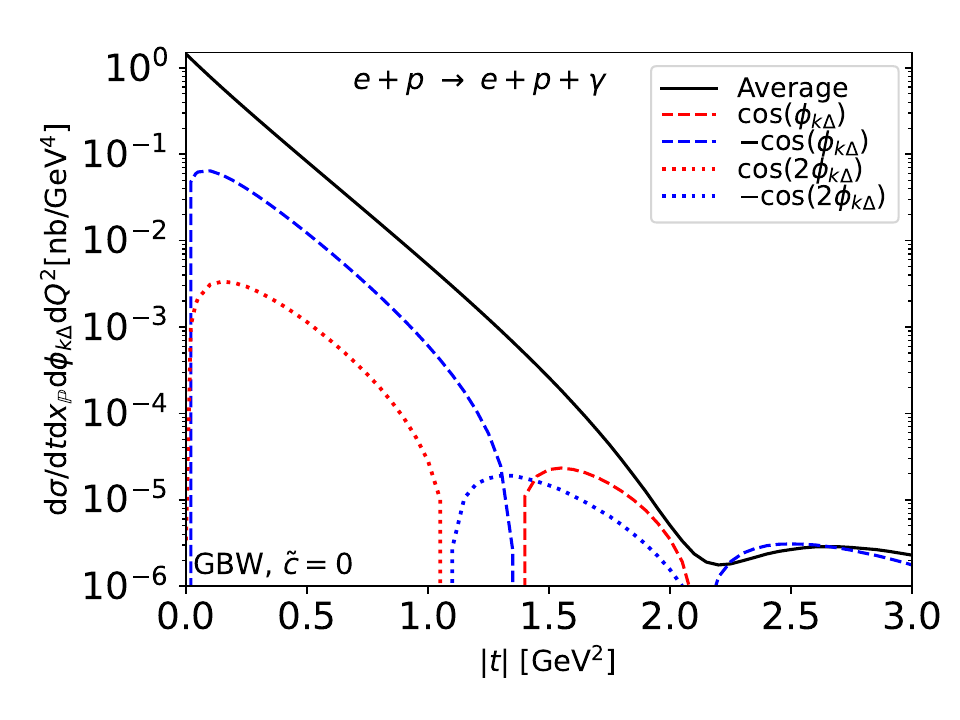}
\label{fig:gbw_dvcs_nomod}
}
\subfloat[DVCS, with angular dependence]{%
\includegraphics[width=0.48\textwidth]{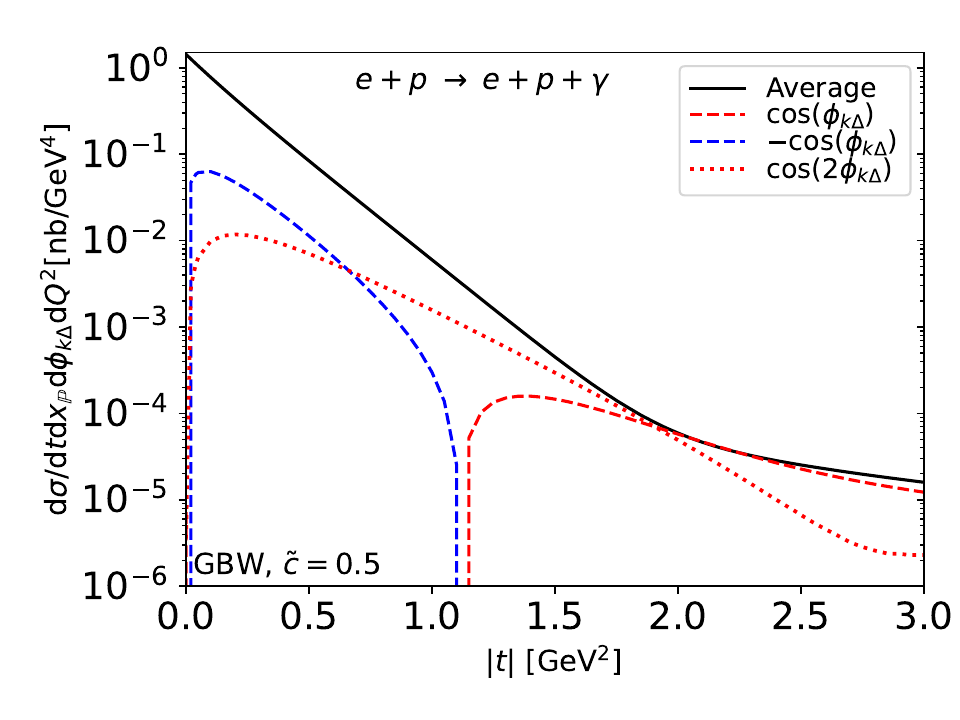} 
\label{fig:gbw_dvcs_mod}
}\\
\subfloat[$\jpsi$, no angular dependence]{%
\includegraphics[width=0.48\linewidth]{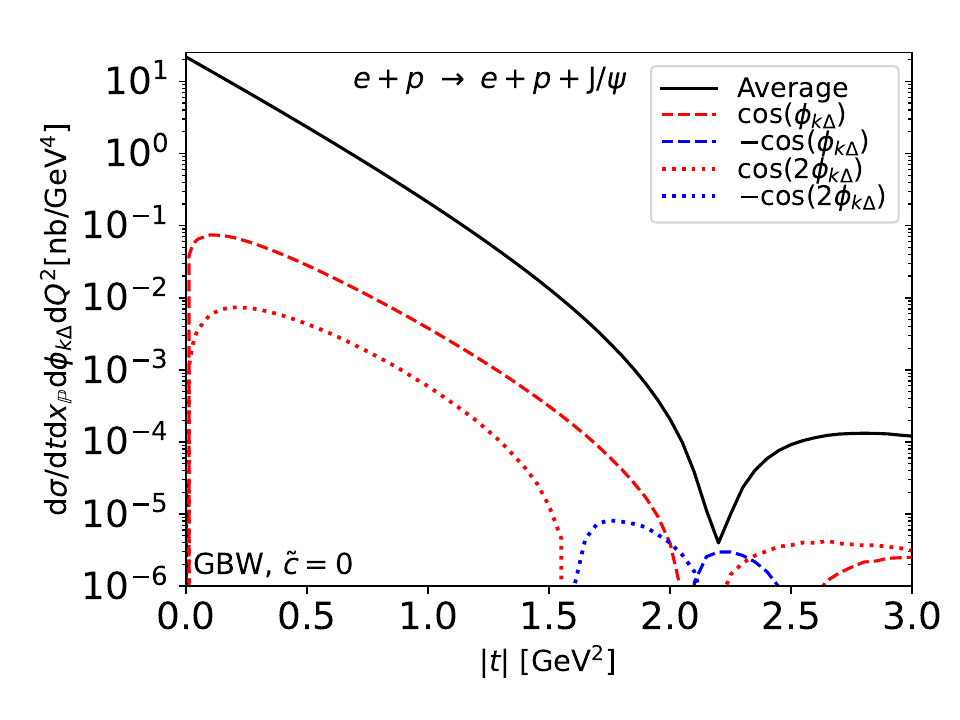}
\label{fig:gbw_jpsi_nomod}
}
\subfloat[$\jpsi$, with angular dependence]{%
\includegraphics[width=0.48\textwidth]{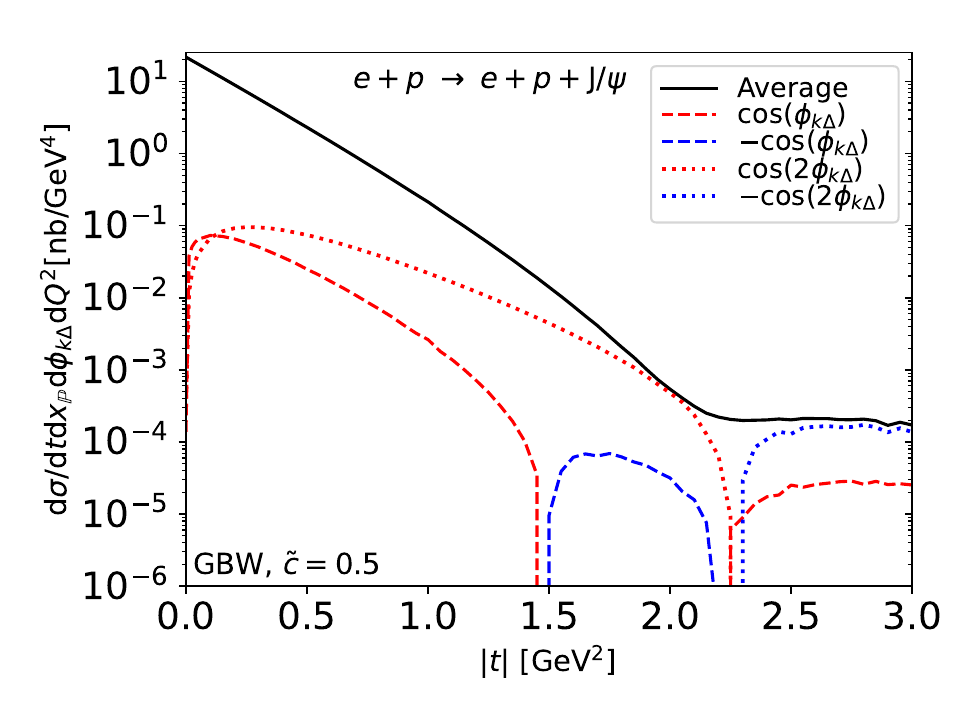} 
\label{fig:gbw_jpsi_mod}
}

\caption{
Different contributions to the DVCS and $\jpsi$ production cross-section using the GBW dipole at $Q^2=5\gev^2$ (DVCS) or $Q^2=2\gev^2$ ($\jpsi$) with and without angular dependence. 
}
\end{figure*}

The exclusive photon production (or DVCS)  cross-section computed using the GBW dipole amplitude as a function of squared momentum transfer $|t|\approx \vect{\Delta}^2$ is shown in Fig.~\ref{fig:gbw_dvcs_nomod} with no angular dependence in the dipole amplitude ($\tilde c=0$) and in Fig.~\ref{fig:gbw_dvcs_mod} when a significant angular modulation is included with $\tilde c=0.5$. We show separately the three different contributions: \emph{Average} refers to the total cross-section averaged over  $\phik$, which is the first line in Eq.~\eqref{eq:xs}.  Additionally, the coefficients of $\cos(\phik)$ and $\cos(2\phik)$ in Eq.~\eqref{eq:xs} are shown. Note that in DVCS the final state photon is real and has only transverse polarization states, and as such $\M{\lambda,0}=0$ for all $\lambda$. As the modulation can change sign, in the logarithmic plots we show separately the positive and negative modulations.

The results are shown at $Q^2=5\gev^2$, and we have checked that the qualitative features depend only weakly on virtuality. We note that as there is no heavy mass scale in DVCS, this process is only marginally perturbative at moderate $Q^2$ especially in the limit $z \to 0,1$ (the so called \emph{aligned jet} configuration), where there is no exponential suppression for the photon splitting into non-perturbatively large dipoles. 
In the GBW model, asymptotically large dipoles scatter with probability one, even when the quarks completely miss the target. Large dipoles also have a large elliptic modulation if $\tilde c\neq 0$, and consequently our GBW model calculation likely overestimates the cross-section and the intrinsic contribution to the modulations. Later when we use a CGC proton or nucleus as a target, non-perturbatively large dipoles (larger than the size of the target) do not contribute significantly. 
However, it is not clear which one of these setups includes a more realistic effective description of confinement scale effects. 
Consequently, in DVCS at small $Q^2$, there may exist a non-perturbative contribution not accurately captured in either of the dipole picture calculations presented here. For a more detailed discussion of the effective implementation of confinement scale effects in the dipole picture, see e.g. Refs.~\cite{Berger:2011ew,Mantysaari:2018zdd,Mantysaari:2018nng,Cepila:2018faq}.

Without the intrinsic contribution ($\tilde c=0$, shown in Fig.~\ref{fig:gbw_dvcs_nomod}) the kinematical contribution results in non-zero $\cos(\phik)$ and $\cos(2\phik)$ modulations, that exhibit a diffractive pattern. The $\cos(\phik)$ modulation is generically larger as expected, as it is mostly sensitive to the polarization changing amplitude $\langle \M{0,\pm 1} \rangle_Y$. On the other hand, the $\cos (2\phik)$ modulation probes the helicity flip amplitude $ \langle \M{\pm 1,\mp1}  \rangle_Y$ which in the dilute limit is suppressed by an additional factor $\Delta_\perp/Q$ relative to $\langle \M{0,\pm 1} \rangle_Y $, as discussed in Section~\ref{sec:off_forward_amp}.

Introducing an elliptic modulation in the GBW dipole significantly increases the  $\cos(2\phik)$ modulation of the cross-section, and renders it even larger than the  $\cos(\phik)$ modulation at intermediate momentum transfer $|t|$, as shown in Fig.~\ref{fig:gbw_dvcs_mod}. The  $\cos(\phik)$ modulation is only weakly modified by the modulations in the dipole amplitude, except at large $|t|\gtrsim 1.5\gev^2$. 
This can be understood based on Eq.~\eqref{eq:xs}. The  $\cos(2\phik)$ contribution is directly proportional to the transverse helicity flip amplitude $\langle \M{\pm 1, \mp 1} \rangle_Y$, which is sensitive to the $\cos (2\phirb)$ modulation in the dipole amplitude. On the other hand, the  $\cos(\phik)$ modulation is dominated by $\langle \M{0,\pm 1} \rangle_Y \langle \M{\pm 1, \pm 1} \rangle_Y$, where neither of the terms are directly sensitive to the angular modulations in the dipole.

Both modulations again become more important at higher momentum transfer where the kinematical contribution is more important and the $\cos(\phik)$ modulation dominates. 
At large $|t|$ the angular modulations change the average cross-section and the $\cos \phik$ modulations because of two reasons. First, the helicity flip amplitude  $\langle \M{\pm 1,\mp1} \rangle_Y$ has a small contribution to both of these components. Additionally, including
the angular dependence in the dipole amplitude also changes the angle averaged interaction, due to the non-linear dependence on $\cos (2\phirb)$ in Eq.~\eqref{eq:modulatedgbw}. 

\begin{figure*}
\centering
\subfloat[Spectra]{%
\includegraphics[width=0.48\textwidth]{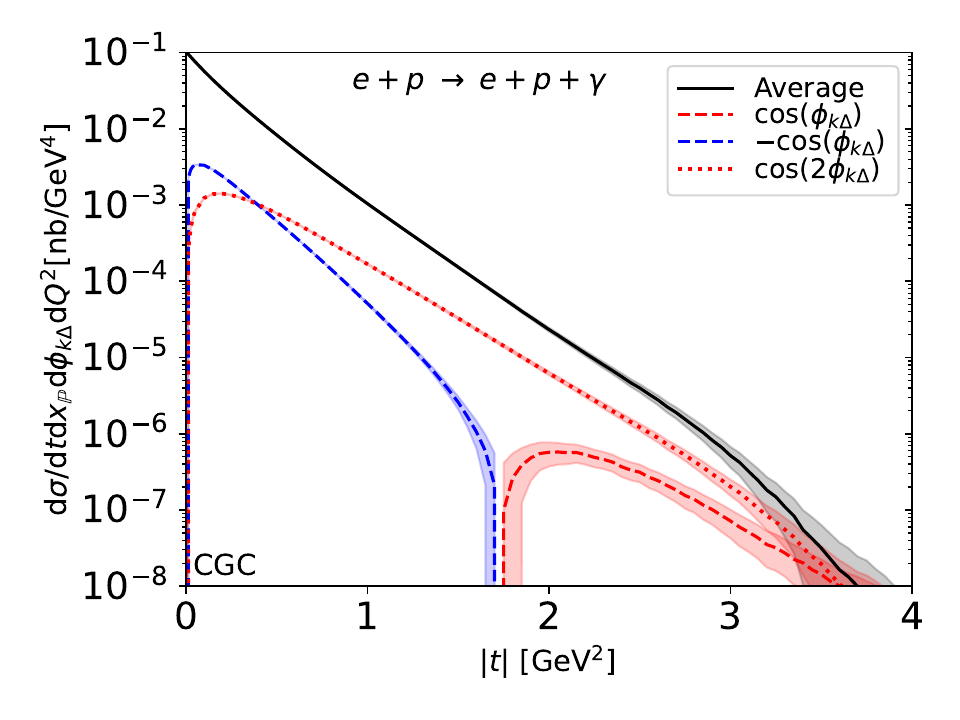}
\label{fig:cgc_dvcs_xp_001}
}
\subfloat[Modulation coefficients]{%
\includegraphics[width=0.48\textwidth]{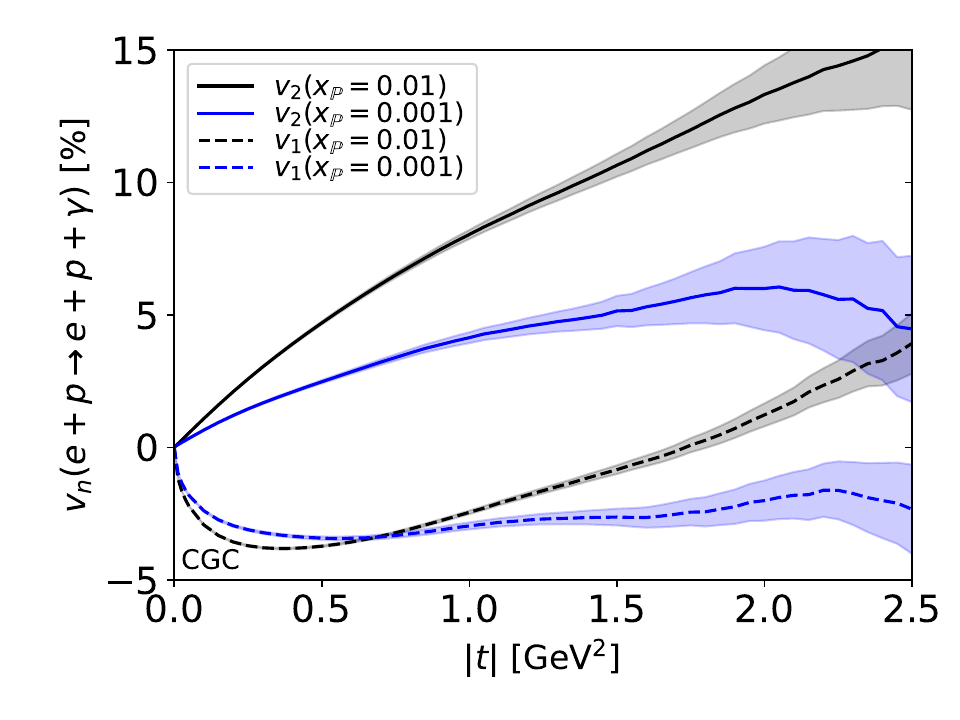} 
\label{fig:cgc_dvcs_vn}
}
\caption{Different contributions to the DVCS production at $Q^2=5\gev^2$, and the  modulation coefficients $v_n$ as defined in Eq.~\eqref{eq:vndef}. The $v_n$'s are shown both at the initial condition ($\xpom=0.01$, black lines) and after the JIMWLK evolution ($\xpom=0.001$, blue lines). The bands show the statistical uncertainty of the calculation resulting from averaging over the fluctuating color charge configurations.
}
\end{figure*}

The cross-sections with azimuthal modulations in  exclusive $\jpsi$ electroproduction at $Q^2=2\gev^2$ are shown in Figs.~\ref{fig:gbw_jpsi_nomod} (no angular dependence in the dipole) and \ref{fig:gbw_jpsi_mod} (with angular dependence). The large mass of the charm quark renders the kinematical contribution small as dipoles are generically smaller, which suppresses the contribution from the off-forward phase $e^{-i\vect{\delta}\cdot\rt}$. Consequently the relative importance of the $\cos \phik$ modulation is significantly smaller than in case of DVCS studied above. The $\cos \phik$ modulation is now positive at small $|t|$, in contrast to the DVCS case. 
The reason for this difference in sign is that in $\jpsi$ production the final state can be longitudinally polarized, and terms $\langle \M{0,0}\rangle_Y \langle \M{\pm 1,0}\rangle_Y$, that are absent in the DVCS case, give a large positive contribution to the $\cos \phik$ modulation.

Similar to $\cos \phik$, the $\cos 2\phik$ modulation is also small at small momentum transfers. If the angular modulations are included in the dipole amplitude, the $\cos 2\phik$ modulation is significant in the $|t|\gtrsim 1\gev^2$ region, similar to the DVCS case studied above. The smallness of the $\cos \phik$ component compared to $\cos 2\phik$ can potentially render the experimental extraction of the $\cos 2\phik$ modulation more precise in $\jpsi$ production compared to the  DVCS case.

We have checked (within the GBW model) that the dependence of the azimuthal anisotropies on charm mass and the saturation scale $Q_s$ is weak for moderate values of $|t|$. We expect this to hold true in the full CGC computation that we outline next.

\subsection{Proton target in the CGC}
\label{sec:CGC_numerics}
We now use a proton target described within the CGC framework, and study coherent DVCS followed by coherent  $\jpsi$ production
in the EIC energy range. As discussed in Section~\ref{sec:cgc_setup} the initial condition for the JIMWLK evolution is constrained such that the coherent $\jpsi$ production cross-section at $W=75\gev$ ($\xpom\approx 10^{-3}$) is compatible with the HERA data. Here we do not include the proton shape fluctuations as we consider the coherent cross-section, which is only sensitive to the average interaction. The incoherent cross-section, which is sensitive to fluctuations, e.g. of the proton shape, is discussed in Section~\ref{sec:incoh_numerics}.

As discussed in Section~\ref{sec:cgc_setup}, the dependence on the dipole orientation is calculated from the CGC framework (assuming an impact parameter dependent MV model initial condition). Consequently, the intrinsic contribution to the azimuthal correlations is a genuine prediction, unlike in the GBW model calculation presented above. Additionally, the dependence on $\xpom$ is a result of the perturbative JIMWLK evolution.

\begin{figure}[tb]
\includegraphics[width=0.5\textwidth]{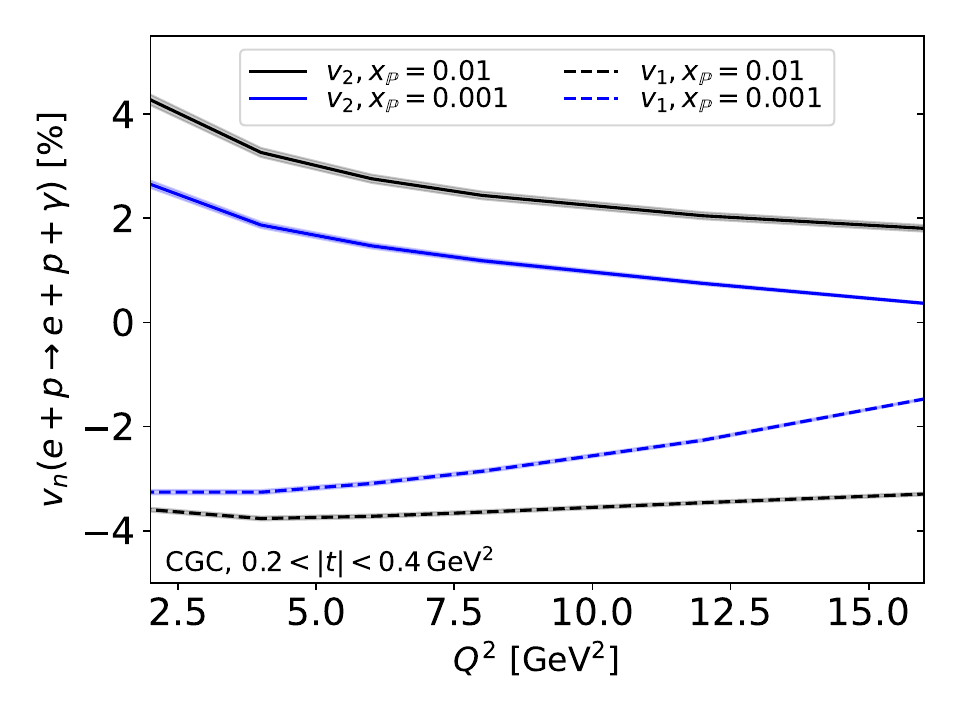}
\caption{Azimuthal modulation coefficients $v_1$ and $v_2$ integrated over $0.2<|t|<0.4\gev^2$ in DVCS as a function of photon virtuality for $\xpom=0.01$ (black lines) and $\xpom=0.001$ (blue lines). 
}
\label{fig:cgc_dvcs_v2_q2dep}
\end{figure}

\subsubsection{DVCS}

The coherent DVCS production cross-section and angular modulations as a function of squared momentum transfer $|t|$ at $\xpom=0.01$ are shown in Fig.~\ref{fig:cgc_dvcs_xp_001}. The results are shown again at moderate $Q^2=5\gev^2$. The qualitative features of our results depend weakly on $Q^2$ (note that our framework is not applicable in the dilute regime of large $Q^2$ whose dynamics is governed by linear evolution equations, namely the Dokshitzer-Gribov-Lipatov-Altarelli-Parisi (DGLAP) equations~\cite{Gribov:1972ri,Lipatov:1974qm,Altarelli:1977zs,Dokshitzer:1977sg}).

\begin{figure*}[t]
\subfloat[$\xpom=0.01$]{%
\includegraphics[width=0.48\textwidth]{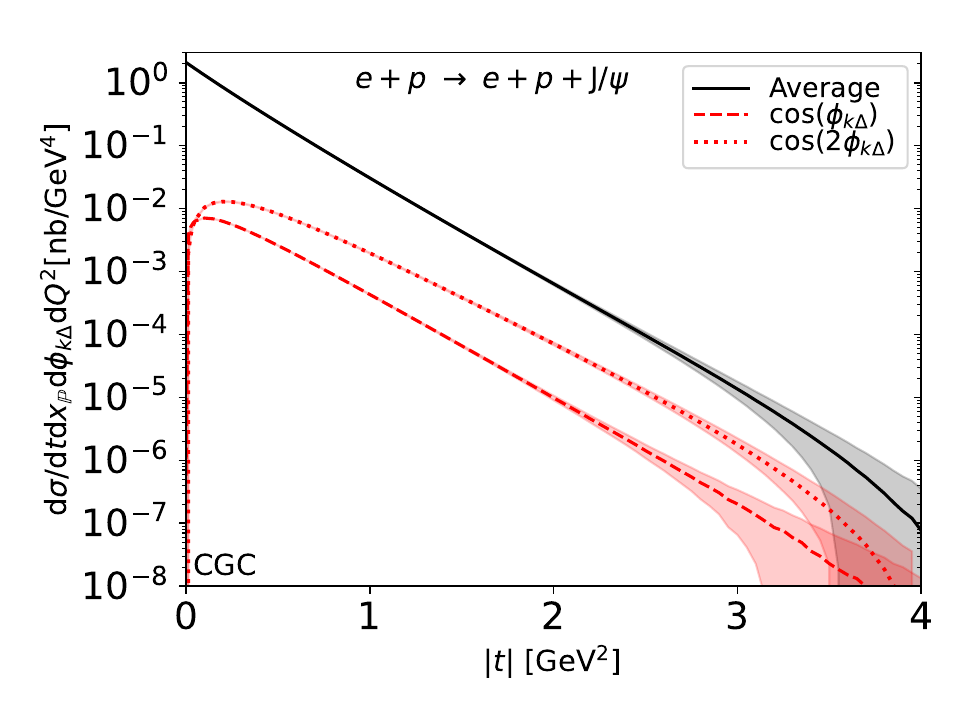}
\label{fig:cgc_jpsi_spectra_xp_001}
}
\subfloat[$\xpom=0.001$]{%
\includegraphics[width=0.48\textwidth]{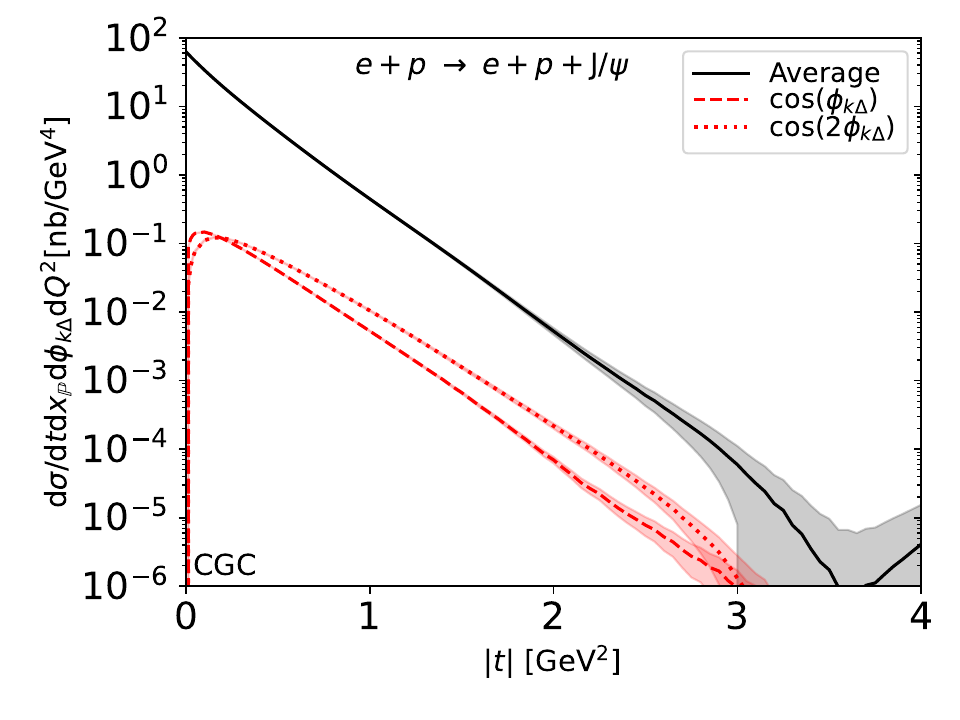}
\label{fig:cgc_jpsi_spectra_xp_0001}
}
\caption{Different contributions to the coherent $\jpsi$ production cross-section at the initial condition $\xpom=0.01$ and after JIMWLK evolution to $\xpom=0.001$. Results are shown at $Q^2=2\gev^2$. }
\end{figure*}

The intrinsic contribution predicted from the CGC calculation results in a sizeable $\cos(2\phik)$ modulation in the DVCS cross-section, similar to the presented GBW model calculation with an artificial angular modulation in the dipole. The $\cos(\phi_{k\Delta})$ modulation, where the intrinsic contribution is subleading, is large in the small $|t|$ region, and suppressed strongly relative to the $\cos(2\phik)$ contribution at larger momentum transfers.

The extracted modulation coefficients $v_1$ and $v_2$, defined as
\begin{equation}
\label{eq:vndef}
    v_n = \frac{\int_0^{2\pi} \der \phik e^{ni \phik} \der \sigma^{e+p\to e+p+V}/\der t \der \phik \der Q^2 \der \xpom } {\int_0^{2\pi} \der \phi_{k\Delta} \der \sigma^{e+p \to e+p+V}/\der t \der \phik  \der Q^2 \der \xpom }\, ,
\end{equation}
 are shown in Fig.~\ref{fig:cgc_dvcs_vn}. Here $V$ refers to the produced particle, in this case $V=\gamma$. The results are shown at the initial condition $\xpom=0.01$, and after the JIMWLK evolution over approximately $2.3$ units of rapidity to $\xpom=0.001$ at fixed $\sqrt{s}=140\gev$ in the EIC kinematics.

Both $|v_1|$ and $v_2$ are found to be sizeable, up to $5-10\%$ in the $|t|$ region where the coherent contribution should be more easily measurable (at very large momentum transfers the incoherent contribution will dominate and render the measurement of the coherent cross-section challenging). The elliptic modulation $v_2$ is suppressed  approximately by a factor $2$ when going from $\xpom=0.01$ to $\xpom=0.001$, mostly due to the JIMWLK evolution (the increase in inelasticity $y$ with decreasing $\xpom$ has a negligible effect here, as demonstrated in Appendix~\ref{appendix:y}). This is expected as the small $x$ evolution results in a larger proton with smaller density gradients. In the CGC framework, these density gradients result in $\cos 2\phirb$ modulation in the dipole-target scattering amplitude (see Appendix~\ref{appendix:mode_decomposition}), and consequently contribute to $\langle \M{\pm 1,\mp 1} \rangle_Y$ defined in Eq.~\eqref{eq:M+-}, which dominates $v_2$. 
Similarly, a small decrease in the magnitude of $v_1$ at small $|t|$ is expected, as the helicity flip amplitude $\langle \M{\pm 1, \mp 1} \rangle_Y$ also contributes to $v_1$. As a result of the small $x$ evolution,  the dip in the $\cos \phik$ spectrum moves to larger $|t|$ which explains why $|v_1|$ increases with decreasing $\xpom$ at large $|t|$.

Dependence on the exchanged photon virtuality is illustrated in Fig.~\ref{fig:cgc_dvcs_v2_q2dep}, where the $v_1$ and $v_2$ coefficients are computed at the initial condition and after JIMWLK evolution as a function of $Q^2$. The results are obtained by using spectra integrated over $0.2<|t|<0.4\gev^2$ in Eq.~\eqref{eq:vndef}. The elliptic modulation is reduced at high virtualities. This is because the characteristic dipole size scales as $r_\perp^2 \sim 1/Q^2$, such that large dipoles, which are most sensitive to the density gradients on the proton size scale, are suppressed at high $Q^2$. These large dipoles also provide the dominant part of the intrinsic contribution to $\langle \M{\pm 1, \mp 1} \rangle_Y$.

The $v_1$ modulation  depends weakly on $Q^2$, and both $v_1$ and $v_2$ are suppressed as a result of small $x$ evolution in the studied $|t|$ range at all $Q^2$. At larger $|t|$ the $Q^2$ dependence is more difficult to interpret because of the presence of a sign change in the $\cos \phik$ modulation. At $\xpom=0.001$, the increase in inelasticity $y$ as a function of $Q^2$ dominates the virtuality dependence in the $Q^2\gtrsim 10\gev^2$ region. We note that the modulations vanish at the kinematical boundary where $y=1$. In DVCS at $\xpom=0.001$, this boundary is at $Q^2=19.6\gev^2$.

\begin{figure}[tb]
\includegraphics[width=0.5\textwidth]{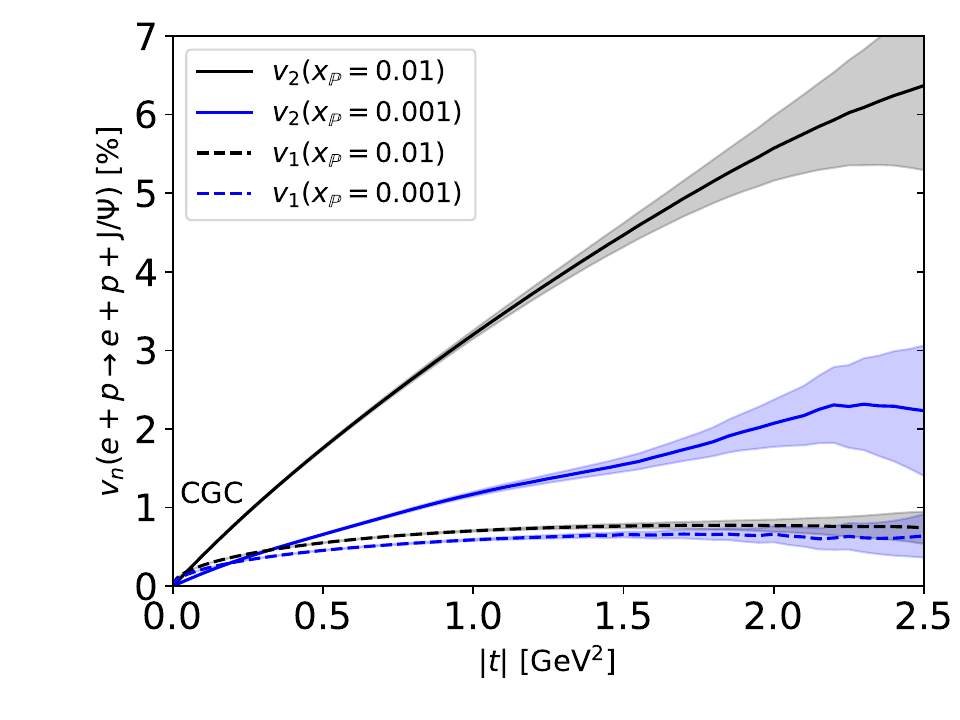}
\caption{Angular modulations in the coherent $\jpsi$ production cross-section at $Q^2=2\gev^2$ with proton targets. 
}
\label{fig:cgc_jpsi_vn}
\end{figure}

\subsubsection{Exclusive $\jpsi$ production}
\label{sec:jpsi_cgc_p}

\begin{figure*}[tb]
\subfloat[DVCS]{%
\includegraphics[width=0.48\textwidth]{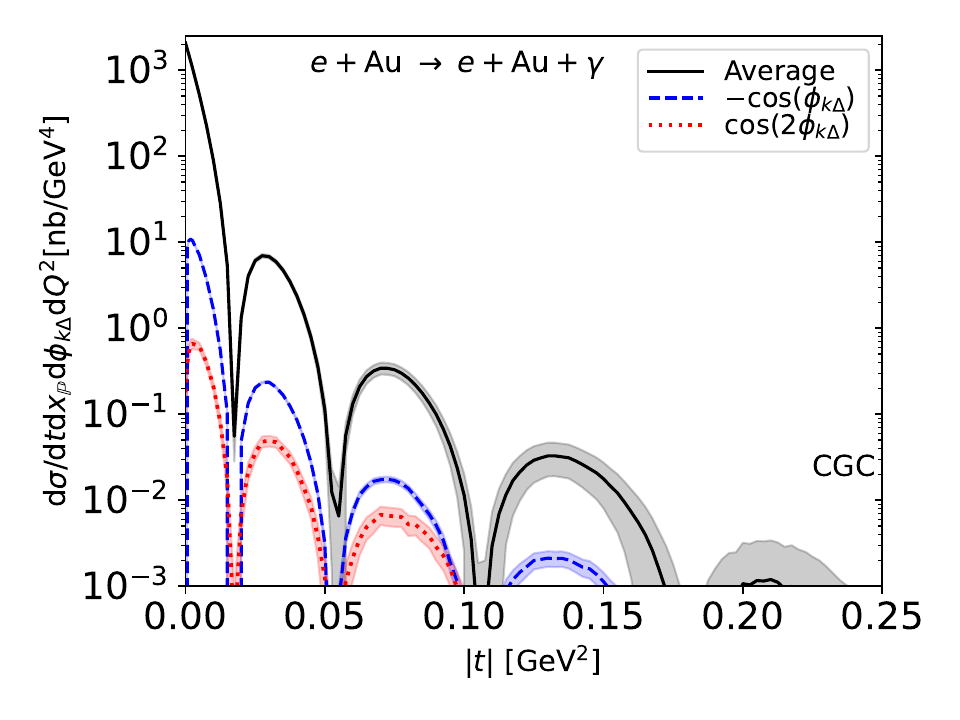}
\label{fig:cgc_Au_dvcs}
}
\subfloat[$\jpsi$]{
\includegraphics[width=0.5\textwidth]{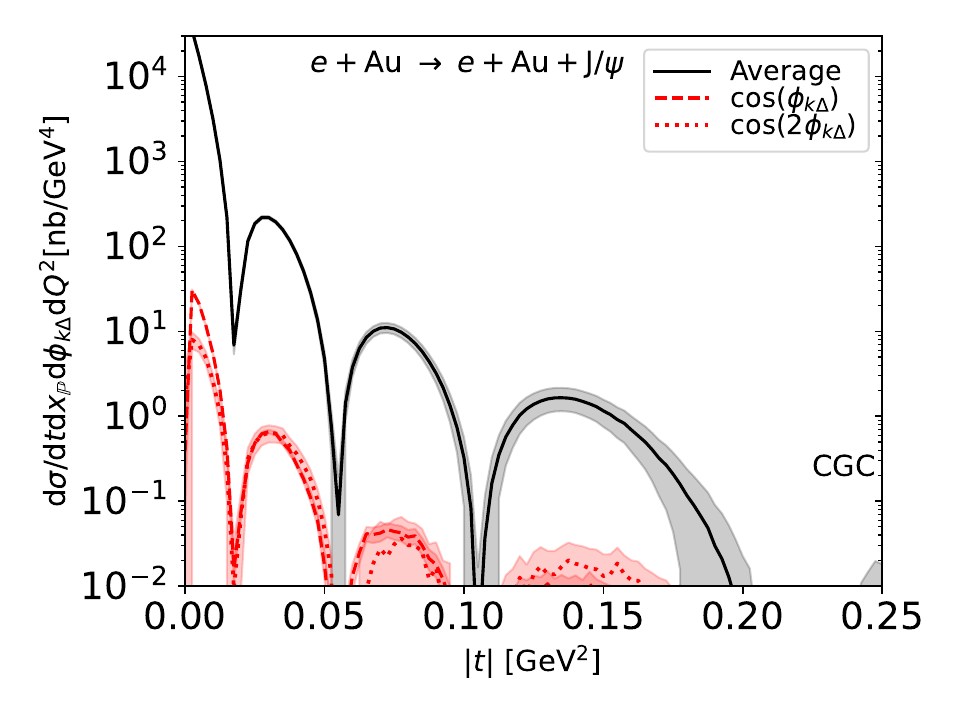}
\label{fig:cgc_Au_jpsi}
}
\caption{Different contributions to DVCS (at $Q^2=5\gev^2$) and $\jpsi$ (at $Q^2=2\gev^2$) production off gold nuclei at $\xpom=0.01$.}
\end{figure*}

We now move to the discussion of $\jpsi$ production. Here, the heavy mass of the charm quark suppresses the kinematical contributions as already discussed in case of the GBW dipole in Section~\ref{sec:gbw_numerics}. 

The coherent $\jpsi$ production cross-section and different azimuthal modulation contributions are shown in Fig.~\ref{fig:cgc_jpsi_spectra_xp_001} at $\xpom=0.01$, corresponding to the initial condition, and in Fig.~\ref{fig:cgc_jpsi_spectra_xp_0001} at $\xpom=0.001$ after the JIMWLK evolution, both for $Q^2=2\,{\rm GeV}^2$. As a result of the growing proton size, the location of the first diffractive minimum in the angle independent coherent cross-section (and in the $\cos(2\phik)$ modulation) moves towards smaller $|t|$. As mentioned above, the heavy quark mass suppresses the kinematical contribution to the modulation, leading to a negligible $\cos(\phik)$ modulation. On the other hand, the $\cos(2\phik)$ modulation remains sizeable, of the order of a few percent, as it is dominated by the intrinsic contribution.  

The $v_n$ coefficients defined in Eq.~\eqref{eq:vndef} as a function of the squared momentum transfer $|t|$ are shown in Fig.~\ref{fig:cgc_jpsi_vn}. The $v_2$ modulation is now approximately one half of the one seen in case of DVCS. The smaller elliptic modulation compared to DVCS is expected for the same reason that larger $Q^2$ suppresses the modulation, as discussed above.

The $v_2$ modulation is again strongly suppressed with decreasing $\xpom$. Now in the $\jpsi$ production case, the change in inelasticity $y$ as a function of $\xpom$ affects the modulation coefficients more compared to DVCS, and approximately one half of the total $\xpom$ dependence seen in Fig.~\ref{fig:cgc_jpsi_vn} is explained by the change in $y$. The other half is a result of the perturbative JIMWLK evolution resulting in a larger proton with smaller density gradients. For a detailed comparison between the JIMWLK effects and the effect of changing $y$, see Appendix~\ref{appendix:y}.  The $v_1$ coefficient is found to be at the per mill level.

The virtuality dependence of $v_2$ (not shown) is qualitatively similar in $\jpsi$ production and in DVCS. Contrary to the DVCS results shown in Fig.~\ref{fig:cgc_dvcs_v2_q2dep}, the $v_1$ in $\jpsi$ production is positive and has a smaller magnitude as already observed above. At $\xpom=0.001$, we note that the kinematically allowed region in $\jpsi$ production is $Q^2<10\gev^2$.

In the following, let us briefly address existing experimental data, that can be compared to our results.
The H1 and ZEUS collaborations have measured~\cite{Aktas:2005xu,Chekanov:2004mw} spin density matrix components in $\jpsi$ electroproduction, that can be directly related to the $v_n$ coefficients studied here. The H1 data~\cite{Aktas:2005xu} is compatible with the $s$-channel helicity conservation (SCHS) assumption, in which case  contributions from the polarization changing and helicity flip amplitudes are zero. For the $t$ integrated $v_n$ coefficients in $\jpsi$ electroproduction at low $Q^2$, the H1 data corresponds to $|v_2| \lesssim 5\%$ and $|v_1|\lesssim 10\%$, compatible with our results. 

The H1 collaboration has also measured the spin density matrix elements in exclusive $\rho$ production, with the results reported in Ref.~\cite{Aaron:2009xp}. As the $\rho$ meson is much lighter, it is sensitive to larger dipoles and especially the kinematical contribution to $v_n$ is larger.
The spin density matrix elements measured by H1 correspond to $t$-integrated $v_1 =3 \dots 15\%$ at low and moderate $Q^2$. 
When $\rho$ production is calculated within our CGC framework using the wave function of Ref.~\cite{Kowalski:2006hc},
we find a comparable modulation.
The $t$ integrated $v_2$ is on the order of a few percent, which can not be accurately compared with the HERA data due to the large experimental uncertainties. 

\subsection{Large nucleus target in the CGC}
\label{sec:large_nucleus}

Let us next consider large nuclear targets, where density gradients are generally smaller compared to the proton, which suppresses the intrinsic contribution of the azimuthal modulations. 

The average DVCS cross-section in $e+\mathrm{Au}$ scattering at $\sqrt{s}=90\gev,Q^2=5\gev^2$, and the $\cos(\phik)$ and $\cos(2\phik)$ modulations are shown in Fig.~\ref{fig:cgc_Au_dvcs}  (see also Ref.~\cite{Goncalves:2015goa}, where this process is studied in the EIC kinematics, without considering angular correlations).  The $v_2$ modulation is found to be extremely small due to the small density gradients, except at very large $|t| >1/R_A^2$ where $R_A$ is the nuclear radius. However, in this region the incoherent contribution dominates, rendering the coherent process difficult to measure. The $v_1$ modulation, which is dominated by the kinematical contribution, is an order of magnitude larger than $v_2$ at small $|t|$, but still at a few percent level.

The cross-section for coherent $\jpsi$ production off gold nuclei at $\xpom=0.01$ and $Q^2=2\,{\rm GeV}^2$ is shown in Fig.~\ref{fig:cgc_Au_jpsi}. As a result of having a small intrinsic contribution and a large quark mass rendering also the kinematical contribution  small, both the $\cos(\phik)$ and $\cos(2\phik)$ modulations are negligible, at the per mill level. In both DVCS and $\jpsi$ production, the azimuthal modulation coefficients exhibit the same diffractive pattern as the average coherent cross-section. In contrast to $e+p$ scattering studied previously, the signs of the modulation terms do not change at the diffractive minima.

\begin{figure}[tb]
\includegraphics[width=0.5\textwidth]{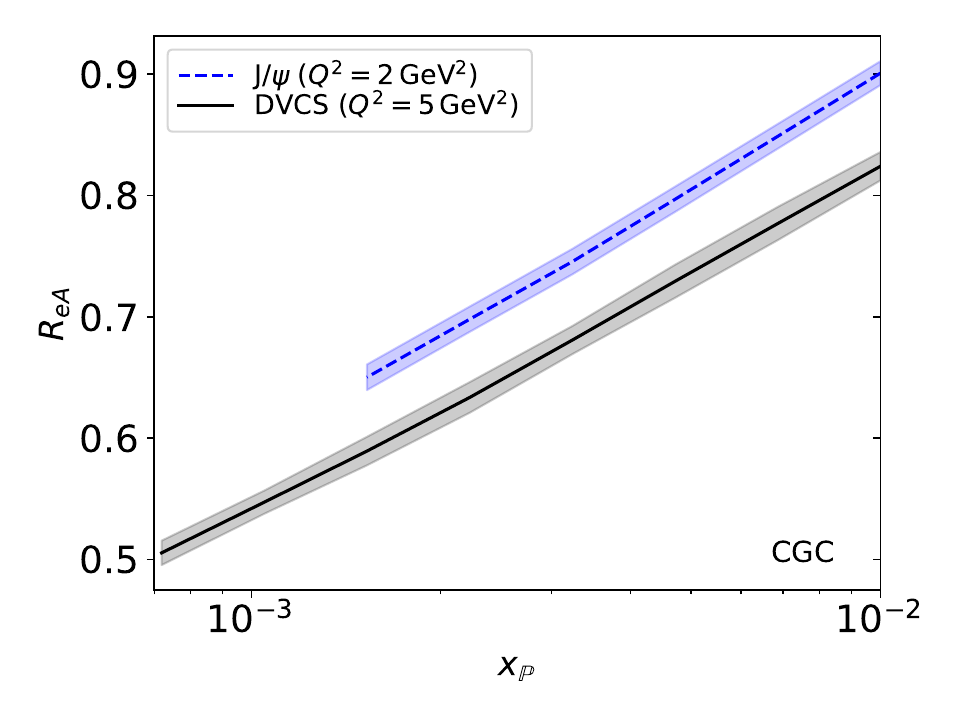} 
\caption{
Nuclear suppression factor as a function of $\xpom$ for exclusive electroproduction of $\jpsi$  ($Q^2=2\gev^2$) and $\gamma$ ($Q^2=5\gev^2$) in the EIC kinematics ($\sqrt{s}=90\gev$) at $t=0$. 
}
\label{fig:nuclear_suppression}
\end{figure}

As exclusive processes are especially powerful in resolving non-linear effects that are enhanced in heavy nuclei, we also compute the nuclear suppression factor at small $|t|$, defined as \begin{equation}
    R_{eA} = \left. \frac{\der \sigma^{e+A \to e+A+V}/\der t \der Q^2 \der \xpom}{A^2 \der \sigma^{e+p \to e+p+V}/\der t \der Q^2 \der \xpom}\right|_{t=0},    
\end{equation}
where $V=\gamma,\jpsim$. The nuclear suppression factors as a function of $\xpom$ computed at $Q^2=2\gev^2$ in case of $\jpsi$ production and at $Q^2=5\gev^2$ in DVCS are shown in Fig.~\ref{fig:nuclear_suppression}. The results are shown in the $\xpom$ range accessible at the EIC with $\sqrt{s}=90\gev$. Note that the coherent cross-section scales as $A^2$ at $t=0$ in the dilute limit~\cite{Ryskin:1992ui,Mantysaari:2017slo}, and consequently $R_{eA}=1$ in the absence of non-linear effects. As discussed in Appendix~\ref{appendix:polflip}, the polarization changing terms have a negligible contribution to the azimuthal angle independent cross-sections entering in $R_{eA}$. 

We predict a significant nuclear suppression in the EIC kinematics, down to $R_{eA}\approx 0.5$ in case of DVCS at the smallest reachable $\xpom$ values. $R_{eA}$ for coherent $\jpsi$ production is approximately $0.05$ above the values for DVCS in the considered kinematics where $Q^2$ values are different. The obtained suppression is comparable to what was found in Ref.~\cite{Mantysaari:2018nng} for $\jpsi$ production using the IPsat parametrization, where the $\xpom$ dependence is parametrized and not a consequence of the perturbative small $x$ evolution (see also Ref.~\cite{Mantysaari:2017slo}). The $\jpsi$ photoproduction data from LHC suggests slightly stronger suppression $R_{eA}\sim 0.4 \dots 0.6$ in this kinematical domain~\cite{Guzey:2020ntc}.

As small $|t|$ also dominates the $t$ integrated coherent cross-section, we  expect non-linear effects of similar magnitude in that case too (see also discussion in Ref.~\cite{Mantysaari:2018nng}). The advantage of examining the ratio at $t=0$ is that in that case there is no need for ($\xpom$ dependent) proton and nuclear form factors used in the impulse approximation to transform the photon-proton cross-section to the photon-nucleus case in the absence of non-linear effects. This would be particularly complicated here, as the effective proton form factor includes a larger contribution from the effective projectile size in the DVCS case compared to the $\jpsi$ production. 

\begin{figure}[tb]
\includegraphics[width=0.5\textwidth]{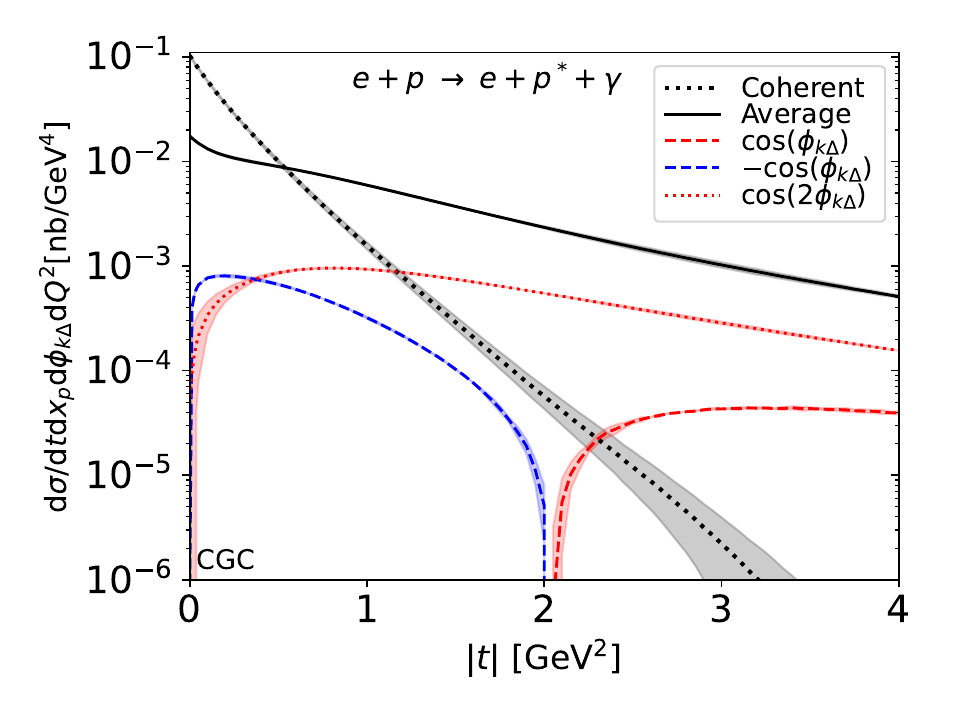}
\caption{Incoherent DVCS cross-section and its modulations at $Q^2=5\gev^2$ as a function of squared momentum transfer. Proton shape fluctuations are included, and the average coherent cross-section is also shown for comparison. 
}
\label{fig:incoh_dvcs_spectra}
\end{figure}

\begin{figure*}[tb]
\subfloat[Incoherent $v_1$]{
\includegraphics[width=0.48\textwidth]{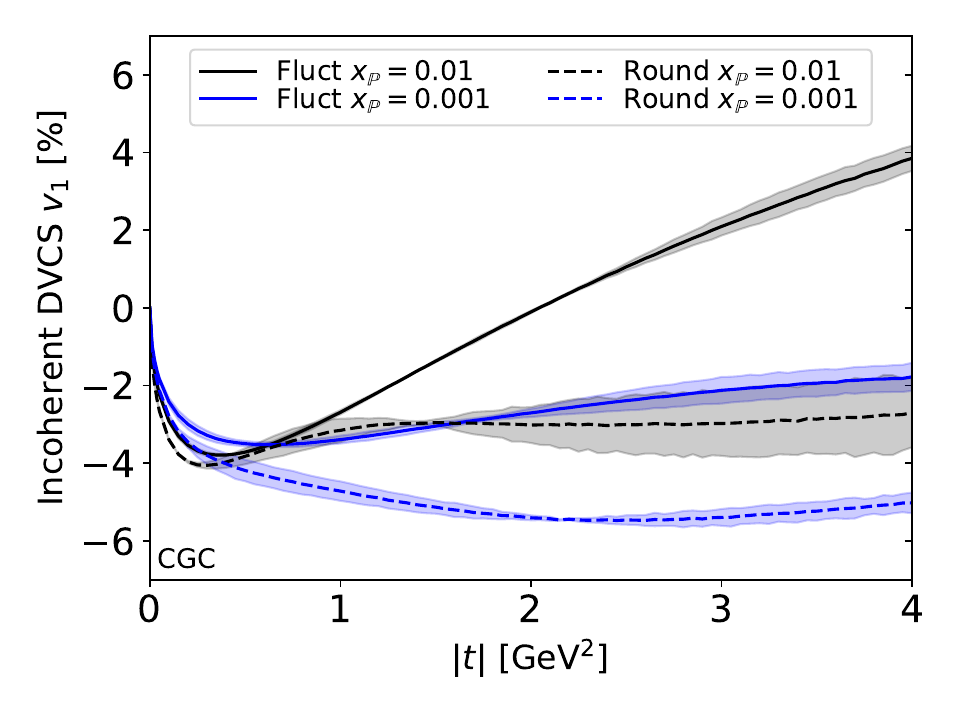}
\label{fig:incoh_v1}
}
\subfloat[Incoherent $v_2$]{%
\includegraphics[width=0.48\textwidth]{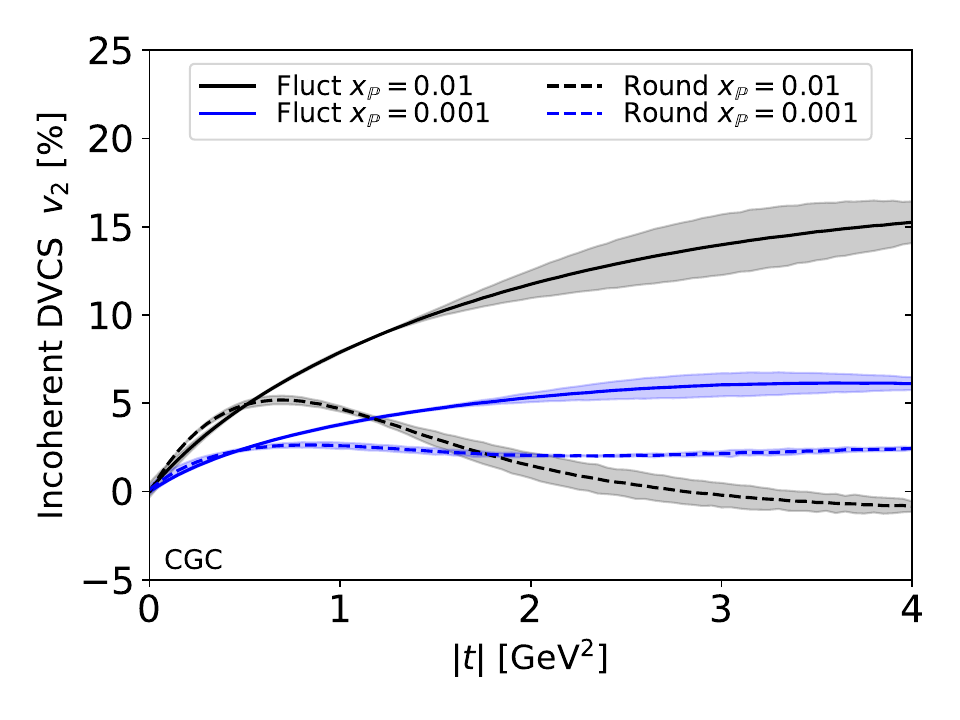}
\label{fig:incoh_v2}
}
\caption{Evolution of the  $v_n$ coefficients in  incoherent DVCS  at $Q^2=5\gev^2$. }
\end{figure*}

\subsection{Incoherent scattering}
\label{sec:incoh_numerics}

The incoherent channel is interesting for two reasons. First, as incoherent processes are sensitive to the fluctuations of the scattering amplitude, angular modulations in the incoherent cross-sections can be sensitive to the details of the fluctuating target geometry. Second, the incoherent cross-section dominates at large $|t|$ where kinematical contributions to the azimuthal modulations are more important, and as such simultaneous analysis of coherent and incoherent cross-sections can potentially be used to distinguish intrinsic and kinematical contributions. 
In order to access fluctuations at different length scales, we consider both $\jpsi$ and DVCS processes.

The cross-section and its modulations for the incoherent DVCS process $e+p \to e+p^*+\gamma$ at $\sqrt{s}=140\gev$ and $\xpom=0.01$ are shown in Fig.~\ref{fig:incoh_dvcs_spectra}. Here, the proton
shape fluctuations constrained by the $\jpsi$ production data are included.
Similar to coherent DVCS, especially the $\cos(2\phik)$ modulation is sizeable, and the $\cos(\phik)$  modulation changes  sign at $|t| \sim 2\gev^2$.

To study the evolution of incoherent $v_1$ and $v_2$ coefficients, and quantify the effect of proton shape fluctuations, these coefficients are shown in Figs.~\ref{fig:incoh_v1} and \ref{fig:incoh_v2} at two different $\xpom$ values within the EIC kinematics, calculated with and without fluctuating proton substructure. We recall that in DVCS the change in $y$ as a function of $\xpom$ has a negligible effect on modulations as shown in Appendix~\ref{appendix:y}.

At small $|t|\lesssim 0.5\gev^2$, where the intrinsic contribution arises from fluctuations in the angular dependence of the dipole scattering amplitude at long distance scales, the substructure fluctuations have only a small effect on the $v_n$ coefficients. 
 At large $|t|$ where one is sensitive to fluctuations in the amplitudes $\M{\lambda,\lambda'}$ at short length scales,  there are significant fluctuations in the angular dependence of the dipole-target scattering amplitude when proton substructure is included, which render incoherent $v_2$ large as shown in Fig.~\ref{fig:incoh_v2}. Smoother density gradients obtained as a result of small $x$ evolution suppress the dependence on the dipole orientation of fluctuations, which results in decreasing $v_2$ with decreasing $\xpom$.

 If proton substructure is not included, the non-zero incoherent cross-section and its azimuthal modulations shown in Figs. \ref{fig:incoh_v1} and \ref{fig:incoh_v2} are consequences of the color charge fluctuations in the target proton. These fluctuations take place at a distance scale $\sim 1/Q_s^2$, which decreases with decreasing $x$. Consequently, since at large $|t|$ these appearing short scale structures can be resolved by the scattering dipole, they result in larger fluctuations in the dependence on the dipole orientation and lead to increasing $|v_n|$ with decreasing $\xpom$ (see also the related discussion in Ref.~\cite{Mantysaari:2019jhh}).

\begin{figure*}[tb]
\subfloat[Spectra, no substructure]{%
\includegraphics[width=0.5\textwidth]{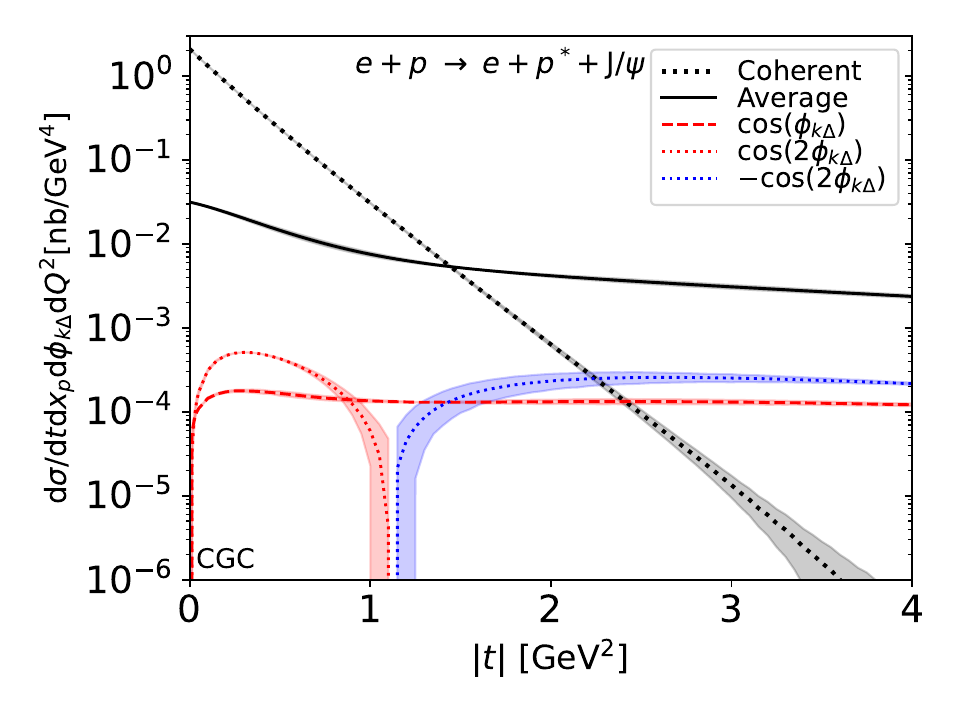}

\label{fig:incoh_jpsi_p_nofluct}
}
\subfloat[Spectra, with substructure]{%
\includegraphics[width=0.5\textwidth]{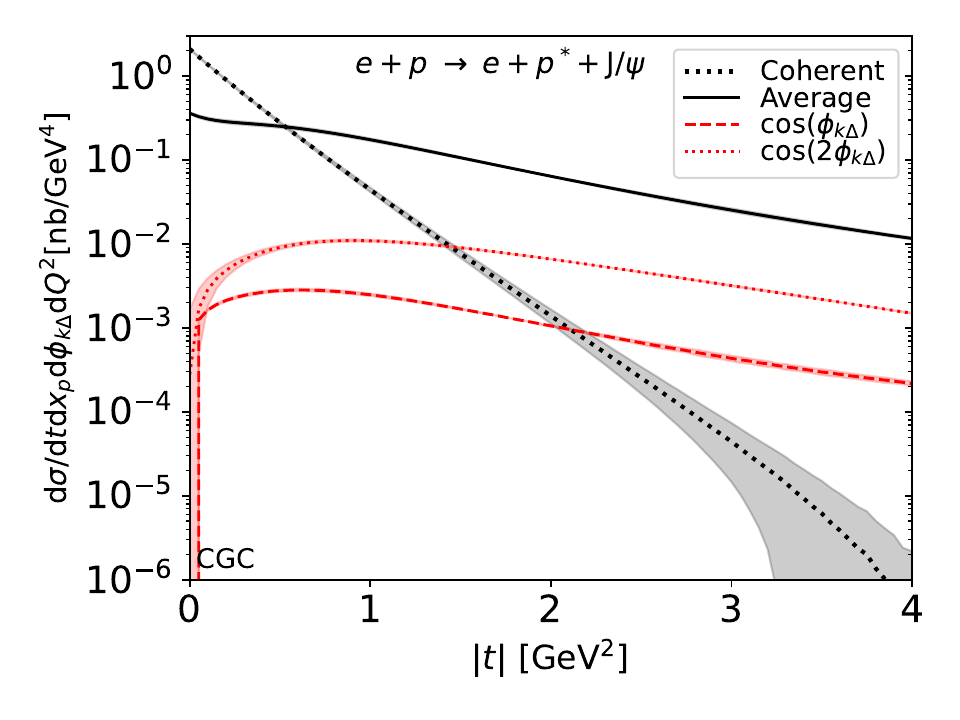}
\label{fig:incoh_jpsi_p_fluct}
}

\subfloat[Incoherent $v_1$]{%
\includegraphics[width=0.5\textwidth]{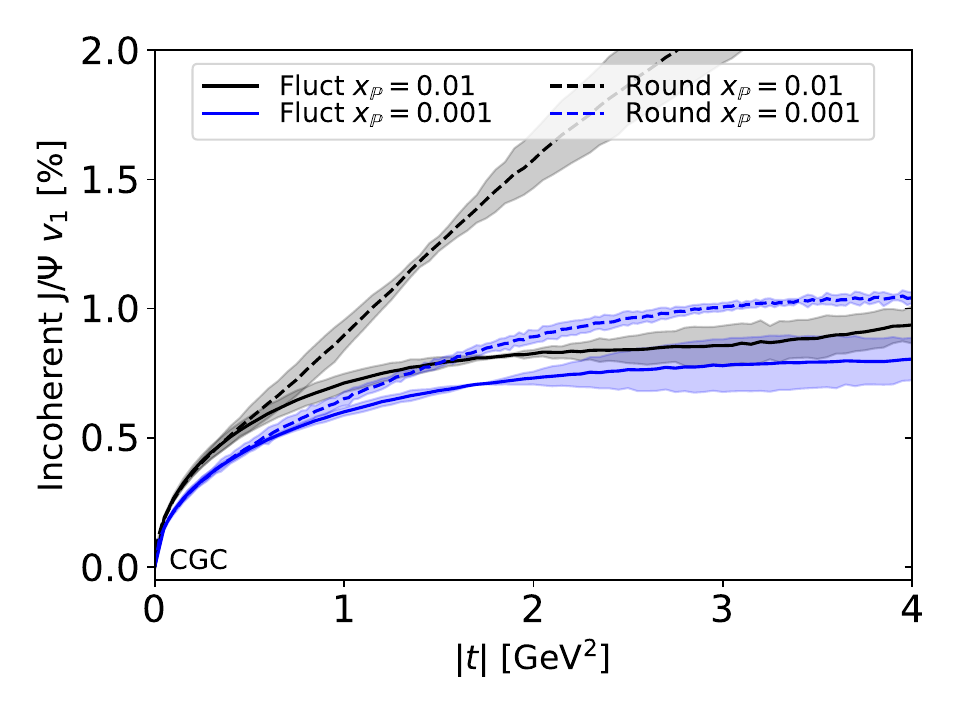}
\label{fig:incoh_jpsi_p_v1_evol}
}
\subfloat[Incoherent $v_2$]{%
\includegraphics[width=0.5\textwidth]{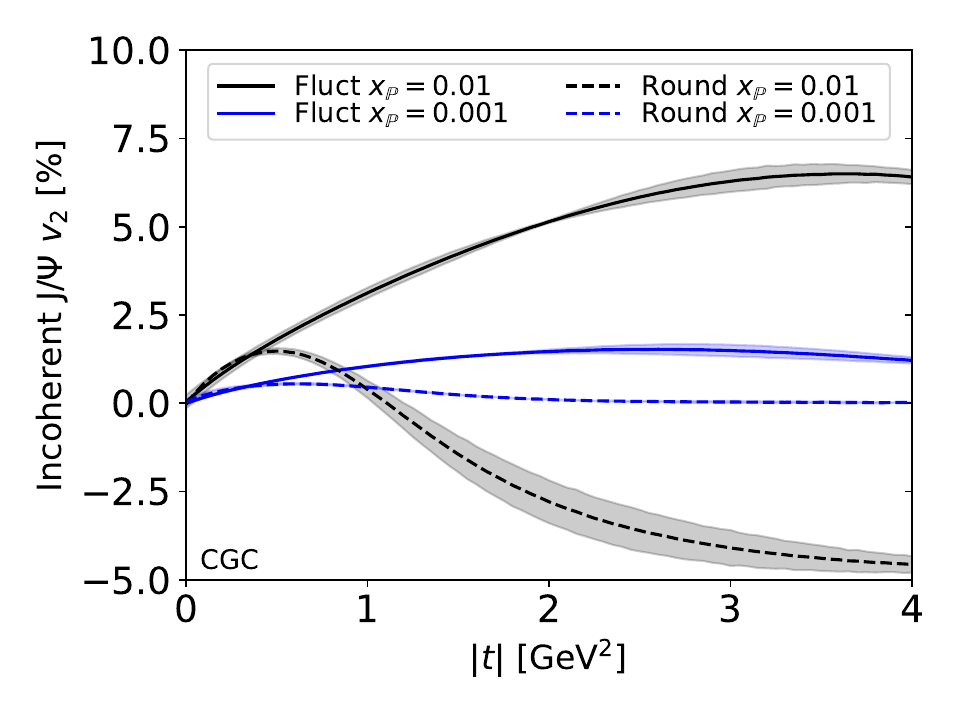}
\label{fig:incoh_jpsi_p_vn_evol}
}
\caption{Upper panel: incoherent $\jpsi$ production cross-section and its modulations with and without proton substructure fluctuations. For comparison the average coherent cross-section is also shown. Lower panel: evolution of the $v_n$ modulation coefficients. All results are at $Q^2=2\gev^2$. }
\end{figure*}

Next we consider incoherent $\jpsi$ production in $e+p \to e+p^*+\jpsim$. The cross-section and its azimuthal modulations are shown in  Fig.~\ref{fig:incoh_jpsi_p_nofluct} (with the spherical proton) and in Fig.~\ref{fig:incoh_jpsi_p_fluct} (with proton shape fluctuations) at the initial $\xpom=0.01$. For comparison we again show the average coherent cross-section studied above.
Although the coherent cross-section probes only the average interaction, it is also slightly altered especially at large $|t|$ by the substructure fluctuations, as the  dependence on the proton density profile given by Eq.\,\eqref{eq:Tpfluct} is non-linear. 
We emphasize that this dependence is an artefact resulting from our choice of the parametrization for the substructure fluctuations, and in principle it would be possible to also construct a fluctuating substructure that results in exactly the same coherent cross-section (or average dipole-target interactions).

The substructure fluctuations significantly increase both the average incoherent cross-section and the magnitude of the azimuthally dependent terms (especially the $\cos (2\phik)$ component) in the studied $|t|$ range.

Additionally, the fluctuations remove the sign change from the incoherent $\cos(2\phik)$ modulation around $|t|\approx 1\gev^2$, and result in an exponential incoherent spectrum instead of a power law (at $|t|\gtrsim 1\gev^2$). 

The extracted modulation coefficients $v_1$ and $v_2$ in incoherent $\jpsi$ production with and without proton shape fluctuations are shown in Figs.~\ref{fig:incoh_jpsi_p_v1_evol} and~\ref{fig:incoh_jpsi_p_vn_evol}.
The large quark mass in $\jpsi$ production suppresses the kinematical contribution to the polarization changing amplitudes, reducing the $v_1$ modulation compared to the DVCS case. Although $\jpsi$ production is generically sensitive to shorter length scales (smaller dipoles) compared to DVCS, we again find that the modulation coefficients begin to be affected by the substructure at $|t| \approx 0.5\gev^2$, similar to the case of DVCS. This is because the momentum transfer region where one is sensitive to the possible substructure fluctuations is mostly controlled by the size of the substructure.

For $|t|\gtrsim 0.5\gev^2$, we predict significantly different $v_2$ with and without substructure fluctuations, even with opposite signs as the substructure fluctuations remove the sign change in $\cos(2\phik)$ spectra.  A clear effect of energy evolution is also visible, with the modulation almost disappearing in the $\xpom$ range accessible at the future EIC (recall that approximately one half of the $\xpom$ evolution is a result of JIMWLK dynamics; see Appendix~\ref{appendix:y}). This suggests that a simultaneous description of the total incoherent cross-section and its $v_2$ modulation could potentially be used to probe more precisely the details of the event-by-event fluctuating substructure, including its $\xpom$ dependence. A more detailed analysis aimed at constraining the substructure geometry and its fluctuations is left for future work.

\section{Conclusions}
\label{sec:conclusions}

We have calculated cross sections and their modulations in azimuthal angle $\phik$ between the exclusively produced particle and the outgoing electron in DVCS and $\jpsi$ production in electron-proton and electron-nucleus scattering. We extracted $\cos(\phik)$ and $\cos(2\phik)$ modulations and identified two separate contributions: one of kinematical origin, and another \emph{intrinsic} component, which is a result of non-trivial angular correlations in the target wave function at small $x$. 
By using dipole models with and without intrinsic contribution to the cross-section, we demonstrated that especially the $\cos(2\phik)$ modulation is a sensitive probe of spatial angular correlations in the gluon distribution, which in the collinear limit can be reduced to the gluon transversity GPD \cite{Hatta:2017cte}.

Using a realistic CGC based setup where the energy (or $\xpom$) dependence of the target can be calculated, we obtained predictions for the azimuthal modulations in DVCS and $\jpsi$ production at the future EIC. Especially for DVCS we predict a significant elliptic modulation, up to $15\%$, and a clear energy evolution in the EIC energy range. The modulations in $\jpsi$ production are roughly an order of magnitude smaller, but potentially still measurable. 

We have also studied incoherent DVCS and $\jpsi$ production, and demonstrated that the momentum transfer dependence of the modulation coefficients can potentially be used to constrain the detailed properties of the target's fluctuating substructure. A more detailed analysis of this possibility is left for future work.

In the scattering of electrons with a heavy nucleus the modulations were found to be highly suppressed as a result of smaller density gradients at the probed distance scales, which strongly suppresses the intrinsic contribution to the modulation coefficients. On the other hand, the angle averaged cross sections for these processes with heavy nuclear targets are sensitive to non-linear effects, and we predict significant nuclear suppression factors as low as $0.5$ in the realistic EIC kinematics.

Our results provide the first explicit predictions from the CGC EFT for the azimuthal modulations measurable at the EIC. To match the expected high precision of future EIC measurements, the presented leading order calculations (which do include a resummation of high energy logarithms $\as \ln 1/x$, to all orders) should be promoted to next-to-leading order (NLO) accuracy. Recently, there has been tremendous progress in the field towards NLO, with the photon wave function and vector meson production impact factors calculated to NLO accuracy~\cite{Lappi:2016oup,Boussarie:2016bkq,Boussarie:2016ogo,Hanninen:2017ddy,Escobedo:2019bxn,Lappi:2020ufv}. Similarly, the small $x$ evolution equations are available at this order in $\as$~\cite{Balitsky:2008zza,Balitsky:2013fea,Lappi:2016fmu,Lappi:2015fma,Lappi:2020srm}, and first phenomenological results have been obtained recently, e.g. for the structure functions~\cite{Beuf:2020dxl}. We will explore performing NLO calculations of the presented processes in future work. Additionally, it is important to study how the Bethe-Heitler contribution neglected in this work affects the azimuthal modulations in DVCS in a realistic EIC setting.   It would also be interesting to compare our results to predictions based on the collinear framework of GPDs, which also incorporates DGLAP evolution.

\section*{Acknowledgments}
We thank T. Lappi for useful discussions.  We are grateful to Emilie Li for pointing out an error in one of the elements of the lepton tensor $\widetilde{L}_{\lambda \bar{\lambda}}$ in an earlier version of this manuscript.
H.M. is supported by the Academy of Finland project 314764, and by the European Research Council project STRONG-2020 (grant agreement No 824093). F.S. and B.P.S. are supported under DOE Contract No.~DE-SC0012704. Computing resources from CSC – IT Center for Science in Espoo, Finland and from the Finnish Grid and Cloud Infrastructure (persistent identifier \texttt{urn:nbn:fi:research-infras-2016072533}) were used in this work. F.S and K.R. were also supported by the joint Brookhaven National Laboratory-Stony Brook University Center for Frontiers in Nuclear Science (CFNS).

\appendix

\section{Importance of the polarization changing contributions}
\label{appendix:polflip}
\begin{figure}[tb]
\includegraphics[width=0.5\textwidth]{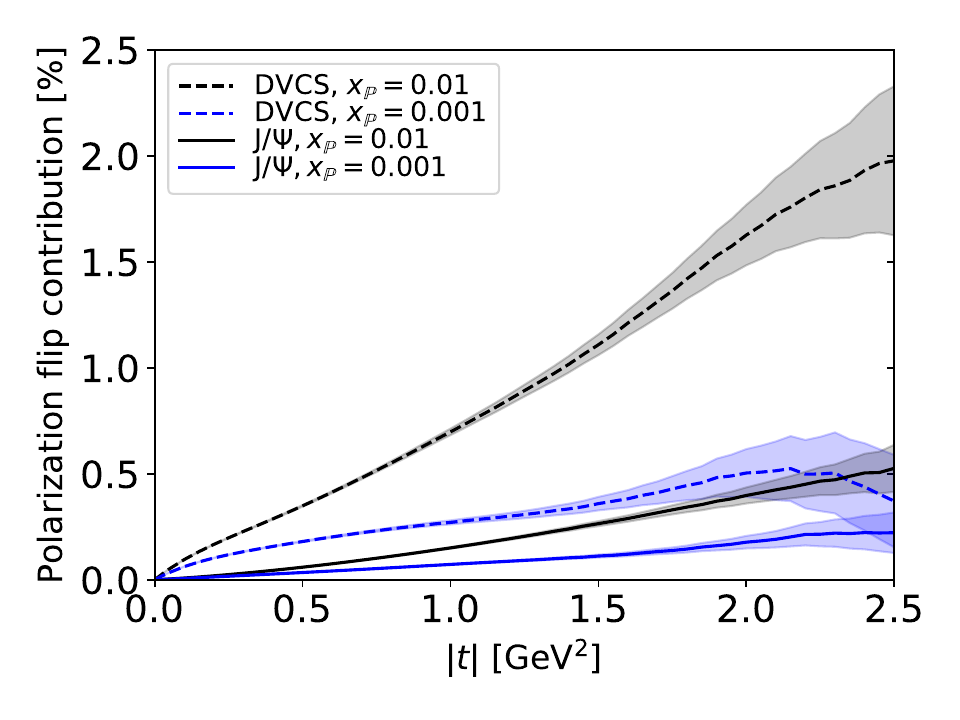}
\caption{Relative importance of the polarization changing terms in the angle averaged cross section for exclusive 
$\jpsi$ production at $Q^2=2\gev^2$ and in DVCS at $Q^2=5\gev^2$.
}
\label{fig:polflip_contrib}
\end{figure}

In this Appendix we determine the importance of the polarization changing contributions, not included in previous phenomenological analyses, for the angle averaged cross sections. To do so, we compute the relative importance of the contributions in electron-proton scattering where the produced particle ($\gamma$ or $\jpsi$) has a different polarization than the incoming photon, allowing $\lambda=0\to \pm 1$, $\lambda=\pm 1\to 0$ and $\lambda=\pm 1 \to \mp 1$  processes. The results at two different $\xpom$ are shown in Fig.~\ref{fig:polflip_contrib}. The polarization changing contributions are suppressed as a result of the small $x$ evolution as expected, as the smaller density gradients in the larger proton suppress the intrinsic contribution especially in $\langle \M{\pm 1, \mp 1} \rangle_Y$ which contributes to the angle independent cross-section, see Eq.~\eqref{eq:xs}. 
In vector meson production these polarization changing contributions are negligible when compared to other uncertainties related to e.g. model dependence in the vector meson wave functions. In DVCS, the correction is also moderate, at most two percent in the studied kinematical domain.

\section{Separating genuine JIMWLK evolution from inelasticity dependence}
\label{appendix:y}
\begin{figure}[tb]
\includegraphics[width=0.5\textwidth]{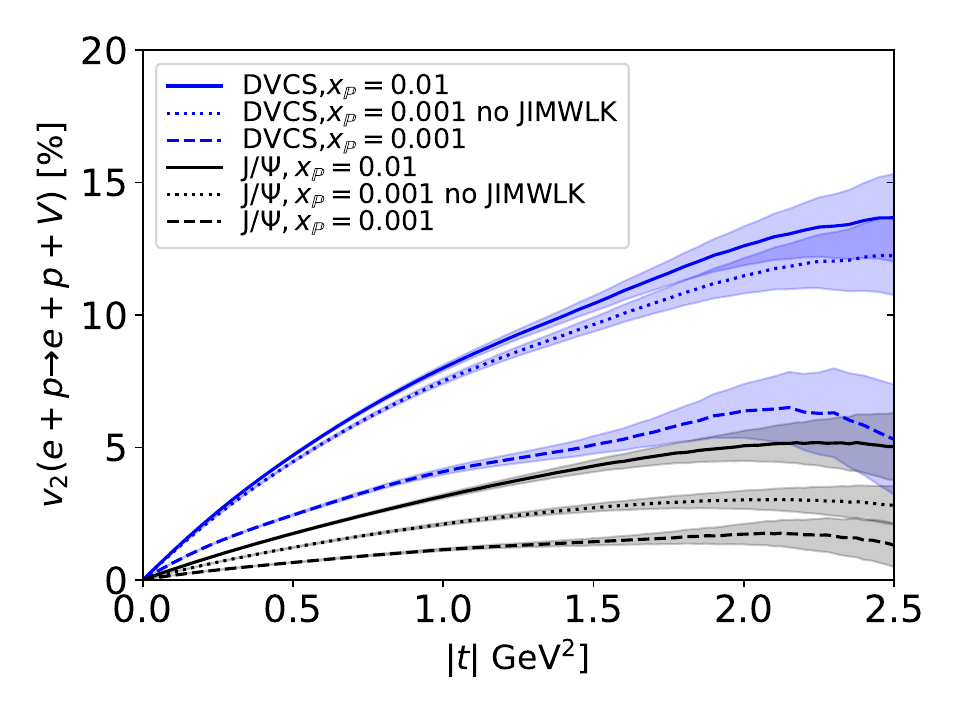}
\caption{Elliptic modulation in DVCS at $Q^2=5\gev^2$ and $\jpsi$ production at $Q^2=2\gev^2$ at $\xpom=0.01$ and $\xpom=0.001$ with and without JIMWLK evolution. }
\label{fig:jimwlk_no_jimwlk}
\end{figure}

The inelasticity $y$ affects the distribution of virtual photon polarization states, and as such the modulation coefficients defined by Eq.\,\eqref{eq:vndef} are a function of $y$, as is apparent from Eq.~\eqref{eq:xs}. In this work we choose to calculate the cross-section and azimuthal modulations at fixed $\xpom$, as it controls the amount of small $x$ evolution. Then, by fixing the incoming photon virtuality $Q^2$ and calculating the cross-section as a function of $|t|$ the kinematics becomes fully determined, and the inelasticity can be written as
\begin{equation}
    y = \frac{M_V^2 + Q^2 - t}{s \xpom},
\end{equation}
where $M_V^2$ is the invariant mass of the produced particle.
Thus, when interpreting our results one has to keep in mind that $y$ depends on both $\xpom$ and $t$, and in particular all modulations vanish in the $y\to 1$ limit. Note that it is not possible to simply fix $y$, as that would correspond to a varying center-of-mass energy $\sqrt{s}$ when selecting independent $t$, $\xpom$, and $Q^2$. 

To demonstrate that the $\xpom$ and $t$ dependencies observed in the modulation coefficients in Section~\ref{sec:results} are genuine probes of small $x$ dynamics and not only effects of trivial $y$ dependence, we study in this Appendix the modulation coefficients by neglecting the JIMWLK evolution. 

The elliptic modulation $v_2$ in DVCS and $\jpsi$ production in electron-proton collisions at $\sqrt{s}=140\gev$ is shown in Fig.~\ref{fig:jimwlk_no_jimwlk}. The results at $\xpom=0.01$ (initial condition for the JIMWLK evolution) and at $\xpom=0.001$ (after the evolution) are identical to those shown in Section~\ref{sec:results}. For comparison, we also calculate the same $v_2$ coefficients at $\xpom=0.001$, but now neglecting the JIMWLK evolution. This corresponds to using the same dipole-target scattering amplitude as at the initial condition ($\xpom=0.01$), but evaluating the process kinematics (namely the inelasticity $y$) at $\xpom=0.001$.

In case of DVCS, the $y$ values are generically smaller, and as such the change in $y$ as a function of $t$ or $\xpom$ in the studied range only slightly alters the $y$ dependent factors in Eq.~\eqref{eq:xs}. If the JIMWLK evolution is neglected, $v_2$ at $\xpom=0.01$ and $\xpom=0.001$ are almost identical. Thus, nearly all the observed $\xpom$ dependence is a result of the genuine JIMWLK dynamics.

In exclusive $\jpsi$ production the large invariant mass renders $y$ larger, and consequently the change in $y$ with decreasing $\xpom$ has a numerically larger effect. As illustrated in Fig.~\ref{fig:jimwlk_no_jimwlk}, the change in $y$ explains approximately one half of the $\xpom$ dependence seen in $v_2$, which suggests that the effects of JIMWLK evolution should still be visible at the EIC. 

\section{Evaluation of Dirac structures}
\label{sec:DiracStructure}
In this Appendix, we provide some useful identities for the evaluation of Dirac traces and derive Eq.\,\eqref{eq:A-Dirac-p-m}.
Some useful standard identities for gamma matrices are:
\begin{align}
   \gamma^- \slashed{p} \gamma^- = 2 p^- \gamma^- , \label{eq:gammapgamma}
\end{align}
\begin{align}
    \Tr \left[ \gamma^i \gamma^j \gamma^m \gamma^n \right] = 4 (\delta^{ij} \delta^{mn} -\varepsilon^{ij} \varepsilon^{mn}) \, ,
\end{align}
\begin{align}
    \Tr[\gamma^{i_1}...\gamma^{i_n} \slashed{p}\gamma^-] = p^-  \Tr[\gamma^{i_1}...\gamma^{i_n}] \, , \label{eq:transverse_minus_identity}
\end{align}
where $i_{1},...,i_{n}=1,2$ are transverse indices.

Another set of useful identities involving contraction with the transverse vector $\vect{\epsilon}^{\lambda,i}$:
\begin{align}
   \Tr &\left[\left( \vect{\epsilon}^{\lambda,i}+c_1\gamma^i \gamma^j \vect{\epsilon}^{\lambda,j} \right) \right] = 4(1-c_1) \vect{\epsilon}^{\lambda,i} \label{eq:polvector-Id1} , 
\end{align}
\begin{align}
   \Tr & \left[\gamma^i \gamma^j \gamma^n \gamma^m \right] \vect{\epsilon}^{\lambda,j} \vect{\epsilon}^{\lambda'*,n} = 4(1+\lambda \lambda') \vect{\epsilon}^{\lambda,i} \vect{\epsilon}^{\lambda'*,m} \label{eq:gamma4} ,
\end{align}
\begin{align}
   \Tr & \left[\left( \vect{\epsilon}^{\lambda,i}+c_1 \gamma^i \gamma^j \vect{\epsilon}^{\lambda,j} \right)\left( \vect{\epsilon}^{\lambda'*,m}+c_2\gamma^n \gamma^m \vect{\epsilon}^{\lambda'*,n} \right) \right] \nonumber \\ &= 4 \Big[1 -c_1 -c_2 +(1+\lambda \lambda') c_1 c_2 \Big] \vect{\epsilon}^{\lambda,i} \vect{\epsilon}^{\lambda'*,m} \label{eq:polvector-Id2} ,
\end{align}
\begin{align}
   \Tr &\left[ \slashed{a} \gamma^i \gamma^- \slashed{b} \gamma^m \gamma^- \right] \vect{\epsilon}^{\lambda,i} \vect{\epsilon}^{\lambda'*,m} = - 4 a^- b^- (1+\lambda \lambda') \, . \label{eq:gamma6}
\end{align}
In deriving some of these identities we used $\varepsilon^{jk} \vect{\epsilon}^{\lambda,k} = i \lambda \vect{\epsilon}^{\lambda,j}$.

We list some useful identities for the computation of the Dirac structure in Eq.\,\eqref{eq:A-Dirac-p-m}. Recall the definition of polarization vectors in Eqs.\,\eqref{eq:PolVecInPhoton} and \eqref{eq:PolVecOutPhoton}.

For the longitudinal polarization:
\begin{align}
   (\slashed{l}+m_f)\slashed{\epsilon}(\lambda=0,q)&(\slashed{l}-\slashed{q}+m_f)\gamma^- \nonumber \\ &= -2Q (1-z) \slashed{l} \gamma^- +C_1 \gamma^- ,
   \label{eq:Dirac-longitudinal-1}
\end{align}
\begin{align}
    (\slashed{l}' -\slashed{\Delta} +m_f) & \slashed{\epsilon}^*(\lambda'=0,\Delta) (\slashed{l}' +m_f)\gamma^- 
    \nonumber \\ &= 2Q' z (\slashed{l}' -\slashed{\Delta}) \gamma^- +C_2 \gamma^- \label{eq:Dirac-longitudinal-2}.
\end{align}
~~~\\
For the transverse polarization:
\begin{align}
    (\slashed{l}+m_f)&\slashed{\epsilon}(\lambda  = \pm 1 ,q)(\slashed{l}-\slashed{q} +m_f)\gamma^- \nonumber \\
    & =  -2 \vect{l}^i  \slashed{l} \gamma^-  \left( \vect{\epsilon}^{\pm 1,i}+\frac{1}{2z}\gamma^i \gamma^j \vect{\epsilon}^{\pm 1,j} \right) \nonumber \\
    &  -  m_f \slashed{q} \gamma^i \gamma^- \vect{\epsilon}^{\pm 1,i} + C_3^i  \gamma^i \gamma^- +C_4 \gamma^- ,
    \label{eq:Dirac-transverse-1}
\end{align}
\begin{align}
    (\slashed{l}' &-\slashed{\Delta}+m_f) \slashed{\epsilon}^*(\lambda'=\pm 1,\Delta)(\slashed{l}' +m_f)\gamma^- \nonumber \\
    &=  -2  \left( \vect{l}'^m - z \vect{\Delta}^m  \right) \left( \vect{\epsilon}^{\pm 1 *,m}+\frac{1}{2z}\gamma^n \gamma^m \vect{\epsilon}^{\pm 1 *,n} \right) \slashed{l}' \gamma^- \nonumber \\
    &   +  m_f \slashed{\Delta} \gamma^m \gamma^- \vect{\epsilon}^{\pm 1 *,m} + C_5^m \gamma^m \gamma^- + C_6 \gamma^-,
    \label{eq:Dirac-transverse-2}
\end{align}
where $C_1$, $C_2$, $C_3^i$, $C_4$, $C_5^m$ and $C_6$ do not contain any spinor structure. Their precise form is not important as they will not contribute to the amplitudes of interest.

\begin{widetext}
Let us compute $A_{+ 1, - 1}$. Inserting Eqs.\,\eqref{eq:Dirac-transverse-1} and \eqref{eq:Dirac-transverse-2} in Eq. \eqref{NVCS} we obtain
\begin{align}
    A_{+1, -1}(\vect{l},\vect{l}',z) = \frac{1}{(2 q^-)^2} \Big \{ 4 \vect{l}^i (\vect{l}'^m - z \vect{\Delta}^m)  & \Tr \left[\slashed{l} \gamma^- \left( \vect{\epsilon}^{+ 1 ,i}+\frac{1}{2z}\gamma^i \gamma^j \vect{\epsilon}^{+ 1 ,j} \right) \left( \vect{\epsilon}^{- 1*,m}+\frac{1}{2z}\gamma^n \gamma^m \vect{\epsilon}^{- 1*,n} \right) \slashed{l}' \gamma^- \right]  \nonumber \\
    - m_f^2 & \Tr \left[ \slashed{q} \gamma^i \gamma^-  \slashed{\Delta} \gamma^m \gamma^- \right] \vect{\epsilon}^{+ 1 *,i} \vect{\epsilon}^{- 1,m} \Big \} \, ,
\end{align}
where we have repeatedly used  $(\gamma^-)^2 = 0$ and that the trace of odd number of gamma matrices vanishes. The second term is zero in virtue of Eq.\,\eqref{eq:gamma6}, and we use identities in Eqs.\,\eqref{eq:gammapgamma} and \eqref{eq:transverse_minus_identity} to obtain an expression for the first term only using transverse gamma matrices
\begin{align}
    A_{+ 1, - 1}(\vect{l},\vect{l}',z) = 2 z^2 \vect{l}^i (\vect{l}'^m - z \vect{\Delta}^m)   \Tr \left[ \left( \vect{\epsilon}^{+ 1 ,i}+\frac{1}{2z}\gamma^i \gamma^j \vect{\epsilon}^{+ 1 ,j} \right) \left( \vect{\epsilon}^{- 1*,m}+\frac{1}{2z}\gamma^n \gamma^m \vect{\epsilon}^{-1*,n} \right) \right] \, .
\end{align}
Finally, we compute the traces of transverse gamma matrices using Eq.\,\eqref{eq:polvector-Id1} and obtain
\begin{align}
    A_{+ 1, - 1}(\vect{l},\vect{l}',z) = -8 z (1-z) \vect{l}^i (\vect{l}'^m - z \vect{\Delta}^m)   \vect{\epsilon}^{+ 1,i} \vect{\epsilon}^{-1*,m} .
\end{align}
Other trace structures $A_{\lambda,\lambda'}$ are computed similarly using the identities provided in this appendix.
\end{widetext}

\section{Useful Integrals}
\label{sec:UsefulIntegrals}
The following two dimensional Fourier transforms are useful:
\begin{align}
    \int_{\vect{l}} \frac{e^{-i \vect{l} \cdot \vect{r}}}{ \left(\vect{l}- \vect{k} \right)^2 + \varepsilon^2} &= \frac{1}{2\pi}  K_0(\varepsilon r_\perp) e^{-i \vect{k} \cdot \vect{r}} \, , \\
    \int_{\vect{l}} \frac{(\vect{l}^i - \vect{k}^i )  e^{-i \vect{l} \cdot \vect{r}}}{  \left(\vect{l}- \vect{k} \right)^2 + \varepsilon^2} &= \frac{-i}{2\pi} \frac{\varepsilon r^i}{r_\perp} K_1(\varepsilon r_\perp) e^{-i \vect{k} \cdot \vect{r}} \, .
\end{align}
The following integral appears when performing the $l^+$ and $l'^+$ integration in the sub-amplitude in Eq.\,\eqref{eq:A-sub-p-m}: 
\begin{align}
    I(l_a,l_b,z) = \lim_{\epsilon \rightarrow 0^+} \int \frac{\der l^+}{(l^+-l_a + \frac{i \epsilon}{z})(l^+-l_b- \frac{i \epsilon}{1-z})} \, .
\end{align}
This integral is easily done via contour integration using Cauchy's residue theorem. Note that when $z<0$ $(z>1)$ both poles sit in the upper (lower) half-plane; and thus the integral is identically zero, by closing the contour in the lower (upper) half plane. 
\begin{align}
I(l_a,l_b,z) =\begin{cases}
    \frac{2\pi i}{(l_b - l_a)} & 0 < z < 1 \, ,  \\
    0 & \mathrm{otherwise} \, .
   \end{cases}
\end{align}

\section{Dipole Fourier modes and non-forward phase contribution to the amplitudes}
\label{appendix:mode_decomposition}

In order to understand the contributions to the polarization changing and helicity flip amplitudes in exclusive production, it is useful to decompose the dipole amplitude in Fourier modes. We do not consider Odderon contributions, and thus we only have even modes:
\begin{align}
    D_Y\left(\vect{r},\vect{b}\right) = \sum^{\infty}_{k=0}  D_{Y,2k}(r_\perp,b_\perp) c_k \cos (2k \phi_{r b}) \, , \label{eq:dipole_modes}
\end{align}
where $c_0=1$ and $c_{k}=2$ for $k>0$.

Using the following identity: \\~~\\
 \begin{align}
    \frac{i^n}{(2\pi)^2} \int_{\vect{b}} \!\!\!\! e^{-i \vect{\Delta} \cdot \vect{b}} \!\! \int_{\vect{r}} \!\!\!\! e^{-i \alpha \vect{\Delta} \cdot \vect{r}} e^{i n \phi_{r\Delta} } D_Y\left(\vect{r},\vect{b}\right) f(r_\perp) \notag \\
    =\sum^{\infty}_{k=0} \int_0^\infty \!\!\!\! b_\perp \der b_\perp J_{2k}(\Delta_\perp b_\perp) \int_0^\infty \!\!\!\! r_\perp \der r_\perp D_{Y,2k}(r_\perp,b_\perp)  \notag\\
     \quad \quad \quad  \times \frac{c_{2k}}{2} \left[ J_{n-2k}(\alpha \Delta_\perp r_\perp) + J_{n+2k}(\alpha \Delta_\perp r_\perp) \right] f(r_\perp) \, ,
    \label{eq:FourierIdentity}
 \end{align}
we arrive at alternative expressions for Eqs.~\eqref{eq:M++} through \eqref{eq:M+-}, which have been CGC averaged at the amplitude level, as required for exclusive production:
\begin{widetext}
Helicity preserving amplitudes:
\begin{align}
    \left\langle \Mgammastar{0,0} \right\rangle_Y &= 16 N_c e^2 q_f^2 \sum_{k=0}^\infty  \int_{0}^\infty \!\!\!\!\! b_\perp \der b_\perp  J_{2k}(\Delta_\perp b_\perp) \! \int_{0}^\infty  \!\!\!\!\! r_\perp \der r_\perp D_{Y,2k}(r_\perp,b_\perp) \! \int_{z} \! c_{k} J_{2k}(\delta_\perp r_\perp) z^2 \bar{z}^2 Q K_0(\varepsilon_f  r_\perp) Q' K_0(\varepsilon_f' r_\perp)  \, , \\
    \left \langle \Mgammastar{\pm 1,\pm 1} \right \rangle_Y &= 4 N_c e^2 q_f^2  \sum_{k=0}^\infty \int_{0}^\infty \!\!\!\!\! b_\perp \der b_\perp  J_{2k}(\Delta_\perp b_\perp) \! \int_{0}^\infty \!\!\!\!\! r_\perp \der r_\perp   D_{Y,2k}(r_\perp,b_\perp) \! \int_{z} \! c_{k} J_{2k}(\delta_\perp r_\perp) \left[ \zeta   \varepsilon_f   K_1(\varepsilon_f  r_\perp) \varepsilon_f' K_1(\varepsilon_f' r_\perp) \right. \nonumber \\  & \left. \quad \quad \quad \quad \quad \quad  \quad \quad \quad \quad \quad \quad \quad \quad \quad \quad \quad \quad  \quad \quad \quad \quad \quad \quad \quad \quad \quad \quad \quad \quad + m_f K_0(\varepsilon_f  r_\perp) m_f K_0(\varepsilon_f' r_\perp) \right] \, .
\end{align}

\begin{figure*}[tb]
\subfloat[Average dipole $ D_Y$ at $x=0.01$]{%
\includegraphics[width=0.5\textwidth]{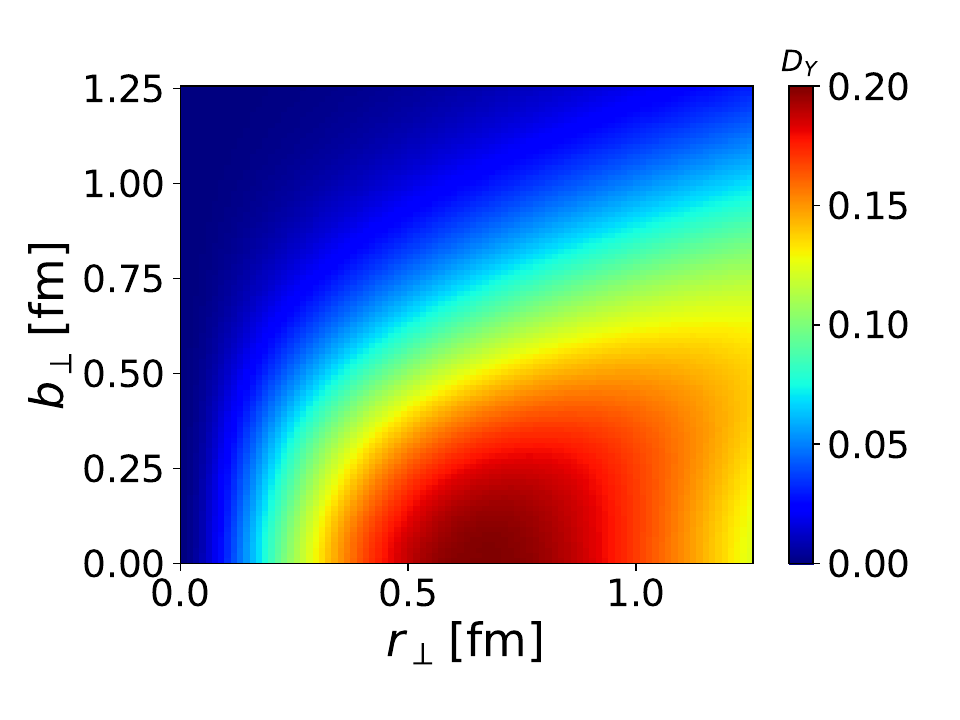}
\label{fig:average_dipole_b_r}
}
\subfloat[Dipole $\bar v_4$ at $x=0.01$]{%
\includegraphics[width=0.5\textwidth]{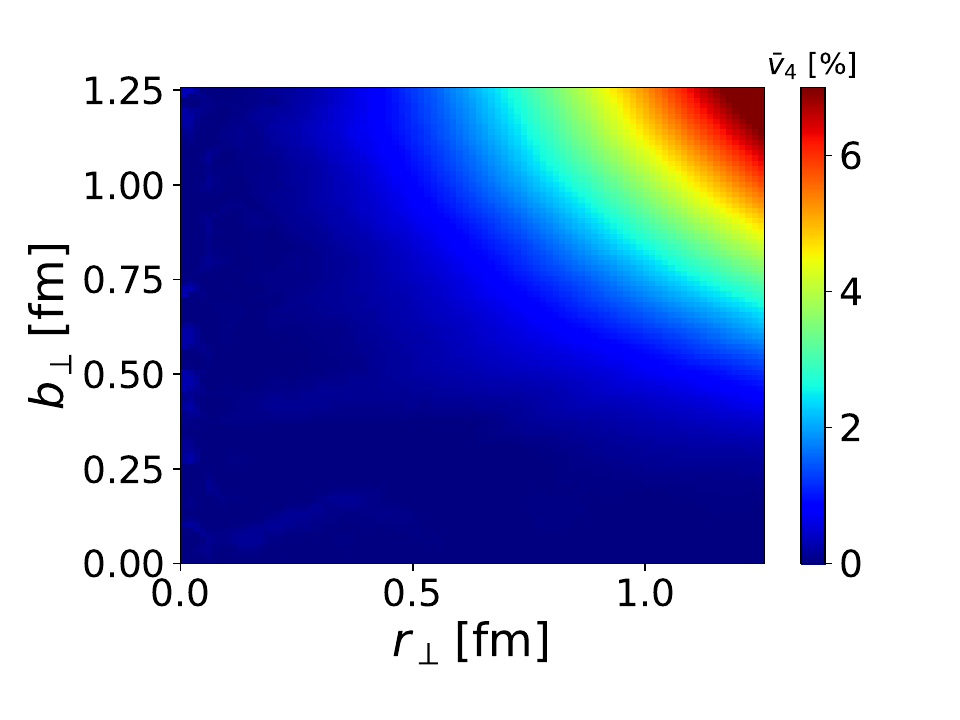}
\label{fig:dipole_v4}
} \\
\subfloat[Dipole $\bar v_2$ at $x=0.01$]{%
\includegraphics[width=0.5\textwidth]{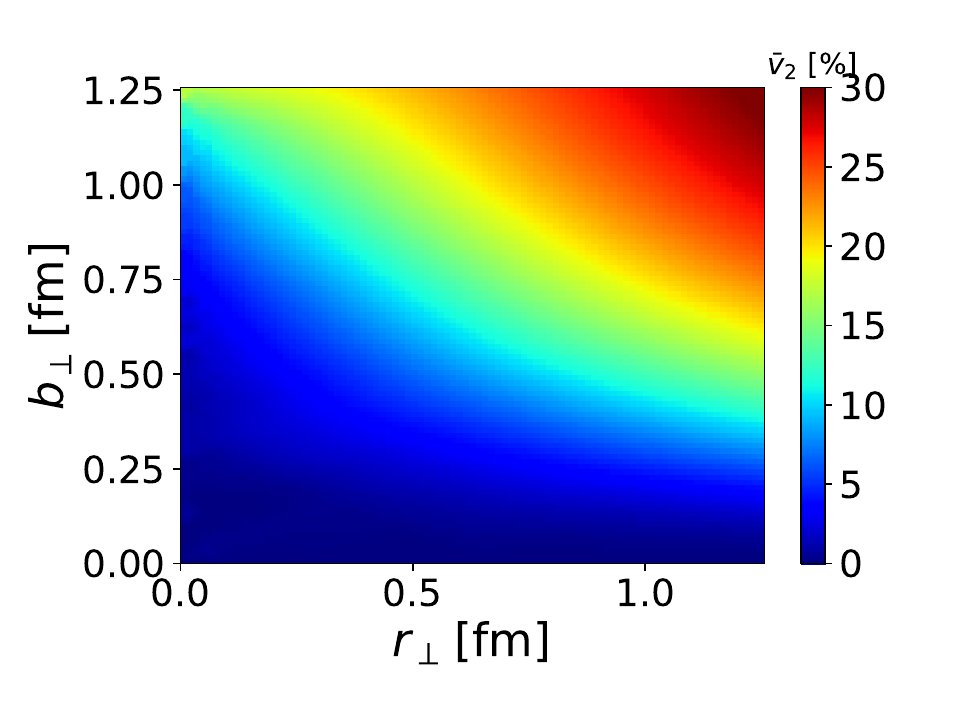}
\label{fig:dipole_v2}
}
\subfloat[Dipole $\bar v_2$ at $x=0.001$]{%
\includegraphics[width=0.5\textwidth]{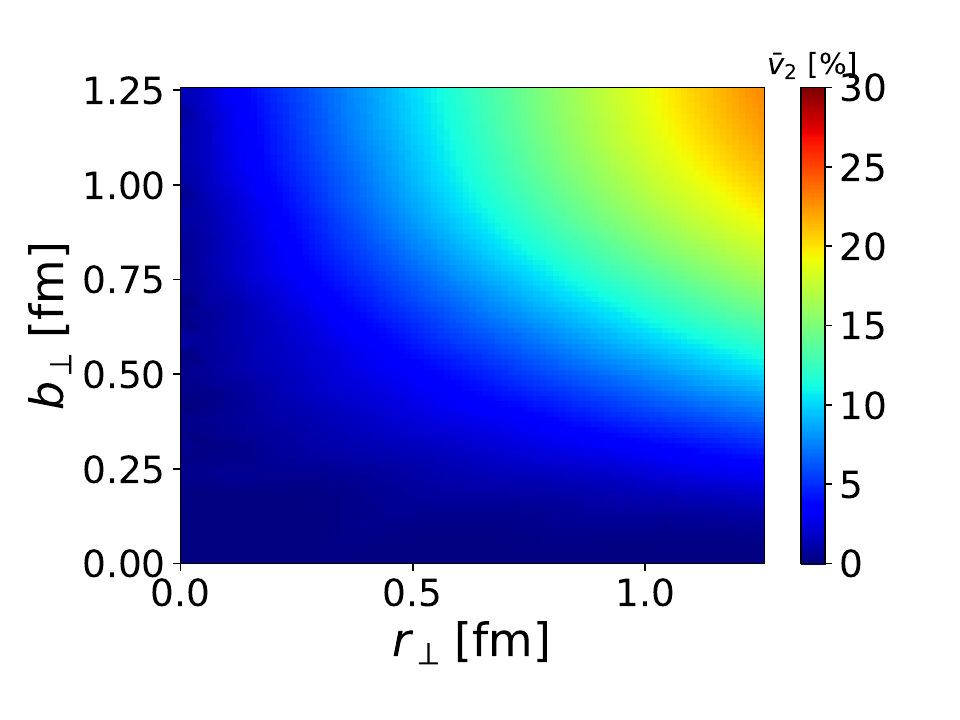}
\label{fig:dipole_v2evol}
}

\caption{Average dipole amplitude for the CGC proton and its spatial modulations as a function of dipole size $r_\perp$ and impact parameter $b_\perp$. }
\end{figure*}

Polarization changing amplitudes ($\rm{T/L \rightarrow L/T}$):
\begin{align}
    \left \langle \Mgammastar{\pm 1,0} \right \rangle_Y & \!\!= 4 \sqrt{2} N_c e^2 q_f^2 \sum_{k=0}^\infty \int_{0}^\infty \!\!\!\!\! b_\perp \der b_\perp  J_{2k}(\Delta_\perp b_\perp) \! \int_{0}^\infty \!\!\!\!  r_\perp \der r_\perp  D_{Y,2k}(r_\perp,b_\perp) \! \int_{z} \! \frac{c_k}{2} \left[ J_{1-2k}(\delta_\perp r_\perp) + J_{1+2k}(\delta_\perp r_\perp) \right] \nonumber \\  & \quad \quad \quad \quad \quad \quad  \quad \quad \quad \quad \quad \quad \quad \quad \quad \quad \quad \quad \quad \quad \quad
    \quad \quad \quad \quad \quad \quad \quad \times z\bar{z} \xi  \varepsilon_f  K_1(\varepsilon_f  r_\perp)
    Q' K_0(\varepsilon_f' r_\perp) \, , \\
    \left \langle \Mgammastar{0,\pm 1} \right \rangle_Y & \!\!= -4 \sqrt{2} N_c e^2 q_f^2 \sum_{k=0}^\infty \int_{0}^\infty \!\!\!\!\! b_\perp \der b_\perp  J_{2k}(\Delta_\perp b_\perp) \! \int_{0}^\infty \!\!\!\!\!   r_\perp \der r_\perp D_{Y,2k}(r_\perp,b_\perp) \! \int_{z} \! \frac{c_k}{2} \left[ J_{1+2k}(\delta_\perp r_\perp) + J_{1-2k}(\delta_\perp r_\perp) \right] \nonumber \\  & \quad \quad \quad \quad \quad \quad  \quad \quad \quad \quad \quad \quad \quad \quad \quad \quad \quad \quad \quad \quad \quad
    \quad \quad \quad \quad \quad \quad \quad \times z\bar{z} \xi Q K_0(\varepsilon_f  r_\perp)
    \varepsilon_f'  K_1(\varepsilon_f'  r_\perp) \, .
\end{align}
Helicity flip amplitude ($\rm{T\pm \rightarrow T\mp}$):
\begin{align}
    \left \langle \Mgammastar{\pm 1, \mp 1} \right \rangle_Y &= 8 N_c e^2 q_f^2 \sum_{k=0}^\infty \int_{0}^\infty  \!\!\!\!\! b_\perp \der b_\perp J_{2k}(\Delta_\perp b_\perp) \! \int_{0}^\infty \!\!\!\!\! r_\perp \der r_\perp  D_{Y,2k}(r_\perp,b_\perp) \! \int_{z} \! \frac{c_k}{2} \left[ J_{2-2k}(\delta_\perp r_\perp) + J_{2+2k}(\delta_\perp r_\perp) \right] \nonumber \\  & \quad \quad \quad \quad \quad \quad  \quad \quad \quad \quad \quad \quad \quad \quad \quad \quad \quad \quad \quad \quad \quad
    \quad \quad \quad \quad \quad \quad \quad\times z \bar{z} \ \varepsilon_f  K_1(\varepsilon_f  r_\perp) \varepsilon_f'  K_1(\varepsilon_f'  r_\perp) \, .
\end{align}
\end{widetext}
We have conveniently used $J_{-m}(z) = (-1)^m J_m(z)$ as needed.

For $\Delta_\perp/Q \lesssim 1$, $J_{0}(\Delta_\perp/Q) \sim 1$ and $J_{2k}(\Delta_\perp/Q) \sim 0$ for $k>0$. Thus at small to moderate values of transverse momentum of the produced vector particle, the helicity preserving and helicity flip amplitudes are most sensitive to isotropic $D_{Y,0}(r_\perp,b_\perp)$ and elliptic $D_{Y,2}(r_\perp,b_\perp)$ modes of the dipole amplitude, respectively. In this limit the polarization changing amplitude is zero. At higher momentum transfers, higher $k>2$ modes of the dipole $D_{Y,k}$ also start to contribute.

The dipole modes in the CGC proton are shown in Fig.\,\ref{fig:average_dipole_b_r} through \ref{fig:dipole_v2evol}, where we normalize the elliptic and quadrangular modes defining:
\begin{equation}
    \bar v_n = \frac{D_{Y,n}}{D_{Y,0}} \, ,
\end{equation}
with $D_{Y,n}$ introduced in Eq.\,\eqref{eq:dipole_modes}. In Fig.~\ref{fig:average_dipole_b_r} the dipole scattering amplitude averaged over the dipole orientation $\langle D_Y\rangle$ is shown as a function of dipole size $r_\perp$ and impact parameter $b_\perp$. We emphasize that when convoluted with the photon and vector meson wave functions, contributions from the largest dipole sizes are exponentially suppressed. As can be seen in Fig.~\ref{fig:average_dipole_b_r}, this also limits the impact parameter range that gives numerically important contributions to the cross-section.

The elliptic modulation $\bar v_2$ at the initial condition and after the JIMWLK evolution is shown in Figs.~\ref{fig:dipole_v2} and~\ref{fig:dipole_v2evol}. Large enough dipoles (compared to the proton size) are required to resolve the density gradients and result in non-negligible modulations. At large impact parameters large dipoles experience strong modulations, as the scattering amplitude is large when one end of the dipole hits the center of the proton and the other one is in vacuum, and vanishes when the dipole is rotated by $90$ degrees when both quarks miss the proton. However, as discussed above, contributions from this range are suppressed when convoluted with the wave functions. 
The modulations also vanish at $b=0$ where the average dipole-target interaction does not depend on the dipole orientation. The JIMWLK evolution significantly suppresses the elliptic modulation in the studied $r_\perp,b_\perp$ range, as gradients are reduced for the larger proton at smaller $x$~\cite{Schlichting:2014ipa,Mantysaari:2019csc}.

As the odderon contribution is not included, odd harmonics vanish in our setup. The higher harmonics that contribute to the cross-section at higher momentum transfers as discussed above, are found to be small, e.g. $v_4\lesssim 1\%$ in the phenomenologically important range $r_\perp,b_\perp<1\fm$, as shown in Fig.~\ref{fig:dipole_v4}.

\bibliographystyle{JHEP-2modlong.bst}
\bibliography{refs}

\end{document}